\documentclass[aps,notitlepage,nofootinbib,longbibliography,twocolumn,superscriptaddress,10pt]{revtex4-2}

\usepackage{bbold}
\usepackage{simplewick}

\usepackage[usenames,dvipsnames]{color}
\usepackage[colorlinks=true,citecolor=PineGreen,linkcolor=blue]{hyperref}

\usepackage{amsmath}
\usepackage{amsthm, amssymb}
\usepackage{physics}
\usepackage{amssymb} 
\usepackage{enumitem}
\usepackage{mathtools}
\usepackage{slashed}
\usepackage{graphicx}
\usepackage{xcolor}
\usepackage{tikz}
\usetikzlibrary{calc,fadings,decorations.pathreplacing,shapes,shapes.multipart,arrows,shapes.misc,intersections,positioning,patterns}
\usepackage{bm}
\usepackage{bbm}
\usepackage[colorlinks=true,citecolor=PineGreen,linkcolor=blue]{hyperref}
\usepackage{dsfont}
\usepackage{changepage}
\usepackage{array}
\usepackage{soul}
\usepackage[capitalise]{cleveref}
\crefname{section}{Sec.}{Secs.}
\Crefname{section}{Sec.}{Secs.}
\usepackage{xifthen}
\usepackage{xargs}
\usepackage[normalem]{ulem}
\usepackage{subcaption}
\usepackage{braket}
\usetikzlibrary{fit,positioning}

\newcommand{\ts}{\theta_{s}}
\newcommand{\hts}{\hat{\theta}_{s}}
\newcommand{\vts}{\vartheta_{s}}
\newcommand{\pc}{\varphi_{c}}

\newcommand{\ps}{\varphi_{s}}
\newcommand{\dx}{\partial_{x}}
\newcommand{\dt}{\partial_{\tau}}
\newcommand{\m}{\{m\}}
\newcommand{\hmU}{\hat{\mathcal{U}}}

\begin{document}
\title{Universal Properties of Critical Mixed States from Measurement and Feedback}
\author{Zhehao Zhang}
			\affiliation{Department of Physics, University of California, Santa Barbara, CA 93106, USA}

\author{Yijian Zou}
			\affiliation{Perimeter Institute for Theoretical Physics, Waterloo, Ontario N2L 2Y5, Canada}
            
                \author{Timothy H. Hsieh}
			\affiliation{Perimeter Institute for Theoretical Physics, Waterloo, Ontario N2L 2Y5, Canada}
            
			\author{Sagar Vijay}
			\affiliation{Department of Physics, University of California, Santa Barbara, CA 93106, USA}

\begin{abstract}

% We explore the universal features of quantum many-body states with quantum-critical correlations, which are obtained from a single round of measurements, followed by local unitary feedback, which is conditioned on non-local measurement outcomes.   mixed states obtained from a single
% round of measurement followed by feedback .
% We characterize the universality classes
% of these mixed states with defect entropy and scaling of entanglement negativity.
% Through a combination of field theory
% and numerical simulations, we discover that these mixed states exhibit a universal
% subleading term in their entropies, governed by the $g$-function. This $g$-function allows
% us to associate the critical mixed states with renormalization group fixed points.
% Specifically, we discover an example of critical mixed state with non-trivial g-function
% and logarithmic negativity scaling. 
% For a family of critical mixed states obtained from interacting spinful fermions, we analytically compute their g-functions and find a
% continuous dependency on the Luttinger parameters.

We explore the universal properties of mixed quantum matter obtained from ``single-shot” adaptive evolution, in which a quantum-critical ground-state is manipulated through a single round of local measurements and local unitary operations conditioned on spatially-distant measurement outcomes.   The resulting mixed quantum states are characterized by altered long-distance correlations between local observables, mixed-state entropy, and entanglement negativity.  By invoking a coarse-grained, continuum description of single-shot adaptation in (1+1) dimensions, we find that the extensive mixed-state entropy exhibits a sub-leading, constant correction ($\gamma$), % — which is related to the Affleck-Ludwig entropy of a boundary conformal field theory — 
while the entanglement negativity can grow logarithmically with sub-region size, with a coefficient ($\alpha$); both constants can attain universal values which are distinct from the expected behavior in any quantum-critical ground-state.  We investigate these properties in single-shot adaptation on ($i$) the critical point between a one-dimensional $Z_{2}\times Z_{2}$ symmetry-protected topological (SPT) order and a symmetry-broken state, and ($ii$) a spinful Tomonaga-Luttinger liquid.  In the former case, adaptive evolution that decoheres one sublattice of the SPT can yield a critical mixed-state in which $\alpha$ attains a universal value, which is half of that in the original state.  In the latter case, we show how adaptation -- involving feedback on the spin degrees of freedom, after measuring the local charge -- modifies long-distance correlations, and determine via an exact replica field-theoretic calculation that $\alpha$ and $\gamma$ vary continuously with the strength of feedback.  Numerical studies confirm these results.

%We explore the universal features of quantum many-body mixed states obtained from a single round of measurements followed by local unitary feedback on quantum-critical ground states. We characterize the universality classes
%of these mixed states with the subleading constant in the entanglement entropy, known as the $g$-function\ZZ{(we didn't use the name g-function in the main text)}, and coefficient $\alpha$ of the logarithmic scaling of entanglement negativity. We first consider a toy example of measurement and feedback on coupled critical Ising chains\ZZ{(maybe call it the critical cluster state?)}, where the measurement and feedback effectively decoheres one of the spin chains. We then consider a more interesting example, in which we measure the charge density and perform unitary feedback on the spin sector of the one-dimensional free \ZZ{(it can be an interacting theory in the UV)} spinful fermion critical ground state. Through a combination of field theory and numerical simulations, we show that both $g$-function and $\alpha$ continuously depend on the parameter in the feedback unitary.
\end{abstract}
\maketitle
\tableofcontents

% \textcolor{red}{1. highlight R\'{e}-1 result 2. figure include corr, g-function, negativtiy and the protocol 3. non-local classical feedback can be described by local field theory in 1d 4. directly study the effect of meas+feedback is hard, so we first purify the mixed state to a coupled-theory and then apply local decoherence to study the reduced density matrix (make this clear in the Ising example)5. different boundary states obtained by imposing the same boundary condition on different initial theory}

\section{Introduction}\label{section:intro}
 Recent advances in quantum simulators \cite{bluvstein2024logical} have opened new avenues for implementing adaptive dynamical protocols \cite{foss2023experimental, iqbal2023topological, iqbal2024non, baumer2024efficient} to prepare highly-entangled and noise-robust quantum states. These techniques, involving measurements and real-time unitary feedback, can efficiently produce certain long-range entangled quantum many-body states that are inaccessible with finite-depth unitary evolution alone. These capabilities have significant implications for quantum information processing and the simulation of quantum many-body physics, facilitating the realization and study of novel topologically-ordered and quantum-critical matter.

Numerous recent proposals \cite{ashvin_2021_measurement, ashvin_hierarchy_2022, bravyi_2022_adaptive, lee2022decoding, lu2022measurement, sahay2024classifying, sala2024quantum, stephen2024preparing, smith2024constant, zhu2022nishimori, zhang2024characterizing, lu2023mixed, kuno2024hierarchy} have investigated the efficient adaptive preparation of long-range-entangled ground-states of gapped, quantum many-body Hamiltonians, starting with a product state, or adaptive evolution to generate novel quantum error-correcting codes \cite{ hastings2021dynamically,zhang2023x,dua2024engineering,davydova2023floquet,davydova2024quantum,ellison2023floquet,zhang2024quantum}. Quantum critical states provide a distinct class of long-range-entangled quantum matter, featuring quantum correlations on all length-scales. How adaptive protocols can re-shape the universal properties of quantum-critical matter remains to be fully understood.  

\indent Some progress has been made to address this question in ``single-shot" adaptive evolution, in which a round of local measurements is performed, followed by a single round of local unitary operations, conditioned on the full set of measurement outcomes; Ref. \cite{lu2023mixed} specifically investigated how such a quantum channel could convert the universal order hidden in highly non-local observables in a quantum phase (e.g. the off-diagonal quasi-long-range order of the composite boson in a quantum Hall state, or ``string order" in certain symmetry-protected topological phases) into order that is detected in local observables.  This perspective produced specific examples of how single-shot adaptation can (i) modify the properties of quantum-critical states in one spatial dimension, and (ii) prepare quantum-critical mixed states starting from a gapped ground-state in two dimensions. %\YZ{Comment on how this approach is related to the coarse-graining approach below }
%A novel feature of these mixed-states was the coexistence of extensive entropy with quantum-critical correlations, and a logarithmic scaling of the entanglement negativity in one spatial dimension.  %The unitary feedback is designed such that the string orders in the input state are `channeled' into local correlation functions in the output mixed-state. An observation we make is that in 1+1d CFT there are local primary operators that correspond to string operators in UV. Thus when an 1d critical state serves as the input to an adaptive quantum circuit, the output would be a critical mixed state which features power-law correlation function inherited from the string operators in the input critical states. The goal of this paper is to understand the entanglement structure of these critical mixed states.
\newline
\indent In this work, we use coarse-grained, continuum descriptions of single-shot adaptive protocols to investigate the universal properties of the resulting quantum-critical mixed-states.  A single-shot, adaptive evolution which can non-trivially modify the long-distance correlations of a quantum many-body state must be a highly non-local quantum channel, in which spatially-distant measurement outcomes are used to determine the local unitary feedback.  Nevertheless, a universal understanding of its effects on quantum critical states in (1+1)-dimensions, which are described by conformal field theory (CFT), remains possible due to the fact that highly non-local operators on the lattice can admit a coarse-grained description as a local primary field in the CFT which emerges in the infrared.  This perspective allows us to study the effect of a family of adaptive protocols on quantum-critical pure states, to construct the purification of the resulting mixed-states, and then study the universal structure of the ($i$) long-distance correlations of certain local observables which are altered by measurements and feedback, ($ii$) mixed-state entropy, and ($iii$) entanglement negativity.  %\YZ{Mention: 1. Critical correlations are rearranged through measurement and feedback 2. The critical mixed state can be understood as local decoherence acting on a purification that is also critical.}

Quantum-critical mixed-states that emerge from adaptive protocols in (1+1)-dimensions can have extensive entropy, which nevertheless coexists with a logarithmic scaling of the entanglement negativity with sub-region size \cite{lu2022measurement}.  A similar scaling is observed when a quantum-critical ground-state is subject to weak, \emph{local} decoherence \cite{lee2023quantum, garratt2023measurements, zou2023channeling, sang2024approximate,yang2023entanglement,liu2024boundary, ma2023exploring,milekhin2024observable,sala2024quantum,weinstein2023nonlocality, murciano2023measurement, chen2024unconventional, myerson2023decoherence, patil2024highly}, though in that case, the long-distance properties of local correlations remain the same as the original quantum-critical wavefunction.  In this situation, a universal understanding of the structure of the entropy and entanglement negativity has been obtained.  The extensive entropy exhibits a universal constant correction ($\gamma$) which is directly related to the Affleck-Ludwig entropy\cite{affleck1991universal} of a CFT with open boundary conditions which are determined by the precise form of the decoherence.  Furthermore, the entanglement negativity exhibits a logarithmic scaling with a coefficient ($\alpha$) which is no longer directly related to the central charge of the CFT.  Quantum-critical mixed-states emerging from adaptive protocols, can be understood as arising from \emph{local} decoherence acting on a quantum-critical purification of this state.  We use this perspective to elucidate the entanglement properties of quantum-critical mixed-states that are produced from adaptation.

\begin{figure}
    \includegraphics[width=\linewidth]{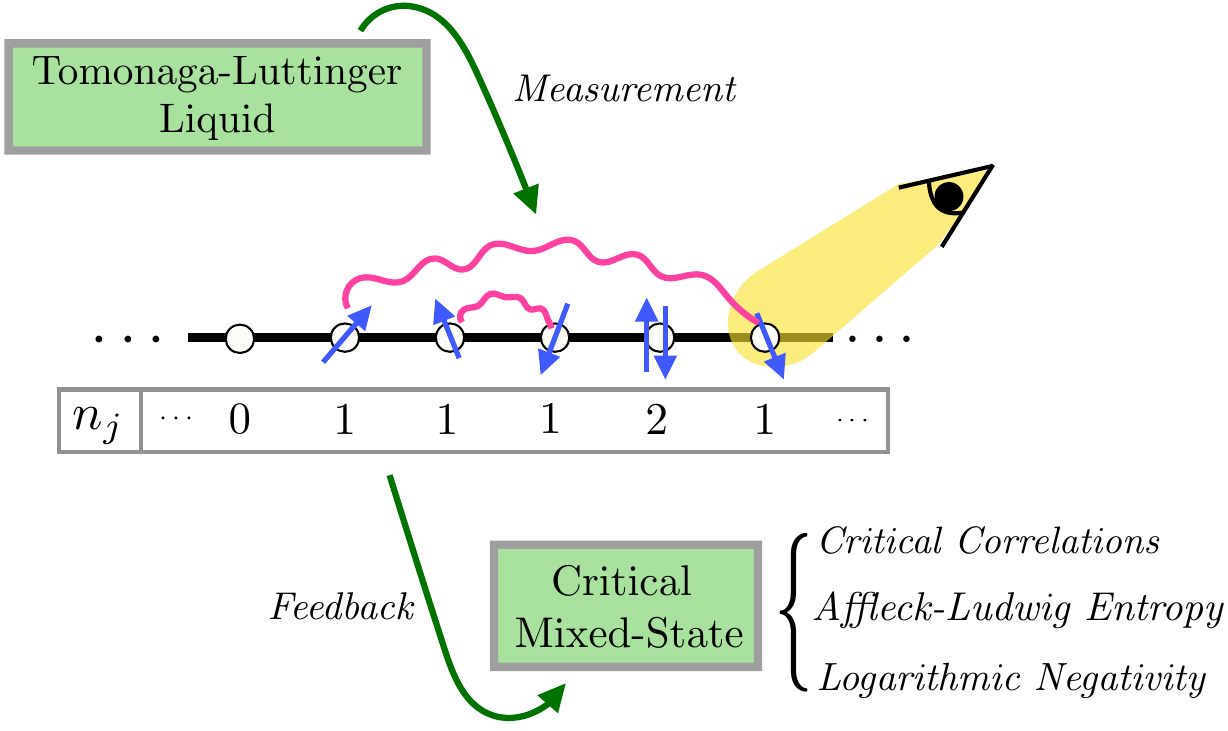}
    \caption{Starting with a critical ground-state (e.g. a spinful Tomonaga-Luttinger liquid, as shown), measurements of the local charge ($n_{j}$), followed by local unitary feedback conditioned on distant measurement outcomes can produce a quantum-critical mixed-state with universal correlations, entropy, and negativity. }
    \label{fig:feedback_TLL}
\end{figure}
        
We primarily focus on the effects of single-shot adaptation on a Tomonaga-Luttinger liquid, a strongly-interacting phase of charged, spin-(1/2) fermions in which the spin and charge degrees of freedom propagate independently.  The adaptive protocols we consider focus on projective measurements of one sector (e.g. charge), followed by unitary feedback on the other (e.g. spin).  A purification of the resulting mixed-state can be interpreted as the wavefunction of a spin-charge-coupled TLL, similar to what might be obtained in the presence of a spin-orbit interaction.  In this case, we are able to exactly determine the Affleck-Ludwig boundary entropy $\gamma$ analytically and the coefficient $\alpha$ of the logarithmic scaling of the negativity semi-analytically. We find that both coefficients continuously depend on the feedback parameter. The only known prior example with this marginal behavior in replica limit is Gaussian channel acting on a free fermion ground state \cite{zou2023channeling}, whereas here we deal with an interacting case. This example contrasts with another which we present, in which a single-shot adaptive protocol on a quantum critical state -- describing the critical point between a $Z_{2}\times Z_{2}$ symmetry-protected topological order and a trivial symmetric state -- yields a critical mixed-state in which $\gamma$ and $\alpha$ attain universal values, which differ from the original critical state.  Specifically, the sub-leading correction to the $n$-th R\'{e}nyi entropy is related to the defect entropy of four copies of the Ising CFT with a symmetry-breaking defect, while the coefficient of the logarithmic scaling of the $n$-th R\'{e}nyi negativity is one-half of that of the input critical state. These predictions are confirmed by numerical calculations.

\indent The paper is organized as follows. Section \ref{sec: reivew} reviews the framework in \cite{lu2023mixed} to characterize the output mixed state from measurement and feedback protocol and also the entanglement quantities we used to characterize critical mixed state. Section \ref{sec: coupled Ising chain} discusses a toy example on qubit chains to illustrate the strategy to describe critical mixed state. Section \ref{sec: coupled Luttinger liquids} discusses how a measurement and feedback protocol on spinful fermions can continuously changing the entanglement structure of the output mixed state. We coarse-grain the lattice protocol into a continuum-field theory description and use the effective action that governs the output mixed state to compute various correlation functions and entanglement quantities. Section \ref{sec: summary} gives a summary of the paper and provides outlook to future directions. Several appendices provide technical details and additional numerical data that complement the main text.

\section{``Single-shot" measurement and feedback} \label{sec: reivew}
Here we briefly review the protocol in \cite{lu2023mixed} for constructing quantum channels based on local quantum operations (measurement and unitary) and non-local classical communication. Given a system with two types of degrees of freedom $A$ and $B$ initialized in the state $\rho_0= \ket{\psi_0}\bra{\psi_0}$, one performs simultaneous, extensive single-site measurement on every degree of freedom in $A$. This leads to  a particular pure state $\ket{\psi_\alpha} = \frac{\hat{P}_\alpha\ket{\psi_0}}{  \sqrt{\bra{\psi_0} \hat{P}_\alpha\ket{\psi_0}}   }$ with probability $p_{\alpha} = \bra{\psi_0 } \hat{P}_{\alpha} \ket{\psi_0} $, where $\alpha$ labels the measurement outcome on $A$ and $\hat{P}_\alpha$ is the projector associated with the measurement.  For each post-measurement state, we apply a unitary $\hat{U}_\alpha$ acting on  $B$ based on the outcome $\alpha$. Note that this may require non-local classical communication since the choice of a local unitary may rely  on measurement outcomes separated by a long distance. The above measurement-feedback protocol leads to a mixed state 
\begin{equation}
\begin{split}
    \rho  &= \sum_{\alpha}  \hat{U}_\alpha \hat{P}_\alpha  \rho_0    \hat{P}_\alpha \hat{U}_\alpha^{\dagger}
\end{split}
\end{equation}
Note that $\rho$ is a classical-quantum state with respect to $A$. We will trace out $A$ and study the reduced density matrix $\rho_B$.

The aforementioned protocol may be viewed as a way to effectively implement a controlled unitary acting on the $AB$ composite system followed by tracing out $A$, which is an extensive bipartition over $AB$.  To see this, we notice that $\rho_B$ admits a purification $\ket{\psi}$ in the Hilbert space of $AB$ given by
\begin{equation}\label{eq:purification}
\rho_B = \Tr_A[\ket{\psi}\bra{\psi}] = \Tr_A[\hat{U}_{\mathrm{ctrl}}\ket{\psi_0}   \bra{\psi_0} \hat{U}_{\mathrm{ctrl}}^{\dagger}],
\end{equation}
where $\hat{U}_{\mathrm{ctrl}} = \sum_{\alpha} \hat{U}_{\alpha}  \hat{P}_{\alpha} $ takes the form of a controlled unitary with $A$ being the control and $B$ being the target.  $\hat{U}_{\mathrm{ctrl}}$ may not be realized as local unitary circuits, therefore it can significant changes the entanglement structure. 
% In particular, $U$ provides a non-local transformation on operators according to Heisenberg evolution. For instance, the expectation of an operator $O_B$ supported on $B$ in the resulting mixed state $\rho_B$ amounts to the expectation of the operator $U^\dagger O_B U$ in the input pure state $\ket{\psi_0}$. As such, this non-local unitary transformation provides a powerful way to convert hidden orders in the input state into long-range order or criticality in the density matrix $\rho_B$, as we will illustrate using various examples.
%This viewpoint of unitary transformation provides a useful way to describe the output $\rho_B$; 
This choice of purification is useful for understanding the correlations and entanglement in $\rho_B$. Since the initial state $\ket{\psi_0}$ and the purified state $\ket{\psi}$ are connected by a unitary $\hat{U}_{\mathrm{ctrl}}$, correlations and entanglement of $\rho_B $ can be understood through $\ket{\psi}$ and its parent Hamiltonian $H$. As we shall discuss in the next section, various entanglement measure of $\rho_B$ can be interpreted as entanglement measure over an extensive bipartition over $\ket{\psi}$.

With an appropriate choice of measurement and unitary feedback, the subsystem $B$ described by a reduced density matrix $\rho_B$ may exhibit various long-range quantum orders and criticality, even though the initial state $|\psi_0\rangle$ is short-range entangled. The key is that the unitary $U_{\mathrm{ctrl}}$ is nonlocal and can map local operators to nonlocal operators.
% \YZ{I don't like the word ``structure of the state". Maybe we want to say ``correlations and entanglement of $\rho_B$ can be analyzed based on the purification $|\psi\rangle$".}
\newline
\indent Throughout this work, we deal with cases where both the input state $\ket{\psi_0}$ and the purified state $\ket{\psi}$ are critical states, whose low energy theories are described by $1+1d$ CFT. In these cases, critical correlations get rearranged in a nonlocal way. We are going to discuss two examples where the reduced density matrix $\rho_B$ exhibits quantum criticality in correlation functions and entanglement properties.

\section{Characterizing a quantum-critical mixed state}\label{sec:Characterization of critical mixed state}
\indent In this section, we will give the definitions of the two entanglement quantities that we will mainly focus on in this paper.\newline
\indent \textit{R\'{e}nyi Entropy:} The simplest entanglement quantity one can consider is the R\'{e}nyi entropy $S^{(n)}(\rho_B) \coloneqq \frac{1}{1 - n} \log \Tr(\rho_B^n)$. By viewing $\rho_B$ as the reduced density matrix of $\ket{\psi}$, $S^{(n)}(\rho_B)$  measures the entanglement between subsystem $B$ and its complement, subsystem $A$. To get more intuition for this quantity, we  replace the partial trace $\Tr_A$ by a maximal depolarization channel \cite{nielsen2002quantum} on  $A$, that is,
\begin{equation}\label{eq: tr as depolarization}
    \frac{\mathbb{1}_A}{d_A} \otimes \rho_B = \mathcal{D}_A[\ket{\psi}\bra{\psi}].
\end{equation}
The partition function $Z^{(n)}(\rho_B)\equiv \Tr(\rho_B^n)$ may now be written in terms of the pure state $\ket{\psi}$ and the depolarization channel $\mathcal{D}_A$ as 
\begin{equation}\label{eq:partition function in terms of boundary state}
\begin{split}
    \frac{1}{d_A^{n-1}}Z^{(n)}(\rho_B) &= \Tr\left[\mathcal{D}_A^{\otimes n}
    [\ket{\psi}\bra{\psi}^{\otimes n}] \mathcal{T} \right] \\
    &= \Tr\left[
    \ket{\psi}\bra{\psi}^{\otimes n} \mathcal{D}_A^{*\otimes n}[\mathcal{T}] \right]\\
    &= \langle\braket{(\psi \otimes \psi^*)^{\otimes n}|\mathcal{D}_A^{*\otimes n}[\mathcal{T}]}\rangle,
\end{split} 
\end{equation}
where $\mathcal{T}$ is an operator that performs a cyclic shift of the replicas\footnote{The operator $\mathcal{T}$ acts on a state of the $n$ replicas  $\ket{i_{1}}\otimes\cdots\otimes\ket{i_{n}}$, as  $\mathcal{T}\ket{i_{1}}\otimes\cdots\otimes\ket{i_{n}} = \ket{i_{n}}\otimes\ket{i_{1}}\otimes\cdots\otimes\ket{i_{n-1}}$.}. To go from the first to the second line, we act the dual channel $\mathcal{D}^{*\otimes n}_A$ on $\mathcal{T}$.\footnote{Notice that $\mathcal{D}^{*\otimes n}_A$ =  $\mathcal{D}^{\otimes n}_A$, so $\mathcal{D}_A^{*\otimes n}[\mathcal{T}]$ at maximal strength is reduced to $\mathbb{1}_A \otimes \mathcal{T}_B$.} To go from the second to the third line we apply the Choi–Jamiołkowski (CJ) map~\cite{choi1975completely, jamiolkowski1972linear} that maps $\ket{\psi}\bra{\psi}^{\otimes n}$ to $2n$ copies of $\ket{\psi}$ and $\mathcal{D}_A^{*\otimes n}[\mathcal{T}]$ to certain boundary state $\ket{\mathcal{D}_A^{*\otimes n}[\mathcal{T}]}\rangle$ in the $2n$-replica Hilbert space. The benefit of introducing the depolarization channel is that one knows its Kraus operators on the lattice, which can be mapped to CFT operators that facilitates the analysis of the boundary state $\ket{\mathcal{D}_A^{\otimes n}[\mathcal{T}]}\rangle$. %Notice that in the above equation we omit the contribution from $\mathbb{1}_A$ which is simply a constant term that scales as $e^{\abs{\mathcal{H}_A}}$. 
Now $S^{(n)}(\rho_B)$ can be conveniently  interpreted as the boundary entropy of $2n$ copies of the CFT. As we take the thermodynamic limit $L \xrightarrow{} \infty$, we expect the R\'{e}nyi entropy has the following scaling \cite{zou2023channeling}
\begin{equation} \label{eq: entropy_universal_scaling_law}
    S^{(n)}(\rho_B) = \beta^{(n)} L + \gamma_n,
\end{equation}
where $\beta^{(n)}$ is the volume-law term coefficient and $\gamma_n$ is a system size independent constant term. $\gamma_n$ is usually referred as the Affleck-Ludwig boundary entropy \cite{affleck1991universal} and depends on the universality class of the conformal boundary states $\ket{\mathcal{D}_A^{*\otimes n}[\mathcal{T}]}\rangle$. $\gamma_n$ is also related to the fixed point value the \textit{g-function}~\cite{casini2016g, cuomo2022renormalization}, which monotonically decreases under boundary renormalization group (RG) flow.   \newline
\indent \textit{Entanglement Negativity:} Another entanglement quantity we will be looking at is the R\'{e}nyi entanglement negativity of a subsystem $B_1$ of $B$, defined through $\mathcal{N}_{B_1}^{(n)}(\rho_B) \coloneqq \frac{1}{1 - n} \log \frac{\Tr(\{\rho_B^{T_{B_1}}\}^n)}{\Tr(\rho_B^n)}$\cite{vidal2002computable}, where $(\cdot)^{T_{B_1}}$ denotes taking partial transpose of the density matrix with respect to the subsystem $B_1$. $\mathcal{N}_{B_1}^{(n)}(\rho_B)$ measures the quantum correlations between subsystem $B_1$ and its complement $\overline{B}_1$. By appealing to the purification $\ket{\psi}$ of $\rho_B$, we can similarly express $\Tr(\{\rho_B^{T_{B_1}}\}^n)$ as 
\begin{equation}
\label{eq:neg_overlap}
    \Tr(\{\rho_B^{T_{B_1}}\}^n) =  \langle \braket{ (\psi \otimes \psi^*)^{\otimes n} | \mathcal{D}_A^{*\otimes n}[\mathcal{T}_{A\overline{B}_1} \mathcal{T}_{B_1}^{-1}]}\rangle,
\end{equation}
where $\mathcal{T}^{-1}_{B_1}$ performs an anti-cyclic shift of the replicas within subsystem $B_1$ and $\mathcal{T}_{A\overline{B}_1}$ performs a cyclic shift of the replicas within the compliment of $B_1$ in $AB$. Note that if we choose $B_1$ to be a single interval in $B$, $B_1$ is not a continuous interval in $AB$. Therefore, Eq.~\eqref{eq:neg_overlap} is not simply the entanglement negativity for a single interval $B_1$ in the CFT ground state $\ket{(\psi \otimes \psi^*)^{\otimes n}}\rangle$. Nevertheless, $ \mathcal{D}_A^{*\otimes n}[\mathcal{T}_{A\overline{B}_1}]$ is still translation invariant if one coarse grain $A$ and $B$. Thus, the state still corresponds to a conformal boundary condition and a boundary condition changing operator must be inserted at the intersection between $B_1$ and $\bar{B}_1$. Therefore, $\mathcal{N}_{B_1}^{(n)}(\rho_B)$ still exhibits the scaling of entanglement negativity for a single interval as in  pure CFT:
\begin{equation}\label{eq: negativity_scaling}
    \mathcal{N}_{B_1}^{(n)}(\rho_B) = \alpha_n \log(\frac{L}{\pi}\sin(\pi \frac{\abs{B_1}}{\abs{B}})),
\end{equation}
where we find the log coefficient $\alpha_n$ corresponds to the scaling dimension of the boundary condition changing operator. Indeed, we find that while $\alpha_n$ is independent of system size,it depends \textit{not} solely on the central charge of $\ket{\psi}$ and the replica index $n$. In contrast, for 1+1d  CFT ground state with central charge $c$, the log coefficient is given by\cite{calabrese2013entanglement}
\begin{equation}
    \alpha_n = \frac{c}{12}(n - \frac{1}{n}),~\text{for $n$ is odd},
\end{equation} which is a function only depends on $c$ and $n$.
\begin{figure}
\centering
    \begin{subfigure}{0.3\textwidth}
        \centering
        \includegraphics[width=\linewidth]{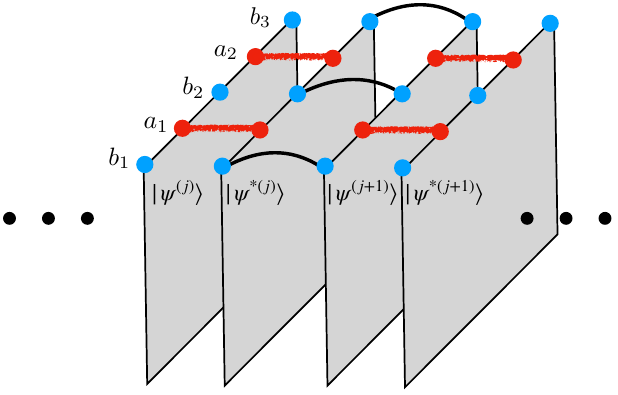}
        \caption{}
        \label{fig:entropy illustration}
    \end{subfigure}
    \begin{subfigure}{0.3\textwidth}
        \centering
        \includegraphics[width=\linewidth]{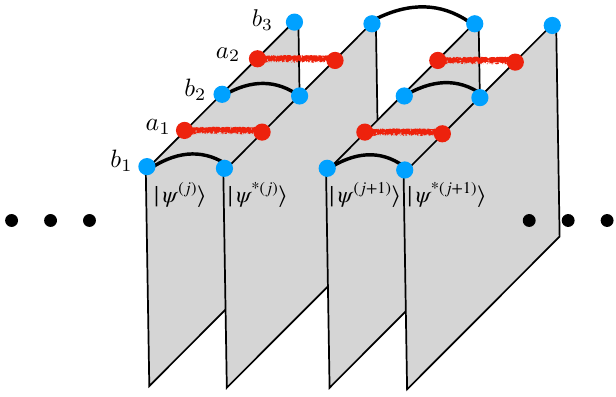}
        \caption{}
        \label{fig:entropy illustration}
    \end{subfigure}
    \caption{The figures illustrate how the (a) R\'{e}nyi entropy in Eq.\eqref{eq:partition function in terms of boundary state} and (b) R\'{e}nyi negativity in Eq.\eqref{eq:neg_overlap} of $\rho_B$ is related to boundary conditions of multiple replicas of the purified state $\ket{\psi}$. The blue (red) dots represent degrees of freedom in $B(A)$ and are labeled by $\{b_i\}(\{a_i\})$. The black lines connecting different replicas denote the degrees of freedom on the two ends are identified. The thick red line denotes $\mathcal{D}_A$, which couples the $A$ degrees of freedom. In the lower figure, we take the subsystem $B_1 = \{b_1, b_2\}$.}
\end{figure}
% \indent Mixed-state $\rho_B$ obtained from measurement and feedback can be viewed as taking partial trace over subsystem $A$ of the purified state $\ket{\psi}$. Alternatively, we can treat partial trace over $A$ as a maximal depolarization channel $\mathcal{D}_A$ over $A$, that is 
% \begin{equation}
%     \mathcal{D}_A(p = 1)[\ket{\psi} \bra{\psi}] = \prod_{i \in A} \mathcal{D}_{A,i}[\ket{\psi} \bra{\psi}] = 1_A \otimes \rho_B,
% \end{equation}
% where $\mathcal{D}_{A,i}$ are depolarization channel acting on local degrees of freedom in $A$. As in \cite{zou2023channeling}, local depolarization channels in the replica field theory can be interpreted as local perturbations acting on the boundary. As we tune the strength $p$ of the channel there are two easily identified RG fixed point, which are $p = 0$ and $p = 1$. The case when $p = 0$ is simply no depolarization and the g-function for a pure state is $0$. The case when $p = 1$, we can think of the strength of the perturbation is turned to infinity and thus would drive the boundary state to an RG fixed point. 

\section{Critical cluster chain} \label{sec: coupled Ising chain}
\subsection{Review: measurement and feedback in the $\mathbb{Z}_2 \times \mathbb{Z}_2$ SPT phase}
Here we review the protocol in \cite{lu2023mixed} that channels the string-order parameter in one-dimensional $\mathbb{Z}_2 \times \mathbb{Z}_2$ SPT into long-range GHZ order in the mixed state. The Hamiltonian describing states in the $\mathbb{Z}_2 \times \mathbb{Z}_2$ SPT phase is given by
\begin{equation}
\begin{split}
    H_{\mathrm{SPT}}(g) = &-\sum_{i} \hat{Z}_{A,i} \hat{X}_{B,i}\hat{Z}_{A,i+1} - \sum_{i}\hat{Z}_{B,i}\hat{X}_{A,i+1}\hat{Z}_{B,i+1}\\
    &+ g \hat{V}
\end{split}
\end{equation}
where the Hamiltonian is defined on a one-dimensional periodic chain. The lower index $A$ ($B$) labels the sublattice and $i$ labels the spatial index on the sublattice. $\hat{V}$ represents perturbations that respect the $\mathbb{Z}_2 \times \mathbb{Z}_2$ symmetry (which is defined below) and $g$ controls the strength of the perturbation. The Hamiltonian $H_{\mathrm{SPT}}(g)$ respects the $\mathbb{Z}_2 \times \mathbb{Z}_2$ symmetry generated by
\begin{equation}
    \hat{U}_A = \prod_i \hat{X}_{A,i},~~~\hat{U}_B = \prod_i \hat{X}_{B,i}.
\end{equation}The groundstate, denoted as $\ket{\psi_0}$, features long-range order in the so-called string order parameter\cite{pollmann2012detection} given by
\begin{equation}
\begin{split}
    &\lim_{\abs{j - i} \xrightarrow{} \infty} \bra{\psi_0} \hat{Z}_{A,i} \left(\prod_{i\leq k < j} \hat{X}_{B,k} \right) \hat{Z}_{A,j} \ket{\psi_0}  > 0\\
    &\lim_{\abs{j - i} \xrightarrow{} \infty} \bra{\psi_0} \hat{Z}_{B,i} \left(\prod_{i < k \leq j} \hat{X}_{A,k} \right) \hat{Z}_{B,j} \ket{\psi_0}  > 0.
\end{split}
\end{equation}
Notably, in the absence of perturbations, the Hamiltonian $H_{\mathrm{SPT}}(g=0)$ reduces to the parent Hamiltonian of the one-dimensional cluster state\cite{raussendorf2001one}.\newline
\indent Given $\ket{\psi_0}$ in the $Z_2 \times Z_2$ phase, we first measure Pauli-X for every site in $A$ sublattice and denote the measurement outcome of $\hat{X}_{A,i}$ by  $\alpha_i$. Defining $\alpha= \{ \alpha_i    \}$ as the collection of outcomes, one obtains a post-measurement state $\left\{ \frac{\hat{P}_\alpha\ket{\psi_0}   }{ \sqrt{  \bra{ \psi_0 } \hat{P}_\alpha  \ket{ \psi_0  }    }     }\right\} $ 	with probability $p_\alpha   = \bra{ \psi_0 } \hat{P}_\alpha  \ket{ \psi_0  }  $, where the projector  $\hat{P}_\alpha \equiv  \prod_{i} \frac{1+ \alpha_i \hat{X}_{A,i}}{2}$. For a post-measurement pure state with outcome $\alpha$, we apply a unitary $\hat{U}_\alpha$ on $B$ sublattice:

	\begin{equation}\label{eq:1d_spt_correct}
		\hat{U}_\alpha = \prod_{ i  }  \hat{X}_{B,i}^{ \frac{1- \prod_{ j=1, 2, \cdots }^i \alpha_j  }{2}}. 
	\end{equation}
	In other words, $\hat{X}_{B,i}$, a Pauli-X on $B$ sublattice, is applied when there is an odd number of outcome $-1$ from the site $(A,1)$ to the site $(A,i)$. Correspondingly, one finds $\hat{U}^{\dagger}_{\alpha}   \hat{Z}_{B,i} \hat{Z}_{B,j} \hat{U}_{\alpha}  =  \hat{Z}_{B,i} \left(\prod_{k=i+1 }^{j}   \alpha_{k}  \right)  \hat{Z}_{B,j}$. The overall measurement and unitary operation lead to the  mixed state
    $\rho  = \sum_{\alpha}  \hat{U}_\alpha \hat{P}_\alpha  \rho_0    \hat{P}_\alpha \hat{U}_\alpha^{\dagger}$. The long-range order can be diagnosed by the two-point $ZZ$ correlation on $B$ sublattice:

	\begin{equation}\label{eq:1d_spt_derivation}
		\begin{split}
			\tr[   \rho  \hat{Z}_{B,i} \hat{Z}_{B,j}    ]  &= \sum_\alpha   \bra{\psi_0}     \hat{P}_\alpha \hat{U}_\alpha^{\dagger}  \hat{Z}_{B,i} \hat{Z}_{B,j} \hat{U}_\alpha \hat{P}_\alpha \ket{ \psi_0} \\
			&    =  \sum_\alpha   \bra{\psi_0}     \hat{P}_\alpha    \hat{Z}_{B,i} \left(\prod_{k=i+1 }^{j}   \alpha_{k}  \right)  \hat{Z}_{B,j}   \hat{P}_\alpha \ket{ \psi_0} \\
			&    = \bra{  \psi_0} \hat{Z}_{B,i}  \left( \prod_{k=i+1 }^{j}   \hat{X}_{A,k} \right)  \hat{Z}_{B,j}  \ket{\psi_0}
		\end{split}
	\end{equation}
	where to go from the first line to the second line we expand the mixed state $\rho$ in terms of the input state $\ket{\psi_0}$ and the measurement-feedback operator. From the second to the third line,  we have used the fact that $\hat{P}_{\alpha} \left(\prod_{k} \alpha_k \right) \hat{P}_{\alpha}= \hat{P}_{\alpha} \left(\prod_{k} \hat{X}_{A,k} \right) \hat{P}_{\alpha}$
    % \YZ{Better to say explicitly: where to go from the second line to the third line, we have used $\sum_{\alpha}P_{\alpha} \prod_{k} \alpha_k = \prod_{k} X_k$. Also mention how to get the second line from the  first line }
    and $\sum_{\alpha} \hat{P}_{\alpha}=1$. Therefore, the two-point function in $\rho$ is exactly the string order in the initial state.\newline
    \indent Our protocol can be viewed as realizing a controlled unitary $\hat{U}_{\mathrm{ctrl}}= \sum_{\alpha} \hat{P}_{\alpha} \hat{U}_{\alpha}$ acting on the $AB$ composite system followed by tracing out $A$. With $\hat{U}_{\mathrm{ctrl}}= \sum_{\alpha} \hat{P}_{\alpha} \hat{U}_{\alpha}$ ($\hat{U}_{\alpha}$ defined in Eq.\eqref{eq:1d_spt_correct}), one derives the transformation rule for operators under the conjugation of $\hat{U}_{\mathrm{ctrl}}$ (see Appendix B of \cite{lu2023mixed} for details).
	
	\begin{equation} \label{eq:U_transform}
		\begin{split}
			&  \hat{X}_{A,i}	  \to   \hat{X}_{A,i},  \quad  \hat{X}_{B,i}	  \to   \hat{X}_{B,i},        \\
			&  \hat{Z}_{A,i} \hat{Z}_{A,i+1}	  \to   \hat{Z}_{A,i}\hat{X}_{B,i} \hat{Z}_{A,i+1} , \\
			& \hat{Z}_{B,i} \hat{Z}_{B,i+1}	  \to   \hat{Z}_{B,i}\hat{X}_{A,i+1} \hat{Z}_{B,i+1}   .
		\end{split}
	\end{equation}
We make the following oberserations: (i) Pauli-X is invariant since $\hat{U}$ is diagonal in X basis, and (ii) neighboring $ZZ$ on one sublattice is attached with a Pauli-X on another sublattice in between two Pauli-Zs. This can be intuitively understood because the unitary feedback is designed to transform the product of two Pauli-Zs on sublattice $B$ with a sign that depends on the product of measurement outcomes on $A$ between these two Pauli-Zs. As a result, the state is transformed into a spontaneous symmetry broken state with $\langle Z_{b,i} Z_{b,j}\rangle = O(1)$.
This is akin to the Kennedy-Tasaki transformation \cite{Kennedy_Tasaki_1992,Kennedy_Tasaki_1992_prb}, which transforms a Haldane spin-1 chain \cite{haldane_spin_chain_1983} with $\mathbb{Z}_2 \times  \mathbb{Z}_2$ SPT order to two spontaneous $\mathbb{Z}_2$ symmetry-breaking orders.\newline

% \YZ{Here should be the end of a subsection, which we could name ``Review of measurement and feedback in the cluster state" }

% \YZ{We begin a new subsection ``Measurement and feedback on the coupled critical Ising chain"}
\subsection{Measurement and feedback in the critical cluster state}
Now we consider the same measurement feedback protocol acting on a closely related critical state. As we will show, the protocol can output a mixed state with volume-law entropy coexisting with critical (algebraic) long-range order. This occurs when applying our measurement-feedback channel to a critical state, whose parent Hamiltonian $H_0$ is obtained by driving the cluster state Hamiltonian $H_{\mathrm{SPT}}(g=0)$ to a critical point:
% \YZ{We add a perturbation to the cluster state Hamiltonian (should have appeared beforehand) and drive it to a quantum critical point,}
\begin{equation}\label{eq: H_critical_cluster_state}
    \begin{split}
        H_0  = & -\sum_{i}\hat{Z}_{A,i} \hat{X}_{B,i} \hat{Z}_{A,i+1 }  -\sum_{i} \hat{Z}_{B,i} \hat{X}_{A,i+1} \hat{Z}_{B,i +1 } \\
			&		    -  \sum_{i}  \hat{Z}_{A,i} \hat{Z}_{A,i+1 }   -  \sum_{i}  \hat{Z}_{B,i}\hat{Z}_{B,i+1 }.
    \end{split}
\end{equation}
The purified Hamiltonian $H$ can be obtained by applying transformations in Eq. \eqref{eq:U_transform} to $H_0$ in Eq.\eqref{eq: H_critical_cluster_state}. And we find that $H = \hat{U}_{\mathrm{ctrl}} H_0 \hat{U}_{\mathrm{ctrl}}^{\dagger} = H_0$, which means that $H_0$ is self-dual under $\hat{U}_{\mathrm{ctrl}}$. To characterize the output reduced density matrix $\rho_B$ from the measurement and feedback channel, it suffices to find the groundstate of $H$ or equivalently $H_0$.\newline
\indent We notice that the $H_0$ (equivalently $H$) in Eq.\eqref{eq: H_critical_cluster_state} can be mapped to two decoupled critical Ising chains on $A$ and $B$ sublattices under a unitary transformation 
\begin{equation}
\begin{split}
    \hat{U}_{\mathrm{CZ}}\, H \,\hat{U}_{\mathrm{CZ}}^{\dagger} &= \hat{U}_{\mathrm{CZ}}\, H_0\, \hat{U}_{\mathrm{CZ}}^{\dagger}\\
    &= -\sum_{i} (\hat{X}_{A,i} + \hat{Z}_{A,i} \hat{Z}_{A,i+1} )\\
    &~~~- \sum_{i}(\hat{X}_{B,i} + \hat{Z}_{B,i} \hat{Z}_{B,i+1}),
\end{split}
\end{equation}
where $\hat{U}_{\text{CZ}}=\prod CZ_{(A,i), (B,i) }CZ_{(A,i), (B,i+1) }$ and $CZ$ is the control-Z gate acting on neighboring sites.
% \YZ{State it here, and write down the output Hamiltonian. Define $|CFT_A\rangle$ and $|CFT_B\rangle$})
If we denote the groundstate wavefunction of the critical Ising chain as $\ket{\mathrm{CFT}}$, the groundstate wavefunction $\ket{\psi}$ of $H$ can be written as	
	\begin{equation}\label{eq:coupled_Ising_cft}
		\ket{\psi} = \hat{U}_{\text{CZ}} \ket{\text{CFT}}_A \otimes \ket{\text{CFT}}_B. 
	\end{equation}  
 $\ket{\text{CFT}}_{A/B}$ denotes the ground state of the critical Ising chain on $A/B$ sublattice. After the measurement and feedback protocol, the corresponding mixed state $\rho_B= \Tr_A \ket{ \psi }\bra{\psi}$ exhibits quantum criticality diagnosed by certain operators. For example, since $\hat{U}_{\text{CZ}}$ commutes with Pauli-Zs, the two-point $ZZ$ function is given by the single Ising critical chain, which exhibits an algebraic decay:  $\Tr[\rho_B \hat{Z}_{B,i}  \hat{Z}_{B,j} ] =   \bra{\text{CFT}}_B \hat{Z}_{B,i } \hat{Z}_{B,j} \ket{\text{CFT}_B } \sim  \frac{1}{\abs{i-j}^{\eta}  }   $ with $\eta=1/4 $ being a critical exponent in 1+1D Ising CFT. On the other hand, following a similar calculation in Eq. \eqref{eq:1d_spt_derivation} we can show that the correlation function of the disorder operator
 % \YZ{This is two-point correlation function of the disorder operator, not the disorder operator itself}
 $\hat{X}_{B,i}\hat{X}_{B,i+1}...\hat{X}_{B,j}$ evaluated with respect to $\rho_B$ maps to
 \begin{equation}
 \begin{split}
     &\Tr[\rho_B \hat{X}_{B,i}\hat{X}_{B,i+1}...\hat{X}_{B,j}]\\
     &= \expval{\hat{Z}_{A,i} \hat{Z}_{A,j+1}}_A   \expval{\hat{X}_{B,i}\hat{X}_{B,i+1}...\hat{X}_{B,j} }_B \\
     &\sim \frac{1}{\lvert i - j \rvert^{2\eta}}.
 \end{split}
 \end{equation} In the second line $\langle \cdot \rangle_{A(B)}$ denotes expectation value evaluated with respect to $\ket{\mathrm{CFT}}_{A(B)}$. To get the third line, we use Krammers Wannier (KW) duality (see e.g. \cite{seiberg2024majorana}) that maps $\hat{X} _{B,i}\hat{X}_{B,i+1}...\hat{X}_{B,j}$ to $\hat{Z}_{B,i} \hat{Z}_{B,j}$.  Importantly, we note that the disorder operator is distinct from a single pure Ising CFT, where $ \expval{\hat{X} _{B,i}\hat{X}_{B,i+1}...\hat{X}_{B,j}}  \sim \frac{1}{\lvert i - j \rvert^{\eta}}$.
 %\YZ{Because it is mapped to ....}. 
\subsection{Field-theoretic argument in the doubled Hilbert space}
In the previous section, we see that $\rho_B$ exhibits different critical exponents compared to Ising CFT, then a natural question one can ask is do they also differ in entanglement properties? Specifically, we will consider the R\'{e}nyi entropy and R\'{e}nyi negativity introduced in Sec.\ref{sec:Characterization of critical mixed state}.\newline 
\indent To answer the above question, we would like to appeal to the doubled Hilbert space picture \cite{jamiolkowski1972linear, choi1975completely} and use field-theoretic argument to study the entanglement properties of $\rho_B$. The purified state density matrix $\rho = \ket{\psi}\bra{\psi}$ (for $\ket{\psi}$ given in \eqref{eq:coupled_Ising_cft}) is mapped to a state $|\rho \rangle \rangle$ in the doubled Hilbert space, which is given by
\begin{equation}\label{eq:double Hilbert space coupled Ising}
    |\rho_0 \rangle \rangle = \hat{U}_{CZ}^{(1)} \hat{U}_{CZ}^{(2)} \ket{\text{CFT}}_A^{(1)} \ket{\text{CFT}}_A^{(2)} \ket{\text{CFT}}_B^{(1)} \ket{\text{CFT}}_{B}^{(2)},
\end{equation}
where for each critical Ising chain wavefunction the lower index $A/B$ labels the sublattice and the upper index $(1)/(2)$ labels the replica index. Similarly, $\hat{U}_{CZ}^{(1)/(2)}$ only couples $A$ and $B$ sublattices with the same replica index. Applying CJ map to both sides of Eq.\eqref{eq: tr as depolarization}, we get
\begin{equation}\label{eq:rho_B in doubled-Hilbert space}
    \ket{\frac{\mathbb{1_A}}{d_A}\otimes\rho_B}\rangle = \hat{\mathcal{D}}_A \ket{\rho_0}\rangle,
\end{equation}
where $\hat{\mathcal{D}}_A \equiv \prod_i \hat{\mathcal{D}}_{A,i}$ is mapped from the maximal depolarization channel $\mathcal{D}_A[\cdot]$ and is given by

\begin{equation}
    \hat{\mathcal{D}}_{A,i} = \exp{\lambda \left(\hat{X}_{A,i}^{(1)}\otimes \hat{X}_{A,i}^{(2)} + \hat{Y}_{A,i}^{(1)} \otimes \hat{Y}_{A,i}^{(2)} + \hat{Z}_{A,i}^{(1)} \otimes \hat{Z}_{A,i}^{(2)}\right)}
\end{equation}
where the parameter $\lambda$ controls the strength of the depolarization and $\lambda \xrightarrow{} \infty$ corresponds to the maximal depolarization channel.\newline
\indent To facilitate our analysis, we would like to view $\hat{\mathcal{D}}_A(\lambda \xrightarrow{} \infty)$ in Eq.\eqref{eq:rho_B in doubled-Hilbert space} as a strong boundary interaction\footnote{$\hat{\mathcal{D}}_A(\lambda)$ has a support on 1+0d. In the path integral picture, if we do a wick-rotation, $\hat{\mathcal{D}}_A$ would map to a 0+1d operator which we can think of as a boundary interaction.} turning on between four coupled Ising CFT and driving to the boundary state $\ket{\frac{\mathbb{1_A}}{d_A}\otimes\rho_B}\rangle$. We can perturbatively analyze the universality class of this boundary state by weakening the strength of $\hat{\mathcal{D}}_A$ to $\lambda \ll 1$. The partition function associated with the boundary perturbation $\hat{\mathcal{D}}_A(\lambda \ll 1)$ is given by
\begin{widetext}
\begin{equation}\label{eq: Ising partition function}
\begin{split}
    Z^{(2)}(\lambda) &\equiv \langle \bra{ \rho_0} \hat{\mathcal{D}}_A^{\dagger} \hat{\mathcal{D}}_A \ket{\rho_0}\rangle \\
    &= \bra{\text{CFT}}^{\otimes 4}  \hat{U}_{CZ}^{(1)} \hat{U}_{CZ}^{(2)} \hat{\mathcal{D}}^{\dagger} \hat{\mathcal{D}} \hat{U}_{CZ}^{(1)} \hat{U}_{CZ}^{(2)} \ket{\text{CFT}}^{\otimes 4} \\
    &= \bra{\text{CFT}}^{\otimes 4} \prod_{i} \exp{2\lambda \left[\hat{Z}_{B,i-1}^{(1)} \hat{Z}_{B,i-1}^{(2)} \left( \hat{X}_{A,i}^{(1)}\hat{X}_{A,i}^{(2)} + \hat{Y}_{A,i}^{(1)}\hat{Y}_{A,i}^{(2)} \right) \hat{Z}_{B,i}^{(1)} \hat{Z}_{B,i}^{(2)} + \hat{Z}_{A,i}^{(1)}\hat{Z}_{A,i}^{(2)} \right]  }\ket{\text{CFT}}^{\otimes 4}.
\end{split}
\end{equation}
\end{widetext}
In the above equation, we use $\ket{\text{CFT}}^{\otimes 4}$ as an abbreviation of the four couples of Ising CFT appeared in \eqref{eq:double Hilbert space coupled Ising}. From the second to the third line, we conjugate the $\hat{U}_{CZ}^{(1)/(2)}$ operators to $\hat{\mathcal{D}}^{\dagger}_A \hat{\mathcal{D}}_A$. $Z^{(2)}(\lambda)$ now looks like the partition function of four decoupled critical Ising chains with a defect operator inserted. One can then analyze the scaling dimension of the defect operator to study the boundary state it drives to.\newline
\indent To proceed, we first apply bosonization to four independent copies of critical Ising chains.  We identify $\ket{\text{CFT}_A^{(1)}} \ket{\text{CFT}_A^{(2)}}$ as the ground state of a free compact boson $\mathbb{Z}_2$ orbifold $\phi_A$ with Luttinger parameter $K_A=1/2$ \cite{klemm1990orbifolds, ginsparg1988applied}. Similarly, $\ket{\text{CFT}_B^{(1)}} \ket{\text{CFT}_B^{(2)}}$ is identified another compact boson $\phi_B$ with Luttinger parameters $K_B = K_A = \frac{1}{2}$. The operator
$\hat{Z}_{A,i}^{(1)}\hat{Z}_{A,i}^{(2)}$ is mapped to the operator $\cos(\phi_A(x,0))$ in the compact boson and has scaling dimension $1/4$.
% Then the third term in the defect operator is identified as
% \begin{equation}
%     \begin{split}
%        &\lambda' Z_{A,i}^{(1)}Z_{A,i}^{(2)} \sim g_1  \cos(\phi_A(x, 0)),
%     \end{split}
% \end{equation}
Such an operator is a relevant perturbation and would gap out the boson $\phi_A$. Indeed, it is the same ferromagnetic defect line in Ising CFT, and would flow to the ferromagnetic boundary condition with a boundary degeneracy of  $2$~\cite{oshikawa1997boundary}. The first term does not alter the boundary universality class since $\langle X^{(1)}_A\rangle = \langle X^{(2)}_A\rangle = \langle Y^{(1)}_A\rangle = \langle Y^{(2)}_A\rangle = 0$ on the ferromagnetic defect.\newline
%\indent The first two terms can be showed to be irrelevant perturbation in the following way. Since the lattice operator $Z_i Z_{i+1}$ is identified as $I - \mathcal{E}$ in Ising CFT, where $\mathcal{E}$ is the energy density primary operator. The leading nontrivial contribution to $\hat{Z}_{B,i-1}^{(1)} \hat{Z}_{B,i-1}^{(2)}\hat{Z}_{B,i}^{(1)} \hat{Z}_{B,i}^{(2)}$ is   
% $ \mathcal{E}_B^{(1)} + \mathcal{E}_B^{(2)}$, which has scaling dimension 1 and is thus a marginal operator. $\left(\hat{X}_{A,i}^{(1)}\hat{X}_{A,i}^{(2)} + \hat{Y}_{A,i}^{(1)}\hat{Y}_{A,i}^{(2)} \right)$ commutes with $\hat{Z}_{B,i-1}^{(1)} \hat{Z}_{B,i-1}^{(2)}\hat{Z}_{B,i}^{(1)} \hat{Z}_{B,i}^{(2)}$ so that the scaling dimension of their product is the sum of each scaling dimension, which would be greater than 1 \YZ{Say precisely what the scaling dimension is}. As a result, the first two terms are irrelevant perturbations and would leave $\phi_A$ and $\phi_B$ unchanged. \newline
\indent Based on the above perturbative analysis, we expect at any finite $\lambda$, the copy $A$ would flow to the ferromagnetic boundary state while the copy $B$ remains unchanged. The ferromagnetic boundary condition on the $A$ sublattice would contribute a universal subleading term to the Affleck-Ludwig entropy $\gamma_2 = -\log 2$ (coming from the boundary degeneracy). The only quantum correlations are within $\ket{\text{CFT}_B^{(1)}} \ket{\text{CFT}_B^{(2)}}$, and thus we expect the log coefficient $\alpha_n$ of the bipartite R\'{e}nyi entanglement negativity is the same as that of one copy of the Ising model ground state, i.e. $\alpha_n = \frac{1}{24}(n - \frac{1}{n})$\cite{calabrese2012entanglement} for $n \geq 3$ and $n$ is odd. We numerically verify the perturbative analysis in this section using tensor network techniques in Appendix. \ref{app:numerics}.\newline
\indent 
At $\lambda \xrightarrow{} \infty$, however, the perturbation theory does not directly apply. Instead, we observe that $\gamma_2$ changes to approximately $0.1$ while $\alpha_3 \approx 1/9$ is the same as finite $\lambda$(see Appendix. \ref{app:numerics}). The discontinuous jump of $\gamma_2$ indicates $\lambda = \infty$ is an unstable fixed point, as also noted by Ref.~\cite{zou2023channeling} which studied the effect of depolarization channel to a single copy of critical Ising chain. Yet,  $\langle X^{(1)}_A\rangle = \langle X^{(2)}_A\rangle = \langle Y^{(1)}_A\rangle = \langle Y^{(2)}_A\rangle = 0$ still holds for the unstable fixed point $\lambda = \infty$, which indicates that the defect has a trivial effect on the subsystem $B$. Thus, we still expect $\alpha_n = \frac{1}{24}(n - \frac{1}{n})$ at the fixed point.
%At any finite strength of the dephasing, $\gamma_2$ would equal to $-\log 2$ in the thermodynamic limit, indicating the state is driven to Ising CFT ferromagntic boundary condition. However, at the maximal Pauli-Z dephasing point, the resulting state is proportional to identity and results in $\gamma_2 = 0$.

\section{Spinful Tomonaga-Luttinger Liquid (TLL)}
\label{sec: coupled Luttinger liquids}
In the previous example, we see that the measurement and feedback protocol simply removes the quantum entanglement in one copy of the Ising CFT. In this section, we will present a more interesting example where the universal entanglement properties of the output mixed state can be tuned continuously. 
Our model involves measurement on the charge degrees of freedom and feedback on the spin degrees of freedom in a Luttinger liquid.
The resulting mixed state in the spin sector can be understood as a reduced density matrix of a Luttinger liquid with spin-charge coupling.
The measurement feedback protocol realizes a continuous family of conformal defects in the purified theory, resulting in continuous changes in correlations and entanglement of the output state.

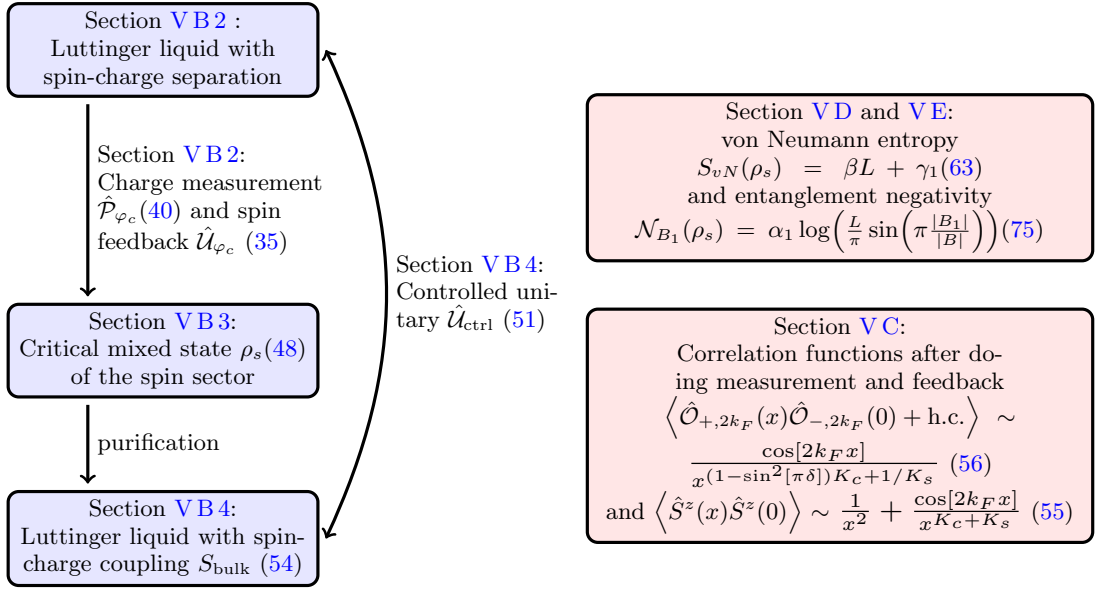
\begin{figure*} \centering \begin{tikzpicture} shft=.5
% \node[rectangle, rounded corners, draw, very thick, fill=blue!10!white, text width=10em, text centered, outer sep=3pt] (lattice_wf) at (-5,4) {Section \ref{sec:fermion_protocol_setup}: \\ Free spinful fermion $\ket{\psi_0}$};
% \node[rectangle, rounded corners, draw, very thick, fill=blue!10!white, text width=10em, text centered, outer sep=3pt] (lattice_RDM) at (-5,0) {Section \ref{sec:fermion_protocol_setup}: \\ Critical mixed state in the charge sector $\rho_s$};
% \node[rectangle, rounded corners, draw, very thick, fill=blue!10!white, text width=8em, text centered, outer sep=3pt] (lattice_purified_H) at (-5,-2.5) {Unknow purified state $\ket{\psi}$ with parent Hamiltonian $H = \hat{U}_{\mathrm{ctrl}} H_0 \hat{U}_{\mathrm{ctrl}}^{\dagger}$};
% \node[draw, very thick,fit=(lattice_wf) (lattice_RDM) (lattice_purified_H), inner sep=10pt, minimum width=6.8cm, minimum height=9cm,xshift=-1.2cm] (BigNode) {};
% \node[above=2mm of BigNode] {\Large Lattice};
\node[rectangle, rounded corners, draw, very thick, fill=blue!10!white, text width=12em, text centered, outer sep=3pt] (cont_bulk_action) at (0,-2.5) {Section \ref{sec:Purification of the spin sector reduced density matrix}:\\Luttinger liquid with spin-charge coupling $S_{\mathrm{bulk}}$ \eqref{eq:pure_bulk_action}};
\node[rectangle, rounded corners, draw, very thick, fill=blue!10!white, text width=12  em, text centered, outer sep=3pt] (cont_RDM) at (0,0) {Section \ref{sec:Field-Theoretic Description of the critical mixed state}:\\Critical mixed state~$\rho_s$\eqref{eq:spin_sector_RDM} of the spin sector};
\node[rectangle, rounded corners, draw, very thick, fill=blue!10!white, text width=12em, text centered, outer sep=3pt] (cont_initial_wf) at (0,4) {Section \ref{sec:Field-Theoretic Description of the Measurement-Feedback Process} :\\
Luttinger liquid with spin-charge separation};
% \\ $\braket{\varphi_c, \vartheta_s|\psi_{\mathrm{LL}}}$ \eqref{eq:boundary_action}};
\node[rectangle, rounded corners, draw, very thick, fill=red!10!white, text centered, text width=20em] (cont_mixed_state_corr) at (9,-1) {Section \ref{sec:correlation functions}:\\ Correlation functions after doing measurement and feedback $\left\langle \hat{\mathcal{O}}_{+, 2k_F}(x)\hat{\mathcal{O}}_{-, 2k_F}(0) + \mathrm{h.c.}\right\rangle \sim$
    $\scalebox{1.25}{$\frac{\cos[2k_F x]}{x^{(1 - \sin^2[\pi \delta]) K_c + 1/K_s}}$}$ \eqref{eq:spin_up_spin_down_corr}\\
    and $\left\langle \hat{S}^{z}(x)\hat{S}^{z}(0) \right\rangle \sim$
    $\scalebox{1.25}{$\frac{1}{x^2} + \frac{\cos[2k_F x]}{x^{K_c + K_s}}$}$ \eqref{eq:ZZ_corr}};
\node[rectangle, rounded corners, draw, very thick, fill=red!10!white, text centered, text width=20em] (cont_mixed_state_entanglement) at (9,2.3) {Section \ref{sec: replica_entropy} and \ref{sec:entropy and negativity semi-analytical}:\\von Neumann entropy\\
$S_{vN}(\rho_s) = \beta L + \gamma_1\eqref{eq: gamma_1}$ \\
and entanglement negativity\\
$\mathcal{N}_{B_1}(\rho_s) = \alpha_1 \log(\frac{L}{\pi}\sin(\pi \frac{\abs{B_1}}{\abs{B}}))$\eqref{eq:trace_norm}};
% \node[draw, very thick,fit=(cont_bulk_action) (cont_RDM) (cont_initial_wf) (cont_mixed_state_corr) (cont_mixed_state_entanglement), minimum height=9.06cm, minimum width=9.5cm, yshift=-0.08cm] (Rightbox) {};
% \node[above=2mm of Rightbox] {\Large Continuum};
% \draw[->, very thick] ([xshift=-1cm]lattice_wf.south) to node [right, shift={(0,0)}, text width=10em] {Section \ref{sec:fermion_protocol_setup}: \\ Charge measurement $\hat{P}_n$ and feedback $\hat{U}_n$ on spin}([xshift=-1cm]lattice_RDM.north);
% \draw[->, very thick] ([xshift=-1cm]lattice_purified_H.north) to node[right, shift={(0,0)}, text width=5em] {$\Tr_c(\cdot)$} ([xshift=-1cm]lattice_RDM.south);
%  \draw[<->, very thick, bend left=25] (lattice_purified_H.west) to node[left, text width=5em]{Section \ref{sec:fermion_protocol_setup}: Controlled unitary $\hat{U}_{\mathrm{ctrl}}$} (lattice_wf.west);
\draw[->, very thick] ([xshift=-1cm]cont_initial_wf.south) to node[right, shift={(0,0)}, text width=10em] {Section \ref{sec:Field-Theoretic Description of the Measurement-Feedback Process}: Charge measurement $\hat{\mathcal{P}}_{\varphi_c}$\eqref{eq: charge_projection_continuum} and spin feedback $\hat{\mathcal{U}}_{\varphi_c}$ \eqref{eq:spin_feedback_unitary}} ([xshift=-1cm]cont_RDM.north);
\draw[<-, very thick] ([xshift=-1cm]cont_bulk_action.north) to node[right, shift={(0,0)}, text width=5em] {purification} ([xshift=-1cm]cont_RDM.south);
\draw[<->, very thick, bend left=25] (cont_initial_wf.east) to node[right, yshift=0pt, text width=10em] {Section \ref{sec:Purification of the spin sector reduced density matrix}:\\ Controlled unitary $\hat{\mathcal{U}}_{\mathrm{ctrl}}$ \eqref{eq: controlled_unitary_continuum}} (cont_bulk_action.east);
% \draw[->, very thick, bend left = 15] (-4.5, 5) to node[pos=0.7, above,text width=10em] {coarse-grain}(0.5, 5);
% \node[rectangle, rounded corners, draw, very thick, fill=red!10!white, text centered, text width=10em, outer sep=3pt] (lattice_initial_corr) at (-6.5,4) {$\left\langle \hat{S}_i^{\alpha} \hat{S}_{j}^{\alpha} \right\rangle \sim \frac{1}{\abs{i - j}^2}$};
% \node[rectangle, rounded corners, draw, very thick, fill=red!10!white, text centered, text width=10em] (lattice_mixed_state_corr) at (-6.5,-2.5) {$\left\langle \hat{S}_i^{+} \hat{S}_{j}^{-} + \mathrm{h.c.} \right\rangle \sim \frac{1}{\abs{i - j}}$ ,\\ $\left\langle \hat{S}_i^{z} \hat{S}_{j}^{z} \right\rangle \sim \frac{1}{\abs{i - j}^2}$};

\end{tikzpicture} 
\caption{Road map of Luttinger liquid measurement and feedback protocol with both the description on lattice and description in the continuum. The blue boxes refer to quantum states at different stages of the protocol. The arrows refer to quantum processes in the protocol that relate different states. The red boxes refer to physical properties of the states.} \label{fig:roadmap} \end{figure*}

In Sec.\ref{sec:fermion_protocol_setup}, we review the measurement and feedback protocol for 1d spinful fermion, which was first introduced in \cite{lu2023mixed}. In Sec.\ref{sec:bosonization_dictionary}, we briefly review the bosonization technique for spinful fermions. In Sec.\ref{sec:Field-Theoretic Description of the Measurement-Feedback Process}, we define a more general protocol where the input fermion state can be described by a LL and we allow the phase factor in the unitary feedback to be continuously tunable. We then use bosonization to give a field-theoretic description of this measurement-feedback process. In Sec. \ref{sec:Field-Theoretic Description of the critical mixed state} and \ref{sec:Purification of the spin sector reduced density matrix}, we give a field-theoretic description of the output critical mixed state and derive the continuum field theory of a spin-charge coupled Luttinger liquid that purifies the spin sector critical mixed state. In Sec.\ref{sec:correlation functions}, \ref{sec: replica_entropy}, and \ref{sec:entropy and negativity semi-analytical}, we compute the correlation functions and entanglement properties of the critical mixed state within the field-theoretic description. Universal data about the critical mixed state are extracted and analyzed. The roadmap of this section is shown in Fig.\ref{fig:roadmap}.

\subsection{Review: measurement and feedback in a free fermion chain} \label{sec:fermion_protocol_setup}
% \YZ{Is this a review? I guess the Luttinger liquid, JW transformation, two pt functions are known; but the measurement feedback is new. Thus, maybe we do two subsections?}
In this subsection we review the measurement-feedback channel based on measuring fermion occupation introduced in \cite{lu2023mixed}.
We consider the initial state $\ket{\psi_0}$ being a free spinful fermion state on a 1d chain. Given the fermion creation and annihilation operator $\hat{c}_{i,\alpha}$ and $\hat{c}_{i,\alpha}^{\dagger}$, where $i$ labels the lattice sites and $\alpha = \uparrow, \downarrow$ labels the spin, the initial state $\ket{\psi_0}$ can simply be the groundstate of a tight-binding Hamiltonian
\begin{equation}\label{eq: free_fermion_H}
    H = -\sum_{i,\alpha} (\hat{c}_{i,\alpha}^{\dagger} \hat{c}_{i + 1,\alpha} + \text{h.c.}) + \mu \sum_{i, \alpha}(\hat{c}_{i,\alpha}^{\dagger} \hat{c}_{i,\alpha} - 1),
\end{equation}
where $\mu$ being the chemical potential. We can define operators $\hat{S}_{i}^{z} \equiv \frac{1}{2}\left[ \hat{n}_{i,\uparrow} - \hat{n}_{i,\downarrow}\right]$, $\hat{S}_{i}^{+} = \hat{c}_{i,\uparrow}^{\dagger} \hat{c}_{i,\downarrow}$, and $\hat{S}_{i}^{-} = \hat{c}_{i,\downarrow}^{\dagger}\hat{c}_{i,\uparrow}$. In the subspace of $\hat{n}_{i,\uparrow} + \hat{n}_{i,\downarrow} = 1$, they act as effective spin-$\frac{1}{2}$ operators and satisfy the correct commutation relation. The ground state $\ket{\psi_0}$ exhibits power-law correlation for the spin operators\cite{giamarchi2003quantum, sachdev1999quantum} 
\begin{equation}
    \expval{\hat{S}_{i}^{z} \hat{S}_{j}^{z}} = \expval{\hat{S}_{i}^{+} \hat{S}_{j}^{-}} \sim \frac{1}{\abs{i - j}^2} + \frac{\cos(2k_F \abs{i - j})}{\abs{i - j}^2},
\end{equation}\newline
where $k_F$ is the Fermi momentum and the equal sign is due to the $SU(2)$ spin rotation symmetry.\newline
\indent Consider measuring the electron charge operator $\hat{n}_{i} \equiv \hat{c}_{i,\uparrow}^{\dagger}\hat{c}_{i,\uparrow} + \hat{c}_{i,\downarrow}^{\dagger}\hat{c}_{i,\downarrow}$ at each lattice site $i$ and label the measurement outcomes by $n = \{n_1, n_2,...\}$. Correspondingly, the post-measurement state becomes $\hat{P}_n \ket{\psi_0}/ \sqrt{ \bra{\psi_0} \hat{P}_n \ket{\psi_0}}$ with probability $p_n = \bra{\psi_0} \hat{P}_n \ket{\psi_0}$.  For each post-measurement state labeled by $n$, we apply a unitary transformation on the spin degrees of freedom, 
\begin{align} \label{eq: example_spin_feedback}
\hat{U}_{n} = \prod_{j} e^{i\pi \hat{S}^{z}_{j}\underset{k<j}{\sum}n_{k}}
\end{align}
The entire ensemble of measurement outcomes is described by a mixed-state density matrix $\rho$, which takes the form of 
\begin{align}
    \rho = \sum_{\{n\}}\hat{U}_n \hat{P}_{n}\ket{\psi_0}\bra{\psi_0}\hat{P}_n \hat{U}_n^{\dagger}.
\end{align}
 The resulting density matrix $\rho$'s correlation function is altered by noticing that the feedback unitary $\hat{U}_n$ transforms $\hat{S}_{i}^{+}\hat{S}_{j}^{-}$ correlator into $\hat{U}_n^{\dagger} \hat{S}_{i}^{+} \hat{S}_{j}^{-} \hat{U}_n = \hat{S}_{i}^{+} (-1)^{\sum_{i\leq k < j}n_k} \hat{S}_{j}^{-}$. $\hat{S}_{i}^{+} \hat{S}_{j}^{-}$ correlator in $\rho$ is thus equal to:
\begin{equation} \label{eq:fermion_enhanced_corr_lattice}
    \begin{split}
        \Tr(\rho\,\hat{S}^{+}_{i}\hat{S}^{-}_{j}) &= \sum_{\{n\}}\bra{\psi_0} \hat{P}_n \hat{U}_{n}^{\dagger} \hat{S}^{+}_{i} \hat{S}^{-}_{j}\hat{U}_{n} \hat{P}_n\ket{\psi_0}\\
    &= \bra{\Psi}\hat{S}^{+}_{i}e^{i\pi \sum_{i\le k < j}\hat{n}_{k}}\hat{S}^{-}_{j}\ket{\Psi} \\
    & \sim \frac{\cos(2k_F |i-j|)}{\abs{i - j}}
    \end{split}
\end{equation}
in the second line we used the fact that the string operator has a slower decaying two-point correlation function~\cite{kruis2004geometry}. \newline
\indent For the case of $\mu=0$, we have $k_F = \pi/2$. In this case, we can understand the string order of the free fermion model in Eq. \eqref{eq: free_fermion_H} in a more straightforward way. Consider the Jordan-Wigner (JW) transformation to each species of the fermion. Each species of fermion is mapped to a spin-$\frac{1}{2}$ XX model, whose Hamiltonian is $H_{\mathrm{XX}} =-i(\hat{X}_i\hat{X}_i +\hat{Y}_i\hat{Y}_i)$. The fermion string operator $\hat{c}^{\dagger}_{i,s} \prod_{l=i+1}^{j-1} (-1)^{\hat{n}_{l,s}}  \hat{c}_{j,s}$ with $s\in \{\uparrow,\downarrow \}$ for each fermion species is mapped to the two-point function of local operators $(\hat{X}_i + i\hat{Y}_i)(\hat{X}_j - i\hat{Y}_j)$ in the XX model. Since it is known that
 the two-point function $\langle \hat{X}_i\hat{X}_j\rangle = \langle \hat{Y}_i\hat{Y}_j\rangle \sim \frac{(-1)^{|i - j|}}{\sqrt{|i - j|}}$ in the XX model \cite{mccoy1968spin,sachdev1999quantum}, the fermion string operator obeys the same
 scaling as well. The scaling of $\Tr(\rho\,\hat{S}^{+}_{i}\hat{S}^{-}_{j})$, which equals to the product of fermion string operator for spin up and spin down fermions, is $\frac{1}{\abs{i - j}}$ and the $(-1)^{\abs{i -j}}$ oscillating factor can be accounted by the Klein factor.
 % \YZ{I see, but this explanation is too packed. Maybe first say: this hidden order can be understood more explicitly through the JW transformation. Then: introduce the JW transformation and the operator map. }
 \newline 
\indent On the other hand, $\hat{S}^z$ commute with the measurement and feedback operators so the $\hat{S}_i^z \hat{S}_j^z$ correlation remains the same as that of the free fermion state
\begin{equation}\label{eq: lattice_ZZ_corr}
    \Tr(\rho\,\hat{S}^{z}_{i} \hat{S}^{z}_{j}) \sim \frac{1}{\abs{i - j}^2} + \frac{\cos(2k_F \abs{i - j})}{\abs{i - j}^2}.
\end{equation}
% This can also be understood from the spin-rotation $SU(2)$ symmetry, where the spin-spin correlations in all directions are the same. \YZ{Confused here. Then why $S^{+} S^{-}$ correlator is different?}

\subsection{Field-theoretic description of single-shot adaptation}\label{sec:Field-theoretic description}
We now discuss the effects of measurements and feedback in a spinful Tomonaga-Luttinger liquid \cite{tomonaga1950remarks}, which provides a  long-wavelength description of certain interacting fermion chains, in which the gapless spin and charge degrees of freedom propagate independently.
%Instead of starting with a free fermion model in Eq.~\eqref{eq: free_fermion_H}, we consider a more general model which includes four-fermion interactions that obey the spin rotation symmetry. Such a model can be bosonized into a Luttinger liquid (LL)~\cite{tomonaga1950remarks}. \ZZ{we consider a generic interacting fermion model whose long-wavelength property is described by a LL.}\YZ{Check if this sentence is correct!}
Below, we give a continuum field-theoretic description of the above measurement and feedback protocol on the Tomonaga-Luttinger liquid. This facilitates our computation of the properties of the resulting mixed states using the field theory.

\subsubsection{Bosonization and correlations in the TLL}\label{sec:bosonization_dictionary}
Recall that the LL fixed point consists of two free bosonic fields $\phi_+$ and $\phi_{-}$, which describe the gapless spin and charge flucutations of interacting one-dimensional electrons. We may represent the electron creation operators in terms of the two boson fields as \cite{senechal2004introduction, giamarchi2003quantum} 
\begin{align}
\hat{c}_{r,\sigma} \sim \frac{{\kappa}_{r,\sigma}}{\sqrt{2\pi a}}e^{ir\sqrt{4\pi}\,{\hat{\phi}}_{r,\sigma}}
\end{align}
where $r = \pm$ for right and left-movers, respectively, while $s = \pm$ for the spin up and down states of the fermion.  Here, $\kappa_{r,\sigma}$ is a Klein factor which encodes the anti-commutation relation of the fermion operators, and $a$ is a short-distance cutoff.  It is convenient to work with the fields $\hat{\phi}_c \equiv \sum_{\sigma, r} \hat{\phi}_{r,\sigma}/\sqrt{2}$,  $\hat{\phi}_s \equiv \sum_{\sigma, r} \sigma\,\hat{\phi}_{r,\sigma}/\sqrt{2}$, $\hat{\theta}_{c} \equiv \sum_{\sigma, r} r\hat{\phi}_{r,\sigma}/\sqrt{2}$, $\hat{\theta}_{s} \equiv \sum_{\hat{\sigma}, r} r\sigma\,\hat{\phi}_{r,\sigma}/\sqrt{2}$, which only act on the charge or spin degrees of freedom, as indicated by their subscripts.\newline
\indent We start with an (interacting) spinful fermion groundstate whose low-energy theory is described by the following LL action
\begin{align}
S_{LL}[\phi_c,\ts] &= \frac{1}{2K_{c}}\int dx\,d\tau\left[ (\dt\phi_c)^{2} + (\dx\phi_c)^{2}\right]\nonumber\\
&+ \frac{K_{s}}{2}\int dx\,d\tau\left[(\dt\theta_{s})^{2} + (\dx\theta_{s})^{2}\right],
\end{align}
with $K_{s,c}$ being the spin and charge Luttinger parameters. Here $\phi_c \equiv \phi_{c}(x,\tau)$ is a scalar field of eigenvalues of $\hat{\phi}_c$ and $\theta_s \equiv \theta_s(x,\tau)$ is a scalar field of eigenvalues of $\hat{\theta}_s$. Below we also give the bosonized form of fermion bilinear operators, which are going to be used in later section. In the continuum description of the LL, the normal-ordered charge density is given by \cite{senechal2004introduction, giamarchi2003quantum} 
\begin{align}\label{eq:bosonized_density}
    :\hat{n}(x): = \sqrt{\frac{2}{\pi}}\dx\hat{\phi}_c + \hat{\mathcal{O}}_{c,2k_{F}}(x)  + \cdots
\end{align}
where $k_F$ is the Fermi momentum. $\hat{\mathcal{O}}_{c,2k_{F}}(x) = \frac{1}{\pi}e^{-i 2k_{F}x}e^{i\sqrt{2\pi}\hat{\phi}_c(x)}\cos[\sqrt{2\pi}\hat{\phi}_s(x)] + \mathrm{h.c.}$ denotes the charge density wave (CDW) order 
and the ellipsis denotes spatially oscillatory corrections at higher wavevectors.
Furthermore, the spin operators are given by\cite{senechal2004introduction,giamarchi2003quantum}
\begin{equation}\label{eq: S_bosonization}
\begin{split}
    &\hat{S}^{\pm}_{j} \sim \kappa_s e^{\pm i\sqrt{2\pi}\hts(x)} + \hat{\mathcal{O}}_{\pm,2k_{F}}(x) + ...,\\
    &\hat{S}^{z}_j \sim \partial_x \hat{\phi}_s + \hat{\mathcal{O}}_{z,2k_{F}}(x) + ...
\end{split}
\end{equation} with $\kappa_s$ being the Klein factor. The second spatially-oscillating terms are the spin density wave (SDW) order parameters in the $xy$-plane and along the $z$ direction. Their bosonized forms are given by \cite{giamarchi2003quantum}
\begin{equation}\label{eq:SDW_order_parameter}
    \begin{split}
        &\hat{\mathcal{O}}_{\pm,2k_{F}}(x) = \frac{1}{\pi} e^{-i2k_F x} e^{\pm i\sqrt{2\pi}\hat{\theta}_s(x)}\cos\left[\sqrt{2\pi}\hat{\phi}_c(x)\right] +\mathrm{h.c.} \\
        &\hat{\mathcal{O}}_{z,2k_{F}}(x) = \frac{i}{\pi} e^{-i2k_F x} e^{i\sqrt{2\pi}\hat{\phi}_c(x)} \sin\left[\sqrt{2\pi}\hat{\phi}_s(x)\right] + \mathrm{h.c.}
    \end{split}
\end{equation}
The correlation functions of SDW order parameters probe universal features of the LL groundstate and they are given by
\begin{equation}\label{eq: LL_groundstate_corr}
    \begin{split}
        \left\langle \hat{\mathcal{O}}_{+, 2k_{F}}(x) \hat{\mathcal{O}}_{-,2k_{F}}(0) + \mathrm{h.c.} \right\rangle
        &\sim \frac{\cos[2k_F x]}{x^{K_c + 1/K_s}} \\
        \left\langle \hat{\mathcal{O}}_{z,2k_{F}}(x) \hat{\mathcal{O}}_{z,2k_{F}}(0)  \right\rangle &\sim \frac{\cos[2k_F x]}{x^{K_c + K_s}},
    \end{split}
\end{equation}
where the leading exponents depend on the Luttinger parameters.
\subsubsection{Field-theoretic description of measurements and feedback} \label{sec:Field-Theoretic Description of the Measurement-Feedback Process}
For the analysis below we slightly generalize our feedback unitary Eq.~\eqref{eq: example_spin_feedback} by allowing the rotation angle to be anywhere between $[-\pi, \pi]$,
%We now outline a more general protocol which can be coarse-grained into continuum, for performing a measurement of the local charges in the system followed by a unitary feedback on the spin degrees of freedom.  On the lattice, we first measure the occupation number $\hat{n}_{i} = \hat{c}^{\dagger}_{i,\uparrow}\hat{c}_{i,\uparrow} + \hat{c}^{\dagger}_{i,\downarrow}\hat{c}_{i,\downarrow}$ at each lattice site $i$ and label the measurement outcome by $n_i$.  Following this strong projective measurement, which we denote as $\prod_i \hat{P}_{n_i}$ \YZ{This is a repetition}, we perform the following general unitary operation, conditioned on the outcome of the occupation number $n$
\begin{align}\label{eq:spin_feedback_unitary}
    \hat{U}_n \equiv \prod_{j} \exp\left[ -i\pi \sin[\pi \delta] \left(\sum_{i<j}{n}_{i} \right)\hat{S}^{z}_{j}\right],
\end{align}
where $\sin[\pi \delta]$ is a parameter that defines the unitary transformation \footnote{Here we want to explain why we choose $\sin[\pi\delta]$ instead of simply $\delta$ to parametrize the unitary feedback in Eq.\eqref{eq:spin_feedback_unitary}. Because charge in quantized on the lattice, $\sum_{k<j}n_k$ is always an integer, so the phase factor appeared in the left hand side of Eq.\eqref{eq:spin_feedback_unitary} is manifestly invariant under $\delta \to \delta + 2$. However, as we bosonize the theory, we would coarse-grain $\sum_{k<j}n_k$ to continuous-valued fields and eventually treat the charge field operator $\hat{\phi}_c$ as noncompact at zero temperature. Therefore in order to keep the unitary transformation invariant under $\delta \to \delta + 2$ in the continuum limit, it is convenient to redefine $\delta$ as $\sin[\pi\delta]$, which is manifestly invariant under $\delta \to \delta + 2$.}. %and we choose it to parametrize the unitary transformation for later convenience to work in the continuum field theory. 
If $\sin[\pi\delta] = 1$, the unitary feedback reduces to that in \eqref{eq: example_spin_feedback}. This unitary commutes with the occupation number $\hat{n}_i$ as well as $\hat{S}^{z}_{i}$ at each lattice site, but transforms $\hat{S}^{\pm}_i$ in the following way
\begin{align}\label{eq:lattice_spin_transformation}
    \hat{S}^{\pm}_{i}\hat{U}_n = \exp\left[\pm i\pi \sin[\pi \delta] \sum_{k<i}{n}_{k}\right] \hat{U}_n \,\hat{S}^{\pm}_{i}.
\end{align} 
We consider the input state $\ket{\psi_{\mathrm{LL}}}$ to be described by a LL. The mixed-state after performing measurements and feedback is then given by
\begin{align} \label{eq: channel_on_lattice}
    \rho \equiv \sum_{\{n\}}\hat{U}_n(\prod_i \hat{P}_{n_i}) \ket{\psi_{\mathrm{LL}}}\bra{\psi_{\mathrm{LL}}}(\prod_i \hat{P}_{n_i}) \hat{U}_n^{\dagger}.
\end{align}
\newline
\indent We wish to understand the nature of the mixed-state $\rho$ by deriving a continuum action that governs its long-wavelength properties. In order to get a field-theoretic description, we perform bosonization on both the measurement operator $\hat{P}_{n}$ and the feedback unitary in Eq. \eqref{eq:spin_feedback_unitary}. First, we show that measuring the charge is described by measuring $\hat{\phi}_c$ in the continuum. %At long-wavelength, the charge measurement operator fixes the eigenvalue of the charge density $n(x) = m(x)$ where $m(x)$ denotes the outcome of a particular set of measurements. Alternatively, we may rearrange $\hat{n}(x)$ in Eq. \eqref{eq:bosonized_density} and write in the eigenbasis of $\hat{\phi}_c$ and $\hat{\phi}_s$ as 
%\begin{align}
%\varphi_c(x) = \varphi_c(-\infty) + \int^{x}_{-\infty}dy \left[\sqrt{\frac{\pi}{2}}m(y) - \mathcal{O}_{c,2k_{F}}(y) + \cdots\right],
%\end{align}
%where $\varphi_c$ is a scalar field of eigenvalues of $\hat{\phi}_c$.
%When the filling is incommensurate ($k_{F} \ne \ell\pi/a$, with integer $\ell$), the highly-oscillatory corrections to the first term in the above expression will provide a negligible contribution after performing the integral.   
Note that Eq.~\eqref{eq:bosonized_density} is equivalent to 
\begin{align}\label{eq:rearrange}
\hat{\phi}_c(x) = \hat{\phi}_c(-\infty) + \int^{x}_{-\infty}dy \left[\sqrt{\frac{\pi}{2}}\hat{n}(y) - \hat{\mathcal{O}}_{c,2k_{F}}(y) + \cdots\right],
\end{align}
We can then neglect the integral of the oscillatory part. Fixing the boundary condition $\hat{\phi}_c(-\infty) = 0$, we obtain 
\begin{equation}
   \hat{\phi}_c(x) = \sqrt{\frac{\pi}{2}} \int^{x}_{-\infty}dy\, \hat{n}(y)
\end{equation}
As a result, a projective measurement of the charge density $\hat{n}(y)$ also fixes the profile of the field $\hat{\phi}_c(x)$ which is consistent with the measurement outcome. Thus, we coarse the measurement operator into the continuum by
\begin{equation} \label{eq: charge_projection_continuum}
    \hat{\mathcal{P}}_{\varphi_c} = \ket{\varphi_c} \bra{\varphi_c},
\end{equation}
where $\varphi_c$ is the eigenvalue of $\hat{\phi}_c$.
\newline
\indent Second, we study the action of the feedback unitary on the field eigenstates. The action of the unitary operator (\ref{eq:spin_feedback_unitary}) can be inferred as follows. First, bosonization tells us that in the scaling limit
\begin{equation}\label{eq:phase_factor_continuum}
\begin{split}
    &\exp\left[i\pi \sin[\pi\delta] \sum_{k<j}n_{k}\right] \\
    &\sim \exp\left[i\sin[\pi\delta] \sqrt{2\pi}\varphi_c(x)\right]
\end{split}
\end{equation}
where $n_k$ is the measurement outcome on the lattice and $\varphi_c$ is the measurement outcome in the continuum. We have again plugged in the expression in \eqref{eq:bosonized_density} and neglected the oscillatory corrections. % We set $\varphi_c(-\infty) = 0$ for the remainder of our discussion, which is justified in a thermodynamically large system 
%  Then \eqref{eq:phase_factor_continuum} becomes
% \begin{equation}
%     \begin{split}
%     \exp\left[i\pi\sin[\pi\delta] \sum_{k<j}n_{k}\right] &\rightarrow  \exp\left[i\pi\sin[\pi \delta]\int_{-\infty}^{x} dy\,n(y)\right]\\
%     \sim \exp& \left[i\sin[\pi \delta] \sqrt{2\pi}\left(\varphi_c(x) - \varphi_c(-\infty)\right)\right]
% \end{split}
% \end{equation}
Then, according to Eq.\eqref{eq:lattice_spin_transformation} and \eqref{eq:phase_factor_continuum}, we expect the unitary feedback operator $\hat{\mathcal{U}}_{\varphi_c}$ in the continuum obeys
\begin{equation}
     \hmU^{\dagger}_{\varphi_c}\hat{S}^{\pm}(x)\hmU_{\varphi_c} \sim \exp\left[\pm i\sin[\pi\delta]\sqrt{2\pi}{\varphi}_c(x)\right]\hat{S}^{\pm}(x).
\end{equation} 
Recall Eq.~\eqref{eq: S_bosonization}, the field eigenstate $|\varphi_c,\vartheta_s\rangle$ is an eigenstate of $\hat{S}^{\pm}$ with eigenvalues $e^{\pm i \sqrt{2\pi} \vartheta_s}$. Thus, the action of $\hat{\mathcal{U}}_{\varphi_c}$ on the field eigenstate is a raising operator specified by 
\begin{align} \label{eq:unitary_feedback_continuum_action}
    \hat{\mathcal{U}}_{\varphi_c}\ket{\varphi_c,\vts} = \ket{\varphi_c,\vts + \sin[\pi\delta]\varphi_c}.
\end{align}
  Using Eq.\eqref{eq: charge_projection_continuum} and \eqref{eq:unitary_feedback_continuum_action}, one can verify that the coarse-grained measurement and feedback operator satisfy the Kraus condition, i.e.$\int d\varphi_c \left(\hat{\mathcal{P}}_{\varphi_c} \hat{\mathcal{U}}_{\varphi_c}^{\dagger}\right) \left(\hat{\mathcal{U}}_{\varphi_c} \hat{\mathcal{P}}_{\varphi_c} \right) = 1$ . 

\subsubsection{Continuum field theory for the critical mixed state}\label{sec:Field-Theoretic Description of the critical mixed state}
\indent We now study the consequences of this measurement and feedback protocol within the continuum field theory of the LL.  The ground-state wavefunction $\ket{\psi_{LL}}$ is given by 
\begin{align}
    \braket{\varphi_c,\vartheta_s|\psi_{LL}}\propto e^{-S_{\partial}[\varphi_c,\vartheta_s]}
\end{align}
where
\begin{align}\label{eq:boundary_action}
S_{\partial}[\varphi_c,\vartheta_s] = \frac{1}{2}\int\frac{dq}{2\pi}\left[\frac{|q|}{K_{c}}|\varphi_c(q)|^{2} + |q|K_{s}|\vartheta_s(q)|^{2}\right].
\end{align}
We derive these equations in Appendix.~\ref{app:derive_bdy_action} using path integral techniques.
From arguments presented in \ref{sec:Field-Theoretic Description of the Measurement-Feedback Process} and , The density matrix after measurement and feedback is
\begin{align} \label{eq:density_matrix}
 \rho &\propto \int\,D\varphi_c\,\hat{\mathcal{U}}_{\varphi_c} \hat{P}_{\varphi_c}\ket{\psi_{LL}}\bra{\psi_{LL}}\hat{\mathcal{P}}_{\varphi_c} \hat{\mathcal{U}}^{\dagger}_{\varphi_c} \nonumber \\
 &\propto \int\,D\varphi_c\,D\vartheta_s\,D\vartheta_s'\,\ket{\varphi_c,\vartheta_s}\bra{\varphi_c,\vartheta_s'}\,e^{-S_{\mathrm{eff}}[\vartheta_s,\vartheta_s',\varphi_c]}
\end{align}
where in the second line we have used Eqs.~\eqref{eq: charge_projection_continuum}, ~\eqref{eq:unitary_feedback_continuum_action} and \eqref{eq:boundary_action}. The effective action $S_{\mathrm{eff}}$ is related to $S_{\partial}$ by
\begin{equation}\label{eq:effective_action}
\begin{split}
    S_{\mathrm{eff}}[\vartheta_s,\vartheta_s',\varphi_c] = &S_{\partial}\left[\varphi_c,\vartheta_s - {\sin[\pi \delta]}\varphi_c\right] \\
    &~~~~~~+ S_{\partial}\left[\varphi_c,\vartheta_s' - {\sin[\pi \delta]}\varphi_c\right].
\end{split}
\end{equation} With such an effective action, we can compute the expectation value of spin-spin correlations and confirm with the result from the lattice protocol. We will present a detailed discussion on correlation functions in Sec.\ref{sec:correlation functions}.

% \indent Although the feedback is applied to the spin sector, we want to comment on the correlation functions in the charge sector. The density correlation, which commutes with the unitary feedback, is unchanged. However, the singlet pairing correlation which is not diagonal in the $\hat{\phi}_c$ eigenbasis would vanish because the density matrix is completely decohered $\hat{\phi}_c$ eigenbasis.

\indent We now trace out the charge sector by integrating over $\varphi_c$. Performing this Gaussian integration yields the reduced density matrix of the spin sector
\begin{equation}\label{eq:spin_sector_RDM}
\begin{split}
    \bra{\vts'}\rho_s\ket{\vts} &\equiv \bra{\vts'}\tr_c \rho\ket{\vts}
    \propto e^{-S_{\mathrm{s,eff}}[\vts, \vts']}
    \end{split}
\end{equation}    
where the spin sector effective action is given by 
\begin{equation} \label{eq:spin_sector_effective_action}
\begin{split}
        S_{\mathrm{s,eff}} &= \frac{1}{2}\int \frac{dq}{2\pi} \abs{q} \Big[ \frac{2K_s + \sin[\pi \delta]^2 K_cK_s^2}{2(1 + K_c K_s \sin[\pi \delta]^2)}\abs{\vartheta_s(q)}^2\\
        &+ \frac{2K_s + \sin[\pi \delta]^2 K_cK_s^2}{2(1 + K_c K_s \sin[\pi \delta]^2)}\abs{\vartheta_s'(q)}^2 \\
        &- \frac{\sin[\pi \delta]^2 K_cK_s^2}{1 + K_c K_s \sin[\pi \delta]^2} \vartheta_s(q)\vartheta_s'(q)^*\Big].
    \end{split}
\end{equation}
Compare with Eq.\eqref{eq:boundary_action}, the spin sector Luttinger parameter $K_s$ has been reset due to the unitary feedback that couples the spin and charge. The coupling between the bra and ket fields $\vartheta_s$ and $\vartheta_s'$ indicates that this density matrix is not pure. \newline
\subsubsection{Purification of the critical mixed-state}\label{sec:Purification of the spin sector reduced density matrix}
 \indent As we mentioned in Sec. \ref{sec: reivew}, the crucial step to study the state under measurement and feedback is by purifying it into a the ground state of another Hamiltonian. For the spin sector density matrix $\rho_s$ studied in this section, we show that a purification is given by the ground state of another LL with spin-charge coupling. A straightforward way to obtain the purified state is to make use of the controlled unitary defined as $\hat{U}_{\mathrm{ctrl}} = \sum_{n}\hat{U}_n \hat{P}_n$. Making use of the field-theoretic description of $\hat{\mathcal{P}}_{\varphi_c}$ and $\hat{\mathcal{U}}_{\varphi_c}$ (in Eq.\eqref{eq: charge_projection_continuum} and \eqref{eq:spin_feedback_unitary}), we can write the continuum description of the controlled unitary $\hat{\mathcal{U}}_{\mathrm{ctrl}}$ as
 \begin{equation}
     \hat{\mathcal{U}}_{\mathrm{ctrl}} = \sum_{\varphi_c} \hat{\mathcal{U}}_{\varphi_c} \hat{\mathcal{P}}_{\varphi_c},
 \end{equation}
 which acts as
 \begin{equation} \label{eq: controlled_unitary_continuum}
     \hat{\mathcal{U}}_{\mathrm{ctrl}}\ket{\pc, \vts} = \ket{\pc, \vts + \sin[\pi \delta] \pc}.
 \end{equation}
Notice that we should not confuse between the above action of $\hat{\mathcal{U}}_{\mathrm{ctrl}}$ and the action of the unitary feedback $\hat{\mathcal{U}}_{\varphi_c}$. In Eq.\eqref{eq:spin_feedback_unitary}, we only use the field $\varphi_c$ to label the measurement outcome and $\hat{\mathcal{U}}_{\varphi_c}$ doesn't have support on $\ket{\varphi_c}$. While in Eq.\eqref{eq: controlled_unitary_continuum}, $\hat{\mathcal{U}}_{\mathrm{ctrl}}$ is a controlled operator acting on $\varphi_c$.
Thus, the matrix elements of the purified density matrix can be obtained from the input LL density matrix by
\begin{equation}
\begin{split} \label{eq: purified_state_boundary action}
     &\bra{\varphi_c, \vartheta_s} \rho_{\text{pure}} \ket{\varphi_c', \vartheta_s'}\\
     &\equiv \bra{\varphi_c, \vartheta_s} \hat{\mathcal{U}}_{\mathrm{ctrl}} \rho_{\text{LL}} \hat{\mathcal{U}}^{\dagger}_{\mathrm{ctrl}} \ket{\varphi_c', \vartheta_s'} \\
      &\propto e^{-S_{\partial}\left[\varphi_c, \vartheta_s - \sin[\pi \delta]  \varphi_c  \right] - S_{\partial}\left[\varphi_c', \vartheta_s' - \sin[\pi \delta]  \varphi_c'  \right]},
     \end{split}
\end{equation}
for $S_{\partial}$ in Eq.\eqref{eq:boundary_action}. The two set of boundary fields obey the same action and are not coupled together, which directly manifests the above density matrix describes a pure state. It is also straightforward to verify that
\begin{equation}
    \tr_c[\rho_{\text{pure}}] \equiv \int \mathcal{D}\varphi_c \, \bra{\varphi_c} \rho_{\text{pure}} \ket{\varphi_c} = \rho_{\text{s}}.
\end{equation}
% Since the unitary feedback Eq.\ref{eq:spin_feedback_unitary} couples the spin and the charge sector with a quadratic interaction, the intuition is to purify the reduced density matrix by also adding the same quadratic spin-charge coupling term as in Eq.\eqref{eq:effective_action} to the LL boundary action. A straightforward choice for the purified density matrix is
% \begin{equation}
% \begin{split}
%     ~~~~~&\bra{\varphi_c, \vartheta_s} \rho_{\text{pure}} \ket{\varphi_c', \vartheta_s'} \\
%     &~~~~~~~\propto e^{-S_{\partial}\left[\varphi_c, \vartheta_s - \sin[\pi \delta]  \varphi_c  \right] - S_{\partial}\left[\varphi_c', \vartheta_s' - \sin[\pi \delta]  \varphi_c'  \right]},
% \end{split}
% \end{equation}
% for $S_{\partial}$ in Eq.\eqref{eq:boundary_action}. The two set of boundary fields obey the same action and are not coupled together, which directly manifests the above density matrix describes a pure state.
% Then it is straightforward to show that
% \begin{equation}
%     \tr_c[\rho_{\text{pure}}] \equiv \int \mathcal{D}\varphi_c \, \bra{\varphi_c} \rho_{\text{pure}} \ket{\varphi_c} = \rho_{\text{s}}.
% \end{equation}
The purified density matrix $\rho_{\mathrm{pure}}$ has the same spin-spin correlations as in Eq. \eqref{eq:ZZ_corr} and \eqref{eq:spin_up_spin_down_corr} because the bosonized form of these spin operators are diagonal in the $\varphi_c$ basis. 
% \YZ{I feel the above derivatio should go first}

The key is that Eq.~\eqref{eq: purified_state_boundary action} describes the ground state of another LL with spin-charge coupling, whose action is denoted as $S_{\mathrm{bulk}}$. As we show explicitly in Appendix. \ref{app:derive_bdy_action},
%\indent The action $S_{\partial}[\varphi_c, \vartheta_s - \sin[\pi \delta] \varphi_c]$ describing the purified state $\ket{\psi_{\mathrm{pure}}}$ should be interpreted as the boundary action of some bulk action $S_{\text{bulk}}$, whose groundstate is exactly $\ket{\psi_{\mathrm{pure}}}$. Given the purified boundary action involves quadratic coupling between charge boundary fields and spin boundary fields. It is natural to reconstruct the bulk action as a LL with quadratic spin density and charge density coupling and also spin momentum and charge momentum coupling. Such a bulk action is given by
\begin{equation} \label{eq:pure_bulk_action}
\begin{split}
    S_{\mathrm{bulk}} = &\frac{1}{2}(\frac{1}{K_c} + \sin^2[\pi \delta] K_s)\int dx\,d\tau\left[ (\dt\phi_c)^{2} + (\dx\phi_c)^{2}\right]\\
    &+ \frac{1}{2}K_s\int dx\,d\tau\left[(\dt\theta_{s})^{2} + (\dx\theta_{s})^{2}\right] \\
    &- \sin[\pi \delta] K_s \int dx d\tau \, \left[(\partial_\tau \phi_c)(\partial_\tau \ts) + (\partial_x \phi_c)(\partial_x \ts) \right],
\end{split}
\end{equation}
where the Luttinger parameters in the bulk action have been modified in order to match the couplings in the boundary action. %We have explicitly verified the boundary action of Eq.\eqref{eq:pure_bulk_action} is indeed, and the detailed calculation is in Appendix \ref{app:derive_bdy_action}. 
Later, we will use this bulk action to compute the von Neumann entropy of the spin sector and also the bipartite Entanglement Negativity within the spin sector.\newline

\subsection{Correlation functions}\label{sec:correlation functions}
As a sanity check, we first compute the correlation functions in the field-theoretic description above and check that they are consistent with the lattice calculation in Sec.\ref{sec:fermion_protocol_setup}.\newline
\indent In order to characterize the mixed state $\rho$ in Eq.\eqref{eq:density_matrix} obtained through measurements and feedback is to compute the expectation value of spin-spin correlations. It also servers as a sanity check for the field-theoretic description to see if the correlations we get are consistent with the lattice calculation in Sec.\ref{sec:fermion_protocol_setup}. For the $\hat{S}^{z}(x)\hat{S}^{z}(0)$ correlation, since $\hat{S}^{z}(x)$ commutes with the feedback unitary in Eq.(\ref{eq:spin_feedback_unitary}), its expectation value remains the same as the input LL ground state in Eq.\eqref{eq: LL_groundstate_corr}. As we have also explicitly calculated in Appendix \ref{app:spin correlation}, the leading terms are 
\begin{align} \label{eq:ZZ_corr}
    \left\langle \hat{S}^{z}(x)\hat{S}^{z}(0) \right\rangle \sim \frac{1}{x^2} + \frac{\cos[2k_F x]}{x^{K_c + K_s}}.
\end{align} 
\indent The correlation that gets enhanced is the SDW order in the $xy$ plane, referred to as $\mathrm{SDW}^{\pm}$ later on. Its order parameter $\hat{\mathcal{O}}_{\pm, 2k_F}$ is given in Eq.\eqref{eq: S_bosonization} and
  its correlation function evaluated with respect to the mixed state $\rho$ in Eq. \eqref{eq:density_matrix} is given by (see the detailed calculation in Appendix. \ref{app:spin correlation})
\begin{equation} \label{eq:spin_up_spin_down_corr}
\begin{split}
    &\left\langle \hat{\mathcal{O}}_{+, 2k_F}(x)\hat{\mathcal{O}}_{-, 2k_F}(0) + \mathrm{h.c.}\right\rangle \sim
    \frac{\cos[2k_F x]}{x^{(1 - \sin^2[\pi \delta]) K_c + 1/K_s}}.
\end{split}
\end{equation}
  Before doing measurement and feedback, the critical exponents of the leading term is $K_c + 1/K_s$ as in Eq.\eqref{eq: LL_groundstate_corr}. Therefore, by tuning the parameter $\sin[\pi \delta]$ in the feedback unitary,  one can enhance the $\mathrm{SDW}^{\pm}$ order by continuously decreasing the exponent in the two-point correlation function.\footnote{ We focus on the $q=2k_F$ component of $\hat{S}^{+}(x)\hat{S}^{-}(0)$ correlation because the $q=0$ contribution involves complications from the Klein factors. The other higher momentum contributions decay faster compared to the $q=2k_F$ contribution.} \newline
\indent In the case when $\sin[\pi \delta] = 1$ and $K_c = K_s = 1$, the protocol corresponds to the free spinful fermion protocol discussed in Sec.\ref{sec:fermion_protocol_setup}. The critical exponents of the $\mathrm{SDW}^{\pm}$ correlation function in Eq. \eqref{eq:spin_up_spin_down_corr} correctly captures the leading singularity of the $\hat{S}^{+}(x)\hat{S}^{-}(0)$ correlation in Eq.~\eqref{eq:fermion_enhanced_corr_lattice}. The unchanged $\hat{S}^{z}(x)\hat{S}^{z}(0)$ correlation also matches the expectation from the lattice model in Eq.~\eqref{eq: lattice_ZZ_corr}. So the comparison to the lattice result serves as a justification for our field theory description of the measurement and feedback process.
\subsection{Replica calculation of the R\'{e}nyi entropy}\label{sec: replica_entropy}

%\indent We can view the charge sector as auxiliary degrees of freedom and the spin sector as physical degrees of freedom that we care about. We can ask how does such a measurement and feedback protocol effectively modify entanglement structure of the spin sector, given that we have already seen the correlation functions in the spin sector are been modified. \YZ{This paragraph is not needed}
In order to characterize the entanglement of the mixed state $\rho_s$, the simplest quantity is the R\'{e}nyi entropy $S^{(n)}(\rho_s) \equiv \frac{1}{1-n}\log \Tr(\rho_s^n)$. Given the bulk action in Eq.\eqref{eq:pure_bulk_action} which purifies $\rho_s$, we can also interpret $S^{(n)}(\rho_s)$ as the R\'{e}nyi entanglement entropy between the spin and charge sector in the spin-charge coupled Luttinger liquid. Since the charge and spin degrees of freedom are locally entangled, the leading contribution to $S^{(n)}(\rho_s)$ is again volume law. The nontrivial part is to compute subleading constant $\gamma_n$. A similar problem is considered in Refs.~\cite{chen2013quantum,furukawa2011entanglement}, which studied the entanglement entropy of coupled Luttinger liquids. Here, we shall follow the techniques used in \cite{furukawa2011entanglement} and derive the subleading term in $S^{(n)}(\rho_s)$. \newline
\indent First we explicitly write down the R\'{e}nyi partition function $Z^{(n)} \equiv \Tr(\rho_s^n)$ for $\rho_s$ given in Eq.\eqref{eq:spin_sector_RDM},
\begin{equation}\label{eq:Z_n in replica field theory}
    \begin{split}
        \Tr[\rho_s^n] &= \frac{1}{Z^n} \int \prod_{j = 1}^n \mathcal{D}\vts^{(j)} \exp{-\sum_{j=1}^n S_{\mathrm{s,eff}}[\vts^{(j-1)}, \vts^{(j)}]} \\
        &= \frac{1}{Z^n} \int \prod_{j = 1}^n \mathcal{D}\vts^{(j)}  \exp{\int \frac{dq}{2\pi} \abs{q}\Theta_s(q)^{\dagger} M_n \Theta_s(q)},
    \end{split}
\end{equation}
where $Z$ is the normalization factor and the upper index $(j)$ labels the replica. To go from the first line to the second line, we plug in the expression for $S_{\mathrm{s,eff}}$ in Eq.\eqref{eq:spin_sector_effective_action}. In the second line of the above equation, the vector $\Theta_s(q)$ and the matrix $M_n$ are given by 
\begin{equation}
\begin{split}
    &\Theta_s(q) = \begin{bmatrix}
        \vts^{(1)}(q) & \vts^{(2)}(q) & \vts^{(3)}(q) & ... & \vts^{(n)}(q)
    \end{bmatrix}^T \\
    &M_n = \begin{bmatrix}
        A & \frac{1}{2}B & 0 &  & &  \frac{1}{2}B \\
        \frac{1}{2}B & A & \frac{1}{2}B & 0 &  & &  \\
        0 & \frac{1}{2}B & A & \frac{1}{2}B &  & & \\
        0 & 0 & \frac{1}{2}B & A & \ddots & \\
        & & & \ddots & \ddots & \frac{1}{2}B \\
        \frac{1}{2}B & & & & \frac{1}{2}B & A
    \end{bmatrix},
\end{split}
\end{equation}
with the matrix elements $A$ and $B$ given by
\begin{equation}
    \begin{split}
         & A = \frac{2K_s + \sin[\pi \delta]^2 K_cK_s^2}{2(1 + K_c K_s \sin[\pi \delta]^2)}, ~~~B = -\frac{\sin[\pi \delta]^2 K_cK_s^2}{2(1 + K_c K_s \sin[\pi \delta]^2)}.
    \end{split}
\end{equation}
% \indent On a circle with length $L$,  the momentum takes value $q_m = \frac{2\pi m}{L}$ for $m = 1, 2, \cdots, \infty$. So we will change the continuous integral over $q$ to a product over $q_m$. Since we are focusing on the ground state of the Luttinger liquid, we only need to consider the winding number zero sector and henceforth treat the boson fields as uncompactified. Firstly, the normalization factor $Z$ can be computed as
% \begin{equation}\label{eq: normalization}
% \begin{split}
%     Z &= \prod_{m=1}^{\infty} \int \mathcal{D}\vts \exp{|q_m| (A + B) |\vts(q)|^2} \\
%     &= \prod_{m=1}^{\infty} \frac{\pi}{|q_m| (A + B)}.
% \end{split}
% \end{equation}
% To get from the first line to the second line of the above equation, we perform the gaussian integral over $\vartheta_s$. Similarly, the R\'{e}nyi entropy is given by
% \begin{equation}\label{eq: Z_n_intermediate}
%     \begin{split}
%         \Tr[\rho_s^n] &= \frac{1}{Z^n} \prod_{m =1}^{\infty} \int \prod_{j = 1}^n \mathcal{D}\vts^{(j)} \exp{ \abs{q_m}\Theta_s(q)^{\dagger} M_n \Theta_s(q)} \\
%         &= \prod_{m=1}^{\infty} \frac{(A + B)^n }{\mathrm{det}M_n},
%     \end{split}
% \end{equation} 
% where to get the second line we plug in the value for $Z$ and perform the integral over $\vartheta_s$. Note that the term inside the product does not depend on $m$. This indicates a straightforward regularization of the R\'{e}nyi entropy,
% \begin{equation}
%     S^{(n)}(\rho_s) = \alpha L + \frac{1}{1-n} \log \frac{(A + B)^n }{\mathrm{det}M_n}
% \end{equation}

As we did the detailed calculation in Appendix.\ref{app:Analytical calculation of the Renyi Entropy}, we regularize Eq.\eqref{eq:Z_n in replica field theory} by introducing a short-distance cutoff $\beta$ and get
\begin{equation} \label{eq:Z_n_final}
    \Tr[\rho_s^n] = e^{-\beta L} \left(\prod_{l = 0}^{n - 1} \frac{\lambda_l}{A + B}\right)^{\frac{1}{2}},
\end{equation}
where $L$ is the system size and $\lambda_l$, the eigenvalues of the matrix $M_n$, are given by
\begin{equation}
    \lambda_l = A + B \cos(\frac{2\pi l}{n}), ~~~\mathrm{for} ~ l = 0, 1, 2, \cdots, n - 1.
\end{equation}
Using Eq.\eqref{eq:Z_n_final}, we can extract the subleading constant $\gamma_n$ of the $n$th R\'{e}nyi entropy as in \eqref{eq: entropy_universal_scaling_law} and is given by
\begin{equation}\label{eq:gamma_n}
\begin{split}
    \gamma_n
    &= \frac{-1}{2(n - 1)} \sum_{l = 0}^{n - 1} \log(\frac{\lambda_l}{A + B}),
\end{split}
\end{equation}
which is a universal function that depends on $K_c$, $K_s$, and $\sin[\pi \delta]$.
% For example, when $n = 2$ it is given by
% \begin{equation}\label{eq: gamma_2}
% \begin{split}
%     \gamma_2
%     &= -\frac{1}{2} \log(1 + K_c K_s \sin[\pi \delta]^2).
% \end{split}
% \end{equation}
If we fix the value of $K_c$ and $K_s$, $\gamma_n$ is a continuous function of $\sin[\pi \delta]$. The physical meaning of $\gamma_n$ can be related to the defect entropy in Eq.\eqref{eq:partition function in terms of boundary state}. As we tune $\sin[\pi \delta]$, we obtain a family of $n$-replica purified state $\langle\bra{(\psi \otimes \psi^*)^{\otimes n}}$. $\gamma_n$ measures the defect entropy of the boundary state $\ket{\mathcal{D}_A^{*\otimes n}[\mathcal{T}]}\rangle$ (specifically to this example $\mathcal{D}_A[\cdot]$ equals $\Tr_c[\cdot]$) with respect to this family of purified states.\newline
\indent One can further take the replica limit $n\rightarrow 1$ using the techniques in \cite{furukawa2013erratum} and obtain the subleading term $\gamma_1$ of the von Neumann entropy $S_{vN}(\rho_s)$. As we have performed the detailed calculation in Appendix.\ref{app: replica limit of entropy}), $\gamma_1$ is given by
\begin{equation} \label{eq: gamma_1}
    \gamma_1 = \ln(1 - z_{-}) - \frac{z_{-}}{z_{-} - 1} \ln(z_{-}),
\end{equation}
where $z_{-}$ is given by
\begin{equation}
    z_{-} = \frac{2 + K_c K_s \sin[\pi \delta]^2 - 2\sqrt{1 + K_c K_s \sin[\pi \delta]^2}}{K_c K_s \sin[\pi \delta]^2}.
\end{equation} We include a plot of $\gamma_1$ as a function of $\sin[\pi \delta]^2$ in Fig.\ref{fig:correlation_matrix_gamma_1}. 

\subsection{Calculation of the entanglement negativity} \label{sec:entropy and negativity semi-analytical}
\indent In the previous section, we use the effective action in \eqref{eq:spin_sector_effective_action} to compute the R\'{e}nyi entropy of the spin sector. However, it is hard to directly generalize the approach to compute R\'{e}nyi negativity because the effective action in Eq. \eqref{eq:spin_sector_effective_action} has long-range interactions if we Fourier-transform to real space. It is then not easy to compute the R\'{e}nyi negativity, which involves partial transpose in the real space. In this section, we propose another semi-analytical approach. First, we regularize the purified action \eqref{eq:pure_bulk_action} on the lattice, which is local in real space. %We would like to regularize that continuum actions as coupled harmonic oscillators on a 1d chain and use semi-analytical method to compute the Entanglement Negativity in that 1d chain. 
We then use the correlation matrix techniques to compute entanglement negativity of a bipartition in the spin sector. As a cross check, we also reproduce the constant term $\gamma_1$ in Eq.\eqref{eq: gamma_1} using this method. The results of the two calculations match perfectly. \newline
% \indent Since the reduced density matrix of $N$ harmonic oscillators takes the form of 
% $\rho_{B_1} \propto \exp{-\sum_{ij}A_{ij} b_i^{\dagger} b_{j}}$. Therefore, by working out all two point correlation functions that are restricted in region $B_1$, we can determine the eigenvalues of $\rho_{B_1}$. Similarly for $\rho^{T_{B_1}}$, one can determine all the eigenvalues by slightly modify the correlation matrices.\newline

\subsubsection{Regularization of $S_{\mathrm{pure}}$}
Given the purified action in \eqref{eq:pure_bulk_action}, we start with the continuum Hamiltonian that give rise to this action
\begin{equation}
\begin{split}
    \mathcal{H} = &\int dx \, \frac{1}{2}\left[\frac{1}{K'_c} (\partial_x \phi_c)^2 + K'_c \Pi_c^2 \right] \\
    &+\int dx \, \frac{1}{2} \left[ K_s (\partial_x \theta_s)^2 + \frac{1}{K_s} \Pi_s^2 \right] \\
    &+ \Delta \int dx \, \left[ (\partial_x \phi_c)(\partial_x \theta_s) - \Pi_c \Pi_s \right],
\end{split}
\end{equation}
where $\phi_c$ and $\theta_s$ are two compact boson fields represent the charge and spin fluctuations. $\Pi_{c(s)}$ are their conjugate momentum respectively. Both $K_c'$ and $\Delta$ are defined through \eqref{eq:pure_bulk_action} and are given by $\frac{1}{K_c'} = \frac{1}{K_c} + \sin^2[\pi \delta] K_s$ and $\Delta \equiv \sin[\pi \delta] K_s$. \newline
\indent We are interested in the bipartite entanglement negativity of the spin sector after we tracing out the charge sector. To make this problem tractable, we first notice that a mode expansion of the compact boson fields $\varphi_c$ and $\vartheta_s$ at zero temperature only contains the finite momentum part. Therefore, we could regularize the theory on a chain by forgetting about the compactness of the fields  and replacing it by a harmonic oscillator at site $n$. A naive attempt is given by
\begin{equation}\label{eq: real_space_coupled_HO}
\begin{split}
    H = &\sum_{n = 0}^{N - 1} \left[\frac{a^2\omega_0^2}{2K_c'} q_{c,n}^2 + \frac{a^2}{2K_c'}(q_{c,n+1} - q_{c,n})^2 + \frac{K_c'}{2a^2} p_{c,n}^2 \right] \\
    &+ \left[ \frac{a^2 K_s \omega_0^2}{2}q_{s,n}^2 + \frac{a^2K_s}{2} (q_{s,n+1} - q_{s,n})^2 + \frac{1}{2a^2 K_s} p_{s,n}^2 \right] \\
    &+\left[a^2 \Delta (q_{c, n + 1} - q_{c,n})(q_{s,n+1} - q_{s,n}) - \frac{\Delta}{a^2}p_{c,n}p_{s,n} \right],
\end{split}
\end{equation}
where the continuum fields are mapped to discrete operators in the following way
\begin{equation}
\begin{split}
    &q_{c,n} = \phi_{c}(x),~~~\frac{p_{c,n}}{a} = \Pi_{c}(x), \\
    &q_{s,n} = \theta_s(x),~~~
    \frac{p_{s,n}}{a} = \Pi_{s}(x),~~~\text{with}~x=na.
\end{split}
\end{equation}
The continuum limit can be restored by taking $N \rightarrow \infty$ and $a \xrightarrow{} 0$.\newline
\indent However, if we transform the above Hamiltonian into momentum space, the Hamiltonian contains couplings between the zero modes from the two sectors. This is inconsistent with the continuum theory, where the zero modes are explicitly dropped out by multiplying with the momentum $k=0$\cite{chen2013quantum, furukawa2011entanglement}. Therefore, we should explicitly remove the couplings between the zero modes to obtain the correct regularized Hamiltonian in the momentum space.  \newline
\indent Taking Fourier transformation
\begin{equation}
\begin{split}
    &q_{c(s),n} = \frac{1}{\sqrt{N}} \sum_{k = 0}^{N-1} \widetilde{q}_{c(s),k} e^{2\pi i kn / N}\\
    &p_{c(s), n} = \frac{1}{\sqrt{N}} \sum_{k=0}^{N-1} \widetilde{p}_{A(B), k} e^{2\pi i kn / N},
    \end{split}
\end{equation}
and removing the couplings between zero modes, we obtain the Hamiltonian in momentum space reads
\begin{equation} \label{eq: k_space_coupled_HO}
    \begin{split}
        \widetilde{H} = &\frac{K_c'}{2a^2} \widetilde{p}_{c,0}^2 + \frac{a^2}{2K_c'} \omega_0^2 \widetilde{q}_{c,0}^2 + \frac{1}{2a^2 K_s} \widetilde{q}_{s,0}^2 + \frac{a^2 K_s}{2} \omega_0^2 \widetilde{q}_{s,0}^2\\
        &+\sum_{k=1}^{N-1} \Big[ \frac{K_c'}{2a^2} \widetilde{p}_{c,k} \widetilde{p}_{c, N - k} + \frac{a^2}{2K_c'} \omega_k^2\widetilde{q}_{c,k} \widetilde{q}_{c, N - k} \\
    &+ \frac{1}{2a^2K_s} \widetilde{p}_{s,k} \widetilde{p}_{s, N - k} + \frac{a^2 K_s}{2} \omega_k^2  \widetilde{q}_{s,k} \widetilde{q}_{s, N - k}\\
    &+ 4\Delta a^2 K \sin^2(\frac{\pi k}{N}) \widetilde{q}_{c,k} \widetilde{q}_{s, N-k} - \frac{\Delta}{K a^2} \widetilde{p}_{c,k} \widetilde{p}_{s, N - k} \Big]. \\
    \end{split}
\end{equation}
% As we have already explained, we dropped the couplings between the two zero modes (i.e. terms involving $\Tilde{p}_{c,0}\Tilde{p}_{s,0}$ and $\Tilde{q}_{c,0}\Tilde{q}_{s,0}$).
We also explicitly add small mass terms $\frac{a^2}{2K_c'} \omega_0^2 \widetilde{q}_{c,0}^2$ and $\frac{a^2 K_s}{2} \omega_0^2 \widetilde{q}_{s,0}^2$ (we shall take $\omega_0 N \ll 1$) for each harmonic oscillator to prevent correlation functions from diverging~\cite{calabrese2013entanglement}. The dispersion relation is defined as $\omega_k = \sqrt{\omega_0^2 + 4 \sin(\frac{\pi k}{N})^2}$. 

\subsubsection{Entanglement entropy and negativity from correlation matrices}
\begin{figure}[h]
    \centering
    % First subfigure
    \begin{subfigure}{0.45\textwidth}
        \centering
        \includegraphics[width=\linewidth]{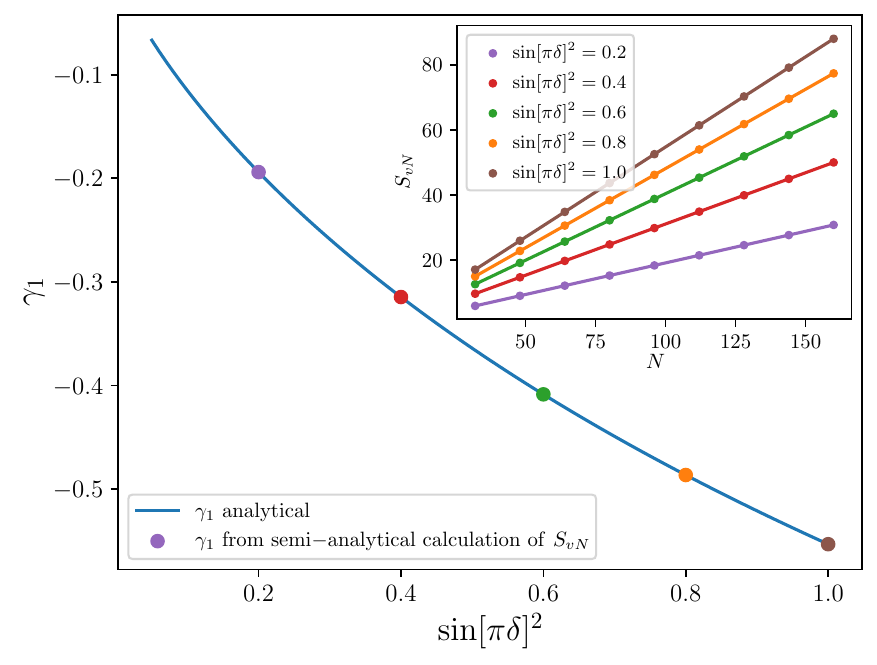}
        \caption{}
        \label{fig:correlation_matrix_gamma_1}
    \end{subfigure}
    \hfill
    % Second subfigure
    \begin{subfigure}{0.45\textwidth}
        \centering
        \includegraphics[width=\linewidth]{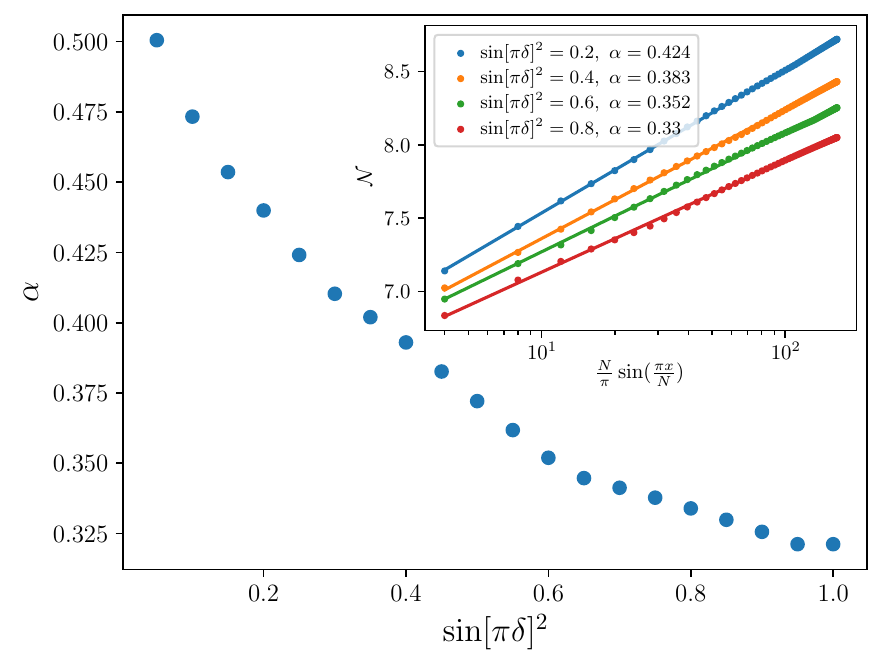}
        \caption{}
        \label{fig:correlation_matrix_N}
    \end{subfigure}
    \caption{(a)The inset shows the von Neumann entropy $S_{vN}$ computed from Eq.\eqref{eq: correlation_matrix_SvN} as a function of the system size $N$. The intercept of the line of best fit to $S_{vN}$ is extracted as $\gamma_1$ (universal constant subleading term from von Neumann entropy), which shown in the main figure is in good agreement with $\gamma_1$ computed from Eq.\eqref{eq: gamma_1}. All the data are computed by fixing $K_c = K_s = 1$. (b)The inset shows the bipartite entanglement negativity $\mathcal{N}$ computed on a periodic chain with varying subsystem size $x$. The data is fitted with the formula for bipartite entanglement negativity of a 1+1d CFT. The log coefficient $\alpha$ is extracted and plotted in the main figure as a function of $\sin[\pi \delta]^2$.}
    \label{fig:entanglement_main}
\end{figure}

Now given our regularized Hamiltonian in \eqref{eq: k_space_coupled_HO}, the spin sector is represented by tensor product of all the harmonic oscillators with subscript $s$. We are interested in the entropy of the spin sector $\rho_s$ and also the bipartite entanglement negativity within the spin sector. The entire Hamiltonian in Eq.~\ref{eq: k_space_coupled_HO} takes a quadratic form, so the reduced density matrix $\rho_s$ is also Gaussian. Following \cite{calabrese2013entanglement}, we can relate the eigenvalues of $\rho_{s}$ to two real space correlation matrices $\mathbbm{Q}_{s}$ and $\mathbbm{P}_{s}$, which are defined by
 \begin{equation}
     [\mathbbm{Q}_{s}]_{mn} = \expval{q_{s,m}q_{s,n}} ~\text{and}~[\mathbbm{P}_{s}]_{mn} = \expval{p_{s,m}p_{s,n}}.
 \end{equation}
 The correlation matrices can easily be computed using the Bogolibov transformation to diagonalize the Hamiltonian and we include the detailed calculation in App.\ref{app: calculation of correlation matrices}.
%  \begin{equation}
%     \begin{split}
%         &\expval{q_{s,m}q_{s,n}} \\
%         &=\frac{1}{N} \sum_{k, k'} e^{2\pi i k m / N} e^{2 \pi i k' n / N} \expval{\widetilde{q}_{s, k} \widetilde{q}_{s, k'}} \\
%         &= \frac{1}{2a^2 K_s \omega_0 N} + \frac{\sqrt{1 + \sin[\pi \delta]^2 K_c K_s}}{2a^2 K_s N}  \sum_{k = 1}^{N - 1} \frac{1}{\omega_k} e^{2\pi i k (m -n ) / N}
%     \end{split}
% \end{equation}
% \begin{equation}
%     \begin{split}
%         &\expval{p_{s,m}p_{s,n}} \\
%         &= \frac{1}{N} \sum_{k, k'} e^{2\pi i k r / N} e^{2 \pi i k' s / N} \expval{\widetilde{q}_{s, k} \widetilde{q}_{s, k'}} \\
%         &=\frac{a^2 K_s \omega_0}{2N} + \frac{a^2 K_s\sqrt{1 + \sin[\pi \delta]^2 K_c K_s}}{2N}  \sum_{k=1}^{N - 1} \omega_k e^{2\pi k i (m -n) / N}.
%     \end{split}
% \end{equation}
 \indent With the correlation matrices $\mathbbm{Q}_{s}$ and $\mathbbm{P}_{s}$, the R\'{e}nyi entropy and von Neumann entropy of $\rho_s$ can be found from\cite{calabrese2013entanglement} 
\begin{equation}
    \mathrm{Spectrum}(\mathbbm{Q}_s \cdot \mathbbm{P}_s) = \{\mu_1^2,...\mu_N^2 \}.
\end{equation}
The R\'{e}nyi entropy is then given by
\begin{equation}\label{eq: correlation_matrix_S2}
    \Tr[\rho_S^n] = \prod_{j=1}^{N}\left[(\mu_j + \frac{1}{2})^n - (\mu_j - \frac{1}{2})^n \right]^{-1}
\end{equation}
and the von Neumann entropy is given by
\begin{equation}\label{eq: correlation_matrix_SvN}
\begin{split}
    S_{vN} = \sum_{j=1}^{N}\Big[(\mu_j + \frac{1}{2})&\ln(\mu_j + \frac{1}{2}) \\
    &- (\mu_j - \frac{1}{2})\ln(\mu_j - \frac{1}{2}) \Big].
\end{split}
\end{equation}
We fix $K_c = K_s = 1$ and numerically compute $S_{vN}$ as a function of system size $N$ for various value of $\sin[\pi \delta]^2$ using \eqref{eq: correlation_matrix_SvN}. The result is shown in the inset of Fig.\ref{fig:correlation_matrix_gamma_1} and we can extract the universal subleading terms $\gamma_1$. We compare the extracted values of $\gamma_1$ to the analytical expression from Eq. \eqref{eq: gamma_1} and plot them in Fig.\ref{fig:correlation_matrix_gamma_1}. From the figure, we can see the values agree very well. \newline
\indent The bipartite entanglement negativity can be found from the two correlation matrices in a similar manner\cite{calabrese2013entanglement}. If we consider a bipartition of the spin sector into $s = s_1 \cup s_2$, the entanglement negativity between $s_1$ and $s_2$ is determined by
\begin{equation}
    \mathrm{Spectrum}(\mathbbm{Q}_{s} \cdot (\mathbbm{R}_{s_2} \mathbbm{P}_{s} \mathbbm{R}_{s_2})) = \{\nu_1^2, ...\nu_N^2 \},
\end{equation}
where the matrix $\mathbbm{R}_{s_2}$ is an $N \times N$ diagonal matrix having $-1$ in correspondence of sites in $s_2$ and $+1$ otherwise. From the eigenvalues $\nu_j$, the trace norm of $\rho_s^{T_{s_2}}$ is given by
\begin{equation} \label{eq:trace_norm}
    ||\rho_s^{T_{s_2}} || = \prod_{j=1}^{N} \left[\abs{\nu_j + \frac{1}{2}} - \abs{\nu_j - \frac{1}{2}} \right]^{-1} = \prod_{j=1}^{N} \max\left[1, \frac{1}{2\nu_j} \right].
\end{equation}
By fixing $K_c = K_s = 1$ and $N = 512$, we numerically compute $\mathcal{N} \equiv \log ||\rho_s^{T_{s_2}} ||$ using the above equation as a function of subsystem size $x$. The result is fitted with the function $\mathcal{N} = \alpha \log(\frac{N}{\pi} \sin(\frac{\pi x}{N})) + \beta$ and we can extract the log coefficient $\alpha$ numerically. We plot  $\mathcal{N}$ and $\alpha$ in Fig.\ref{fig:correlation_matrix_N}. 

We see that the log coefficient $\alpha$ has a continuous dependence on $\sin[\pi \delta]$. In the limit of no feedback ($\sin[\pi \delta] = 0$), the spin and charge sector in the purified state are not coupled. So the entanglement negativity of the spin sector simply equals to the R\'{e}nyi-$\frac{1}{2}$ entanglement entropy of a compact boson. The logarithmic coefficient would be $\alpha = \frac{c}{2} = \frac{1}{2}$\cite{calabrese2013entanglement}. This is consistent with the data point at $\sin[\pi \delta]^2 = 0$ in Fig.\ref{fig:correlation_matrix_N}. Similar to how $\gamma_1$ depends on $\sin[\pi \delta]$, as we increase $\sin[\pi \delta]$, which we can think of the charge and spin sector get more coupled to each other in the purified state, $\alpha$ continuously decreases. Because entanglement negativity measures the amount of quantum correlations in $\rho_s$, as we trace out the charge sector, some degrees of freedom of the spin sector has also been taken away, which results in less amount of quantum correlations in the system. Also, unlike the critical cluster state example in Sec.\ref{sec: coupled Ising chain} where the entanglement properties are the same as that of the Ising CFT, in this example as long as the unitary feedback is nontrivial ($\sin[\pi \delta] \neq 0$), the universal entanglement properties of $\rho_s$ is different from that of the compact boson.

\section{Summary and Discussion} \label{sec: summary}
\indent In this work we have studied universal properties of mixed quantum states generated through measurements and applying feedback unitaries on a 1+1d critical ground state. The quantities we focus on are correlation functions, bipartite Entanglement Negativity, and R\'{e}nyi entropy after tracing out the measured degrees of freedom. We have shown that these mixed state are `critical' in the sense that not only they have power-law decaying correlation functions but they also feature logarithmic scaling of bipartite Entanglement Negativity. By tracing out the measured degrees of freedom, we have also shown the reduced density matrix of the critical mixed state exhibit volume-law scaling of the R\'{e}nyi entropy with a constant universal subleading term. Moreover, by purifying the reduced density matrix, we obtain a pure critical state which gives the mixed state by acting with depolarization channels. The correlation and entanglement properties can be computed from the purification. The example we focus on in this paper is a measurement and feedback protocol on spinful fermions, where the universal entanglement properties of the resulting critical mixed state can be continuously changed through tuning parameters in the feedback unitary.\newline
\indent The measurement and feedback protocol introduced in Sec.\ref{sec: coupled Ising chain} can be easily generalized to bosonic $\mathbb{Z}_n \times \mathbb{Z}_n$ SPT in 1d\cite{chen2013symmetry}, where the channel can convert state in the SPT phase to a mixed state with $\mathbb{Z}_n \times \mathbb{Z}_n$ SSB order. One future direction could be studying how such a channel can alter the entanglement properties of the critical point separating the $\mathbb{Z}_n \times \mathbb{Z}_n$ SPT phase and the $\mathbb{Z}_n \times \mathbb{Z}_n$ SSB phase for $n=2,3,4$\cite{verresen2017one, tsui2017phase}. To make progress in answering this question, one can first identify how the operators in the qu-nit depolarization channel map to primary operators in the CFT governing the SPT critical point, and then analyze the scaling dimension of the defect operators, which are mapped from the qu-nit depolarization. More interestingly, when $n \geq 5$ there is a critical phase separating the SPT phase to the SSB phase. If we input the critical phase to the measurement and feedback protocol, would we obtain a series of critical mixed states whose entanglement properties are continuously changing? We leave the answers to these questions for future work. \newline 
\indent One promising direction for feature work is to study the robustness of the measurement and feedback protocol in reshaping the entanglement properties of critical states with additional noise. In Appendix.\ref{app:charge weak measurement} we have shown that if the charge projective measurement in the Luttinger liquid protocol is changed to charge weak measurement, then the resulting mixed state has exponentially decaying spin-spin correlation function. Does it implies the spin sector would not have a logarithmic scaling entanglement negativity? It will be interesting to verify this conjecture using numerics. \newline
\indent Another interesting direction to pursue is to consider measurement and feedback protocol acting on higher dimensional quantum states. In higher dimensions one don't necessarily need to start with a critical state, one can obtain critical states by doing a single-shot measurement with post-selection~\cite{zhang2024long, li2023measuring} on gapped states with discrete symmetry or measurement followed by feedback~\cite{lu2023mixed} on gapped states with continuous symmetry. Therefore, those protocols are more straightforward to realize in experiment because preparing the initial gapped state is far more easier than preparing a higher dimensional critical state. It will thus be interesting to study the entanglement properties of those output states which have already been shown to have power-law correlations. 
	
	\acknowledgements{We thank Matthew Fisher, Shang Liu, Tsung-Cheng Lu, Andreas Ludwig, Kaixiang Su, Cenke Xu, Zi-Yue Wang, Zack Weinstein for useful discussions. ZZ, TH, and SV thank Tsung-Cheng Lu for collaboration on previous related work. ZZ especially thanks Nayan Myerson-Jain for explaining Ref.\cite{myerson2023decoherence}. Use was made of computational facilities purchased with funds from the National Science Foundation (CNS-1725797) and administered by the Center for Scientific Computing (CSC). The CSC is supported by the California NanoSystems Institute and the Materials Research Science and Engineering Center (MRSEC; NSF DMR 2308708) at UC Santa Barbara. YZ and TH acknowledge the support of the Perimeter Institute for Theoretical Physics (PI) and the Natural Sciences and Engineering Research Council of Canada (NSERC). ZZ and SV acknowledge support from a grant from the W. M. Keck Foundation. Research at PI is supported in part by the Government of Canada through the Department of Innovation, Science and Economic Development Canada and by the Province of Ontario through the Ministry of Colleges and Universities. 
	}
	

	%------------------------------------------------------------
	\newpage
\onecolumngrid
\appendix
\section{Numerical results: critical cluster chain}\label{app:numerics}
 We use matrix product state (MPS) techniques to numerically compute the second R\'{e}nyi entropy $S^{(2)}[\rho]$ and third R\'{e}nyi negativity $\mathcal{N}^{(3)}$, where $\rho = \mathcal{D}_A[\ket{\psi}\bra{\psi}]$ for $\ket{\psi}$ in Eq.\eqref{eq:coupled_Ising_cft}. Each local deplorization channel $ \mathcal{D}_{A,i}$ admits the following Kraus operator decomposition
 \begin{equation}
     K_0 = \sqrt{1 - \frac{3p}{4}}I,~K_1 = \sqrt{\frac{p}{4}}X_i,~K_2 = \sqrt{\frac{p}{4}}Y_i,~\mathrm{and}~ K_3 = \sqrt{\frac{p}{4}}Z_i,
 \end{equation}
where $p \in [0, 1]$ controls the strength of the channel and $p = 1$ represents the maximal depolarization channel. The numerics are done using the Julia package ITensor\cite{itensor}.
 
\begin{figure}[h]
   \begin{subfigure}{0.45\textwidth}
       \includegraphics[height=6.5cm, width=\textwidth]{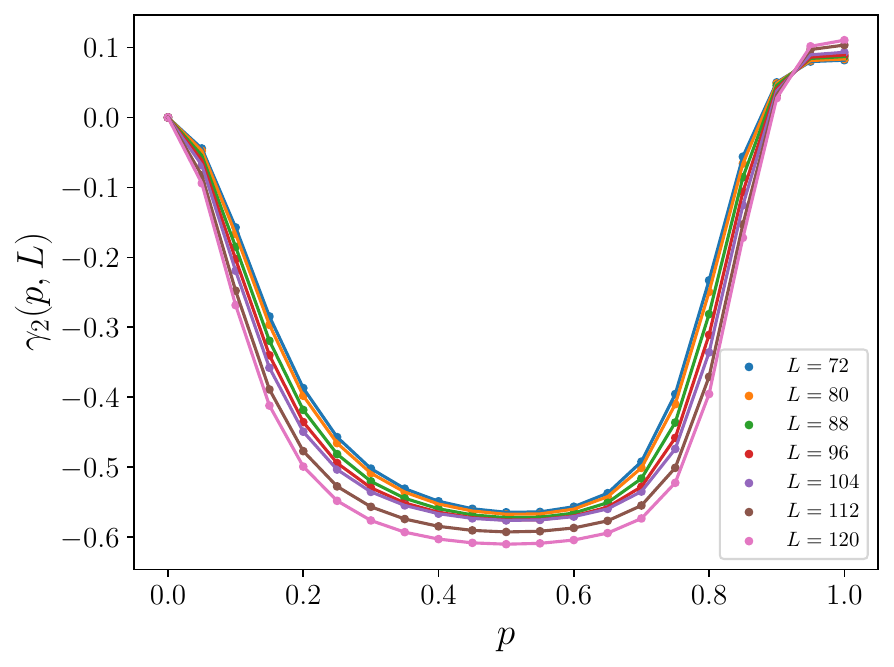}
       \label{fig: Ising_gamma}
   \end{subfigure}
   \hfill
   \begin{subfigure}{0.45\textwidth}
       \includegraphics[height=6.5cm, width=\textwidth]{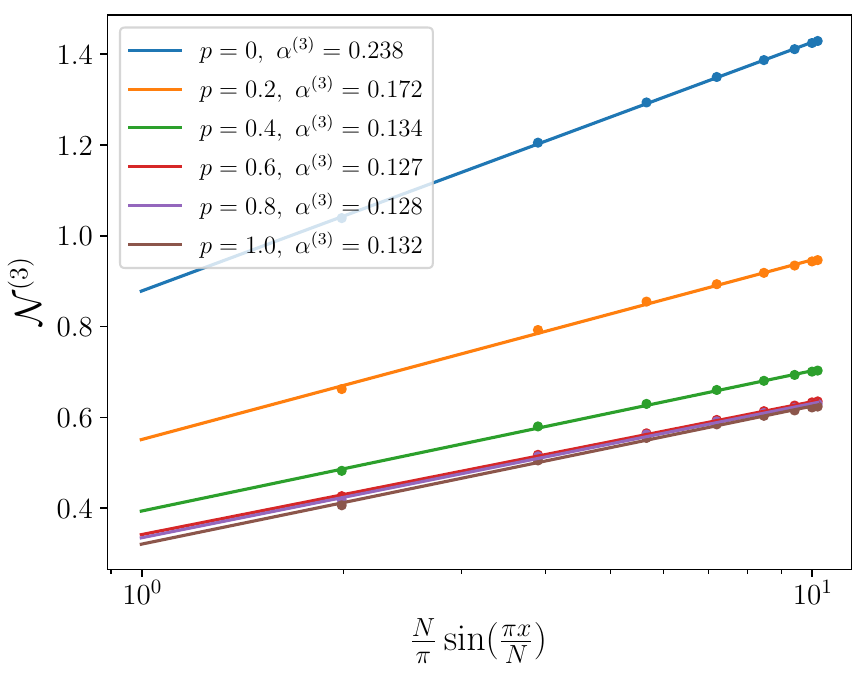}
       \label{fig: Ising_N}
   \end{subfigure}
    \caption{(a) We numerically compute the second R\'{e}nyi entropy in Eq.\eqref{eq: Ising partition function} as a function of system size $L$ and noise strength $p$. We then extract the subleading term $\gamma_2$, which is a function of $p$ and $L$. (b) We numerically compute the third R\'{e} negativity through Eq.\eqref{eq:neg_overlap} (taking $n=3$) as a function of subsystem size $x$ for a fixed total system size $L = 32$ and various noise strength $p$. We fitted the data with the formula in Eq.\eqref{eq: negativity_scaling} and extracted the log coefficient $\alpha^{(3)}$. The theoretical value we expect at $p=0$ is $\alpha^{(3)} = \frac{2}{9}$ and for any other values of $p$ is $\alpha^{(3)} = \frac{1}{9}$. }
\end{figure}

\section{Boundary action of coupled Luttinger liquids}\label{app:derive_bdy_action}
The thermal density matrix of a Luttinger liquid $\rho_{\beta} \equiv e^{-\beta H}/\Tr(e^{-\beta H})$ is given in the basis of $\hat{\phi}_c$ and $\hat{\theta}_s$ by performing the following path integral
\begin{align}\label{eq:thermal_density_matrix}
\braket{\varphi_c',\vartheta_s'|\rho_{\beta}|\varphi_c,\vartheta_s} \propto \int\,D\phi_c D\theta_{s} \,e^{-S_{LL}[\phi_c,\theta_{s}]}
\end{align}
where the integration is performed with the fixed boundary conditions $\phi_c(x,0) = \varphi_c(x)$, $\phi_c(x,\beta) = \varphi_c'(x)$, $\ts(x,0) = \vartheta_s(x)$, $\ts(x,\beta) = \vartheta_s'(x)$. We can then integrate out the fluctuations at $\tau \neq 0, \beta$ and obtain a 1d boundary action.\newline
\indent In this section, we would like to give a detail derivation of the boundary action of two coupled Luttinger liquids whose action is given by
\begin{align}
S_{LL}[\phi_c,\ts] &= \frac{1}{2 K_{c}}\int dx\,d\tau\left[ (\dt\phi_c)^{2} + (\dx\phi_c)^{2}\right]\nonumber\\
&+ \frac{K_{s}}{2}\int dx\,d\tau\left[(\dt\theta_{s})^{2} + (\dx\theta_{s})^{2}\right] \\
&- \Delta \int dx\,d\tau \, \partial_x \phi_c \partial_x \ts - \Delta \int dx \, d\tau \, \partial_\tau \phi_c \partial_\tau \theta_s
\end{align}
with $K_{s,c}$ the spin and charge Luttinger parameters and $\Delta$ is the parameter of the coupling between two Luttinger liquids. For boundary action, we mean to fix the field configurations at temporal boundaries $\tau = 0$ and $\tau = \beta$ and integrate out the bulk field configuration. We would then obtain a $1+0 ~ d$ action $S_{\partial}$ in terms of the boundary fields. More explicitly, the boundary action can be written as
\begin{align}
 e^{-S_{\partial}} \sim \int\,D\phi_c \bigg\vert_{\phi_c(x,0) = \varphi_c(x)}^{\phi_c(x,\beta)=\varphi'_c(x)} D\theta_{s} \bigg\vert_{\theta_s(x,0)=\vartheta_s(x)}^{\theta_s(x,\beta)=\vartheta'_s(x)} \,e^{-S_{LL}[\phi_c,\theta_{s}]}.
\end{align}
We may alternatively choose enforce these boundary conditions by defining the action
\begin{align}
\begin{split}
    S_{LL}'[\phi_c,\theta_s, \lambda_c, \lambda'_c, \lambda_s, \lambda'_s]  = S_{LL}[\phi_c,\theta_{s}] &+ i\int\,dx \,\lambda_{c}(x)\left[\phi_c(x,0) - \varphi_c(x)\right]\\ &+ i\int\,dx \,\lambda_{c}'(x)\left[\phi_c(x,\beta) - \varphi_c'(x)\right]\\
    &+ i\int\,dx \,\lambda_{s}(x)\left[\ts(x,0) - \vartheta_s(x)\right]\\
    &+ i\int\,dx \,\lambda_{s}'(x)\left[\ts(x,\beta) - \vartheta_s'(x)\right]
\end{split}
\end{align}
where we introduce four Lagrangian multipliers $\lambda_c$, $\lambda'_c$, $\lambda_s$ and $\lambda'_s$ to impose the boundary conditions.
By performing the path integral over these fields with free boundary conditions at $\tau = 0,\beta$,  we can obtain the boundary action in terms of $\varphi'_c$, $\varphi'_c$, $\vts$, and $\vts'$. It is convenient to integrate out the fields in momentum space, so by performing Fourier transformation, we have
\begin{align}
    &S'_{LL} =  \int \frac{dq\, d\omega}{(2\pi)^2} \Big[ \frac{1}{2K_c}(\omega^2 + q^2) \abs{\phi_c}^2 + \frac{K_s}{2} (\omega^2 + q^2) \abs{\theta_s}^2 - \Delta ( q^2 + \omega^2) \phi_c \theta_s^*  \Big] \\
    &+ \int\frac{dq\,d\omega}{(2\pi)^{2}}\left[J_{c}(q,\omega)\phi_c(q,\omega)^{*} + J_{s}(q,\omega)\ts(q,\omega)^{*}\right]   \\
    &+i \int\frac{dq}{2\pi}\left[\lambda_{c}(q)\varphi_c(q)^{*} + \lambda_{c}'(q)\varphi_c'(q)^{*}\right]\nonumber
     -i \int\frac{dq}{2\pi}\left[\lambda_{s}(q)\vartheta_s(q)^{*} + \lambda_{s}'(q)\vartheta_s'(q)^{*} \right]\nonumber\\ 
\end{align}
where
\begin{align}
    J_{c}(q,\omega) = i\left[\lambda_{c}(q) + \lambda'_{c}(q)e^{i\omega\beta}\right]\\
    J_{s}(q,\omega) = i\left[\lambda_{s}(q) + \lambda'_{s}(q)e^{i\omega\beta}\right]
\end{align}
Then we can integrate out the bulk fields $\phi_c$ and $\theta_s$ and obtain the boundary action. Integrating out $\phi_c$ yields
\begin{equation}
    \begin{split}
        S'_{LL} &= \int \frac{dq\,d\omega}{(2\pi)^2} -\frac{\left[J_c(q,\omega) -\Delta (\omega^2 + k^2) \theta_s(q, \omega) \right] \left[ J_c(-q, -\omega) - \Delta (\omega^2 + q^2) \theta_s(-q, -\omega) \right]}{\frac{\textcolor{black}{2}}{K_c}(\omega^2 + q^2)} \\
        &+ \int \frac{dq\, d\omega}{(2\pi)^2} \left[ \frac{K_s}{2} (\omega^2 + q^2) \abs{\theta_s}^2 +  J_{s}(q,\omega)\ts(q,\omega)^{*}\right] \\
        &+i \int\frac{dq}{2\pi}\left[\lambda_{c}(q)\varphi_c(q)^{*} + \lambda_{c}'(q)\varphi_c'(q)^{*}\right]\nonumber
     -i \int\frac{dq}{2\pi}\left[\lambda_{s}(q)\vartheta_s(q)^{*} + \lambda_{s}'(q)\vartheta_s'(q)^{*} \right] \\
     &= \int \frac{dq \, d\omega}{(2\pi)^2} \left[ \textcolor{black}{\frac{1}{2}} (K_s - \Delta^2 K_c) (\omega^2 + q^2) \abs{\theta_s}^2 + \textcolor{black}{\frac{1}{2}}(\Delta K_c J_c(-q, -\omega) + J_s(-q,-\omega)) \theta_s + \textcolor{black}{\frac{1}{2}}(\Delta K_c J_{c}(q,\omega) + J_s(q,\omega)) \theta_s^* \right] \\
     &- \int \frac{dq\,d\omega}{(2\pi)^2} \frac{J_c(q,\omega)J_c(-q,-\omega)}{\frac{\textcolor{black}{2}}{K_c}(\omega^2 + k^2)} +i \int\frac{dq}{2\pi}\left[\lambda_{c}(q)\varphi_c(q)^{*} + \lambda_{c}'(q)\varphi_c'(q)^{*}\right] -i \int\frac{dq}{2\pi}\left[\lambda_{s}(q)\vartheta_s(q)^{*} + \lambda_{s}'(q)\vartheta_s'(q)^{*} \right],
    \end{split}
\end{equation}
where we regroup the terms in the second equation. Then integrating out $\theta_s$ yields,
\begin{equation}
    \begin{split}
        S'_{LL} &= \int \frac{dq\,d\omega}{(2\pi)^2} -\frac{\left[ \Delta K_c J_{c}(q,\omega) + J_s(q,\omega)\right] \left[ \Delta K_c J_c(-q, -\omega) + J_s(-q,-\omega) \right]}{\textcolor{black}{2}(K_s - \Delta^2 K_c)(\omega^2 + q^2)} \\
        &~~~~- \int \frac{dq\,d\omega}{(2\pi)^2} \frac{J_c(q,\omega)J_c(-q,-\omega)}{\frac{\textcolor{black}{2}}{K_c}(\omega^2 + k^2)} +i \int\frac{dq}{2\pi}\left[\lambda_{c}(q)\varphi_c(q)^{*} + \lambda_{c}'(q)\varphi_c'(q)^{*}\right] -i \int\frac{dq}{2\pi}\left[\lambda_{s}(q)\vartheta_s(q)^{*} + \lambda_{s}'(q)\vartheta_s'(q)^{*} \right] \\
        &= \int \frac{dq\,d\omega}{(2\pi)^2}  \frac{1}{\textcolor{black}{2}(\omega^2 + q^2)} \Bigg\{ \left(\frac{\Delta^2 K_c^2}{K_s - \Delta^2 K_c} + K_c \right) \left[ \lambda_c(q) \lambda_c(-q) + \lambda'_c(q) \lambda'_c(-q)\right] \\
        &~~~~~~~~~~+ \frac{1}{K_s - \Delta^2 K_c} \left[\lambda_s(q) \lambda_s(-q) + \lambda'_s(q) \lambda'_s(-q) \right]\\
        &~~~~~~~~~~+  \frac{\Delta K_c}{K_s - \Delta^2 K_c} \left[\lambda_c(q)\lambda_s(-q) + \lambda_c(-q)\lambda_s(q) + \lambda'_c(q) \lambda'_s(-q) + \lambda'_c(-q) \lambda'_s(q) \right] \Bigg\} \\
        &~~~~+i \int\frac{dq}{2\pi}\left[\lambda_{c}(q)\varphi_c(q)^{*} + \lambda_{c}'(q)\varphi_c'(q)^{*}\right] -i \int\frac{dq}{2\pi}\left[\lambda_{s}(q)\vartheta_s(q)^{*} + \lambda_{s}'(q)\vartheta_s'(q)^{*} \right] \\
        &= \int \frac{dq}{2\pi} \frac{1}{\textcolor{black}{8} \abs{q}} \Bigg\{ \frac{K_c K_s}{K_s - \Delta^2 K_c} \left[ \lambda_c(q) \lambda_c(-q) + \lambda'_c(q) \lambda'_c(-q)\right] + \frac{1}{K_s - \Delta^2 K_c} \left[\lambda_s(q) \lambda_s(-q) + \lambda'_s(q) \lambda'_s(-q) \right]\\
        &~~~~~~~~~~+  \frac{\Delta K_c}{K_s - \Delta^2 K_c} \left[\lambda_c(q)\lambda_s(-q) + \lambda_c(-q)\lambda_s(q) + \lambda'_c(q) \lambda'_s(-q) + \lambda'_c(-q) \lambda'_s(q) \right] \Bigg\} \\
        &~~~~+i \int\frac{dq}{2\pi}\left[\lambda_{c}(q)\varphi_c(q)^{*} + \lambda_{c}'(q)\varphi_c'(q)^{*}\right] -i \int\frac{dq}{2\pi}\left[\lambda_{s}(q)\vartheta_s(q)^{*} + \lambda_{s}'(q)\vartheta_s'(q)^{*} \right]
    \end{split}
\end{equation}
Notice that in the second equation we omit the terms with a $e^{i\omega \beta}$ or $e^{-i\omega \beta}$ factor because such terms would vanish upon doing the integral over $\omega$, i.e. $\lim_{\beta \xrightarrow{} \infty} \int_{0}^{\infty} \frac{e^{\pm i \omega\beta}}{\omega^2 + k^2} d\omega = 0$. In the third line, we perform the integral over $\omega$ using the formula $\int_{0}^{\infty} \frac{d\omega}{2\pi} \frac{1}{\omega^2 + q^2} = \frac{1}{4\abs{q}}$. At this point we have integrated out all the bulk fields and integrating out the remaining Lagrange multipliers would give us the desired boundary action. Integrating out $\lambda_c$ and $\lambda'_c$ gives
\begin{equation}
    \begin{split}
        S'_{LL} &= -\int \frac{dq}{2\pi} \frac{\left[\frac
        {1}{\textcolor{black}{8}\abs{q}}\frac{\Delta K_c}{K_s - \Delta^2 K_c} \lambda_s(q) +  \textcolor{black}{\frac{1}{2}} i \varphi_c(q)\right] \left[ \frac{1}{\textcolor{black}{8}\abs{q}} \frac{\Delta K_c}{K_s - \Delta^2 K_c} \lambda_s(-q) +   \textcolor{black}{\frac{1}{2}} i \varphi_c(-q) \right]}{\frac{1}{\textcolor{black}{8}\abs{q}}\frac{K_c K_s}{K_s - \Delta^2 K_c}} \\
        &~~+ \int \frac{dq}{2\pi} \left[ \frac{1}{\textcolor{black}{8}\abs{q}} \frac{1}{K_s - \Delta^2 K_c}\lambda_s(q)\lambda_s(-q) - i\lambda_s(q) \vartheta_s(-q) - i\lambda_s(-q)\vartheta_s(-q) \right] + (...)'\\
        &= \int \frac{dq}{2\pi} \left[\frac{1}{\textcolor{black}{8}\abs{q}} \frac{1}{K_s} \lambda_s(q)\lambda_s(-q) +  i \textcolor{black}{\frac{1}{2}}\left( \frac{\Delta}{K_s} \varphi_c(-q) - \vartheta_s(-q) \right)  \lambda_s(q) + i \textcolor{black}{\frac{1}{2}}\left( \frac{\Delta}{K_s} \varphi_c(q) - \vartheta_s(q) \right) \lambda_s(-q) \right] \\
        &+ \int \frac{dq}{2\pi} \textcolor{black}{2} \abs{q} \frac{K_s - \Delta^2 K_c}{K_c K_s} \abs{\varphi_c}^2 + (...)'
    \end{split}
\end{equation}
the $(...)'$ part means replacing all the previous terms in the action by field variables and Lagrange multiplier with the $'$. Finally, integrating out $\lambda_s$ and $\lambda'_s$ gives
\begin{equation}
    \begin{split}
        S'_{LL} &= \int \frac{dq}{2\pi} \textcolor{black}{2}\abs{q} K_s\left(\frac{\Delta}{K_s} \varphi_c(q) + \vartheta_s(q)\right) \left(\frac{\Delta}{K_s} \varphi_c(-q) + \vartheta_s(-q)\right) + \int \frac{dq}{2\pi} \textcolor{black}{2}\abs{q} \frac{K_s - \Delta^2 K_c}{K_c K_s} \abs{\varphi_c}^2 + (...)' \\
        &= \int \frac{dq}{2\pi} \textcolor{black}{2}\abs{q} \left[ K_s \abs{\vartheta_s(q)}^2 +  \frac{1}{K_c} \abs{\varphi_c}^2 +  \Delta \varphi_c(q) \vartheta_s(-q) + \Delta \varphi_c(-q)\vartheta_s(q) \right] + (...)',
    \end{split}
\end{equation}
where the factor of $2$ can be taken away by a rescaling of the fields.

\section{Detailed calculation of correlation functions in the spin sector}\label{app:spin correlation}
In this section, we provide a detailed calculation of various correlation functions for the mixed state density matrix $\rho$ in Eq.\eqref{eq:density_matrix}. The quantities we are interested in are spin density wave orders $\mathrm{SDW}^{\pm}$ and $\mathrm{SDW}^{z}$ whose order parameters are given by \cite{giamarchi2003quantum}
\begin{equation}\label{eq:SDW_order_parameter}
    \begin{split}
        &\hat{\mathcal{O}}_{\pm,2k_{F}}(x) = \frac{1}{\pi} e^{-i2k_F x} e^{i\sqrt{2\pi}\hat{\phi}_c(x)} e^{\pm i\sqrt{2\pi}\hat{\theta}_s(x)} +  \frac{1}{\pi} e^{-i2k_F x} e^{-i\sqrt{2\pi}\hat{\phi}_c(x)} e^{\pm i\sqrt{2\pi}\hat{\theta}_s(x)} \\
        &\hat{\mathcal{O}}_{z,2k_{F}}(x) = \frac{i}{\pi} e^{-i2k_F x} e^{i\sqrt{2\pi}\hat{\phi}_c(x)} \sin\left[\sqrt{2\pi}\hat{\phi}_s(x)\right] + \mathrm{h.c.}
    \end{split}
\end{equation}
Their quadratic correlations probe non-universal features, which depend on the interactions, of the Luttinger Liquid.\newline
\indent The $\mathrm{SDW}^{\pm}$ order correlation functions can be computed as
\begin{equation}\label{eq:SDW_x_corr}
    \begin{split}
        &\left\langle \hat{\mathcal{O}}_{+, 2k_{F}}(x) \hat{\mathcal{O}}_{-,2k_{F}}(0) + \mathrm{h.c.} \right\rangle \\
        &\propto \cos[2k_F x] \Tr \Big[ \big( \sum_{\sigma =\pm 1} e^{i\sqrt{2\pi}(\hat{\phi}_c(x) - \hat{\phi}_c(0) + \sigma \hat{\theta}_s(x) - {\sigma}\hat{\theta}_s(0))} + \mathrm{h.c.} \big) \rho \Big] \\
        &\propto \cos[2k_F x] \sum_{\sigma= \pm 1} \int \mathcal{D}{\pc}\mathcal{D}\vts \big( e^{i\sqrt{2\pi}({\pc}(x) - {\pc}(0) + \sigma {\vartheta}_s(x) - {\sigma}\vartheta_s(0))} + \mathrm{h.c.} \big) e^{-S_{\partial}[\pc, \vts + \frac{\sin[\delta\pi]}{\pi}\pc]}
    \end{split}
\end{equation}
where to go from the second to third line, we express $\rho$ in terms of the Luttinger liquid boundary action $S_{\partial}$ in Eq. \eqref{eq:boundary_action}.
We then do a change of variable $\vartheta_s \xrightarrow{} \vartheta_s - \sin[\delta\pi] \pc$ and get 
\begin{equation}
    \begin{split}
        &\left\langle \hat{\mathcal{O}}_{+, 2k_{F}}(x) \hat{\mathcal{O}}_{-,2k_{F}}(0) \right\rangle \\
        & \propto \cos[2k_F x] \sum_{\sigma =\pm 1} \int \mathcal{D}\pc \, e^{i \sqrt{2\pi} (1 - {\sigma \sin[\delta\pi]})(\pc(x) - \pc(0)} \int \mathcal{D}\vartheta_s \, e^{i\sigma \sqrt{2\pi}(\vartheta_s(x) - \vartheta_s(0))} e^{-S_{\partial}[\varphi_c, \vartheta_s]} \\
        &\sim \sum_{\sigma = \pm 1} \frac{\cos[2k_F x]}{x^{(1 +\sigma \sin[\delta\pi])^2 K_c + 1/K_s}}
    \end{split}
\end{equation}
 where in the second line the integral can be interpreted as evaluating observables with respect to the Luttinger liquid groundstate.  Keeping the leading singularity of the above result, we reach our result in Eq.\eqref{eq:spin_up_spin_down_corr} in the main text. \newline
\indent Before evaluating $\hat{\mathcal{O}}_{z, 2k_{F}}^{\dagger}(x) \hat{\mathcal{O}}_{z,2k_{F}}(0)$ correlation, it will be instructive to calculate a simpler correlation $\partial_x \hat{\phi}_s(x) \partial_x \hat{\phi}_s(0)$ first. Notice that the $\hat{\phi}_s$ operators are not diagonal in the $\hat{\phi}_c$ and $\hat{\theta}_s$ eigenbasis, which is the basis we express $\rho$ in Eq.\eqref{eq:density_matrix}, so we want to insert an resolution of identity in the $\hat{\phi}_s$ eigenbasis. The $\partial_x \hat{\phi}_s(x) \partial_x \hat{\phi}_s(0)$ correlation reads
\begin{equation}
    \begin{split}
        &\left\langle \partial_x \hat{\phi}_s(x) \partial_x \hat{\phi}_s(0)  \right\rangle \\
        &\propto \Tr\left[\int \mathcal{D}\{\pc, \vts, \vts', \ps\} \left( \partial_x \ps(x) \partial_x \ps(0) \right) \ket{\ps}\bra{\ps} e^{-\frac{1}{2}S_{\partial}[\pc, \vts + \sin[\delta\pi] \pc] - \frac{1}{2}S_{\partial}[\pc, \vts' + \sin[\delta\pi] \pc]}\ket{\pc, \vts}\bra{\pc, \vts'}\right] \\
        &\propto \int \mathcal{D}\{\pc, \vts, \vts', \ps\}\left(\partial_x \ps(x) \partial_x \ps(0)\right) e^{-\frac{1}{2}S_{\partial}[\pc, \vts + \sin[\delta\pi] \pc] - \frac{1}{2}S_{\partial}[\pc, \vts' + \sin[\delta\pi] \pc] +  i \int dx \, \partial_x \ps (\vartheta_s - \vartheta_s')} 
    \end{split}
\end{equation}
where in the second line we insert the resolution of identity $1 = \int \mathcal{D} \varphi_s \ket{\varphi_s} \bra{\varphi_s}$. To go from the second to the third line, we use the fact that $\braket{\varphi_s|\vartheta_s} = e^{i \int dx \, (\partial_x \varphi_s) \vartheta_s}$. Then we can make the transformation $\vartheta_s \xrightarrow{} \vartheta_s - \sin[\delta\pi] \pc$ and $\vartheta_s' \xrightarrow{} \vartheta_s' - \sin[\delta\pi] \pc$ to simplify the expression. We get
\begin{equation}
    \begin{split}
        &\left\langle \partial_x \hat{\phi}_s(x) \partial_x \hat{\phi}_s(0)  \right\rangle \\
        & \propto \int \mathcal{D}\{\pc, \vts, \vts', \ps\}\left(\partial_x \ps(x) \partial_x \ps(0)\right) e^{-\frac{1}{2}S_{\partial}[\pc, \vts] - \frac{1}{2}S_{\partial}[\pc, \vts'] +  i \int dx' \partial_x \ps (\vartheta_s - \vartheta_s')} \\
        & \propto \int \mathcal{D}\pc \mathcal{D}\ps \left(\partial_x \ps(x) \partial_x \ps(0)\right) e^{-\Tilde{S}_{\partial}[\pc, \ps]} \\
        &\sim \frac{1}{x^2}
    \end{split}
\end{equation}
Going from the second to the third line, we integrate over $\vts$ and $\vts'$ and obtain the dual description of the Luttinger liquid boundary action $\Tilde{S}_{\partial}[\pc, \ps]$ in terms of $\pc$ and $\ps$. The third line can simplify to interpreted as the $\partial_x \hat{\phi}_s(x) \partial_x \hat{\phi}_s(0)$ correlation with respect to the Luttinger liquid groundstate. In the last line, we obtain the scaling $\frac{1}{x^2}$, which corresponds to the uniform part of Eq. \eqref{eq:ZZ_corr}.\newline
\indent The $\hat{\mathcal{O}}_{z, 2k_{F}}^{\dagger}(x) \hat{\mathcal{O}}_{z,2k_{F}}(0)$ correlation, which also contains $\hat{\mathcal{\phi}}_s$ can be evaulated in a similar way. We first want to insert a resolution of identity in the $\hat{\phi}_s$ basis. Then we make the transformation $\vartheta_s \xrightarrow{} \vartheta_s - \sin[\delta\pi] \pc$ and $\vartheta_s' \xrightarrow{} \vartheta_s' - \sin[\delta\pi] \pc$, and integrating out $\vts$ and $\vts'$. Finally the correlation function can be converted into correlation functions in the Luttinger liquid groundstate. The detailed calculation is written below:
\begin{equation}\label{eq:SDW_z_corr}
    \begin{split}
        &\left\langle \hat{\mathcal{O}}_{z, 2k_{F}}^{\dagger}(x) \hat{\mathcal{O}}_{z,2k_{F}}(0) \right\rangle \\
        &\propto \cos[2k_F x] \sum_{\sigma= \pm 1} \int \mathcal{D}\{\pc, \vts, \vts', \ps\} \\
        &~~~~~~~~~\big( e^{i\sqrt{2\pi}({\pc}(x) - {\pc}(0) + \sigma {\varphi}_s(x) - {\sigma}\varphi_s(0))} + \mathrm{h.c.} \big) e^{-\frac{1}{2}S[\pc, \vts + \frac{\sin[\delta\pi]}{\pi}\pc]-\frac{1}{2}S[\pc, \vts' + \frac{\sin[\delta\pi]}{\pi}\pc]} \braket{\vts|\ps}\braket{\vts'|\ps} \\
        &\propto \cos[2k_F x] \sum_{\sigma= \pm 1} \int \mathcal{D}\{\pc, \vts, \vts', \ps\} \\
        &~~~~~~~~~\big( e^{i\sqrt{2\pi}({\pc}(x) - {\pc}(0) + \sigma {\varphi}_s(x) - {\sigma}\varphi_s(0))} + \mathrm{h.c.} \big) e^{-\frac{1}{2}S[\pc, \vts + \frac{\sin[\delta\pi]}{\pi}\pc]-\frac{1}{2}S[\pc, \vts' + \frac{\sin[\delta\pi]}{\pi}\pc] +  i \int dx' \nabla \ps (\vartheta_s - \vartheta_s')} \\
        &\propto \cos[2k_F x] \sum_{\sigma= \pm 1} \int \mathcal{D}\{\pc, \vts, \vts', \ps\}
        \big( e^{i\sqrt{2\pi}({\pc}(x) - {\pc}(0) + \sigma {\varphi}_s(x) - {\sigma}\varphi_s(0))} + \mathrm{h.c.} \big) e^{-\frac{1}{2}S[\pc, \vts]-\frac{1}{2}S[\pc, \vts'] +  i \int dx' \nabla \ps (\vartheta_s - \vartheta_s')}\\
        &\propto \cos[2k_F x]\sum_{\sigma= \pm 1} \int \mathcal{D}\pc \mathcal{D}\ps \big( e^{i\sqrt{2\pi}({\pc}(x) - {\pc}(0) + \sigma {\varphi}_s(x) - {\sigma}\varphi_s(0))} + \mathrm{h.c.} \big) e^{-S[\pc, \ps]} \\
        &\sim \frac{\cos[2k_F x]}{x^{K_c + K_s}}.
    \end{split}
\end{equation}
 In the last line, we obtain the scaling $\frac{\cos[2k_F x]}{x^{K_c + K_s}}$, which corresponds to the oscillating part of Eq. \eqref{eq:ZZ_corr}.

\section{Detailed replica calculation of the R\'{e}nyi entropy} \label{app:Analytical calculation of the Renyi Entropy}
In this section, we apply the techniques introduced in \cite{furukawa2011entanglement} and analytically compute the R\'{e}nyi partition function $\Tr[\rho_s^n]$. \newline
\indent Given the spin sector reduced density matrix in \eqref{eq:spin_sector_RDM}, we have
\begin{equation}
    \begin{split}
        \Tr[\rho_s^n] &= \frac{1}{Z^n} \int \prod_{j = 1}^n \mathcal{D}\vts^{(j)} \exp{-\sum_{j=1}^n S_{\mathrm{s,eff}}[\vts^{(j-1)}, \vts^{(j)}]} \\
        &= \frac{1}{Z^n} \int \prod_{j = 1}^n \mathcal{D}\vts^{(j)}  \exp{\int \frac{dq}{2\pi} \abs{q}\Theta_s(q)^{\dagger} M_n \Theta_s(q)},
    \end{split}
\end{equation}
where 
\begin{equation}
    \Theta_s(q) = \begin{bmatrix}
        \vts^{(1)}(q) & \vts^{(2)}(q) & \vts^{(3)}(q) & ... & \vts^{(n)}(q)
    \end{bmatrix}^T
\end{equation} and the matrix $M_n$ is given by
\begin{equation}
\begin{split}
    &M_n = \begin{bmatrix}
        A & \frac{1}{2}B & 0 &  & &  \frac{1}{2}B \\
        \frac{1}{2}B & A & \frac{1}{2}B & 0 &  & &  \\
        0 & \frac{1}{2}B & A & \frac{1}{2}B &  & & \\
        0 & 0 & \frac{1}{2}B & A & \ddots & \\
        & & & \ddots & \ddots & \frac{1}{2}B \\
        \frac{1}{2}B & & & & \frac{1}{2}B & A
    \end{bmatrix} \\
    & A = \frac{2K_s + \sin[\pi \delta]^2 K_cK_s^2}{2(1 + K_c K_s \sin[\pi \delta]^2)}, ~~~B = -\frac{\sin[\pi \delta]^2 K_cK_s^2}{2(1 + K_c K_s \sin[\pi \delta]^2)}.
\end{split}
\end{equation}
We would like to compactify the boson fields on a length $L$ ring, so the momentum takes value $q_m = \frac{2\pi m}{L}$ for $m = 1, 2, \cdots, \infty$\footnote{The discretization of the momentum is consistent with the regularization scheme we use in the later semi-analytical calculation.}. Since we are focusing on the groundstate of the Luttinger liquid (this refers to the purified state in Eq.\eqref{eq: purified_state_boundary action}), we only need to consider the winding number zero sector. Finally, the R\'{e}nyi reduced density matrix can be written as
\begin{equation} \label{eq: Z_n in appendix}
    \begin{split}
        \Tr[\rho_s^n] = \frac{1}{Z^n} \prod_{m =1}^{\infty} \int \prod_{j = 1}^n \mathcal{D}\vts^{(j)} \exp{ \abs{q_m}\Theta_s(q)^{\dagger} M_n \Theta_s(q)}.
    \end{split}
\end{equation}
The normalization factor $Z$ can be computed as
\begin{equation}\label{eq: normalization}
\begin{split}
    Z &= \prod_{m=1}^{\infty} \int \mathcal{D}\vts \exp{|q_m| (A + B) |\vts(q)|^2} \\
    &= \prod_{m=1}^{\infty} \frac{2\pi}{|q_m| (A + B)}.
\end{split}
\end{equation}
Plugging in the above expression of $Z$ to Eq.\eqref{eq: Z_n in appendix} and performing the Gaussian integrals over $\vts^{(j)}$, we get
\begin{equation}\label{eq: infinite product over m}
\begin{split}
     \Tr[\rho_s^n] &= \frac{1}{Z^n} \prod_{m=1}^{\infty} \Big[(\frac{2\pi}{\abs{q_m}})^n \frac{1}{\mathrm{det}M_n} \Big] \\
     &= \prod_{m=1}^{\infty} \frac{(A + B)^n }{\mathrm{det}M_n}.
\end{split}
\end{equation}
The matrix $M_n$ has the same form of the Hamiltonian of a 1D tight-binding model, so its eigenvalues are given by
\begin{equation}
    \lambda_l = A + B \cos(\frac{2\pi l}{n}), ~~~\mathrm{for} ~ l = 0, 1, 2, \cdots, n - 1
\end{equation}
The determinant of the matrix is thus given by
\begin{equation}
    \mathrm{det}M_n = \prod_{l = 0}^{n - 1} \lambda_l.
\end{equation}
Following the techniques introduced in \cite{furukawa2011entanglement}, we regularized the infinite product over $m$ in \eqref{eq: infinite product over m} in the following way. We introduce a short-distance cutoff $a_0$ of the order of the lattice spacing. We rewrite the product as $\prod_{m \neq 0} C^{-\frac{1}{2}}$, where $m$ runs over $L / a_0 - 1$ modes by considering the exclusion of the zero mode (which is not present in the infinite product in \eqref{eq: infinite product over m}). Therefore the product scales as $C^{\frac{1}{2}} e^{-\alpha L}$ (with $\alpha = (\log C) / (2a_0)$). The prefactor $C^{\frac{1}{2}}$ gives a cutoff-independent (and thus universal) constant.
\begin{equation}
    \Tr[\rho_s^n] = e^{-\alpha L} (\prod_{l = 0}^{n - 1} \frac{\lambda_l}{A + B})^{\frac{1}{2}},
\end{equation}
where $\alpha$ is a cutoff dependent constant. The subleading constant $\gamma_n$ is obtained as
\begin{equation}
    \gamma_n = \frac{-1}{2(n - 1)} \log(\prod_{l = 0}^{n - 1} \frac{\lambda_l}{A + B}) = \frac{-1}{2(n - 1)} \sum_{l = 0}^{n - 1} \log(\frac{\lambda_l}{A +B}).
\end{equation}
\subsection{Replica limit ($n\rightarrow 1$)} \label{app: replica limit of entropy}
Here we apply the method in \cite{furukawa2013erratum} to compute the replica limit of $\Tr[\rho_s^n]$. We first focus on the quantity 
\begin{equation}
    \Tilde{\gamma}_n \equiv \sum_{l=0}^{n-1}\log(\frac{\lambda_l}{A+B}) = n\log(1 + \frac{1}{2}K_cK_s\sin[\pi\delta]^2) + \sum_{l=0}^{n-1}\log(1 - \frac{K_c K_s\sin[\pi \delta]^2 }{2 + K_cK_s \sin[\pi \delta]^2}\cos(\frac{2\pi l}{n})),
\end{equation}
and we would like to obtain the von Neumann entropy in the weak-coupling limit, that is when $\frac{K_c K_s\sin[\pi \delta]^2}{2 + K_c K_s\sin[\pi \delta]^2} \ll 1$. For convinience, we shall define $\kappa \equiv \frac{K_c K_s\sin[\pi \delta]^2}{2 + K_c K_s\sin[\pi \delta]^2}$ and $\Tilde{\gamma}_n$ written in terms of $\kappa$ is given by
\begin{equation}
\Tilde{\gamma}_n = n \log(\frac{1}{1 - \kappa}) + \sum_{l = 0}^{n - 1} \log(1 - \kappa \cos(\frac{2\pi l}{n}))    
\end{equation}
Differentiating $\Tilde{\gamma}_n$ with respect to $\kappa$, we get
\begin{equation}
    \frac{d \Tilde{\gamma}_n}{d \kappa} = \frac{n}{1 - \kappa} - \sum_{l=0}^{n-1} \frac{\cos(\frac{2\pi l}{n})}{1 - \kappa \cos(\frac{2\pi l}{n})} = n f(0) - \sum_{l=0}^{n-1} f(\frac{2\pi l}{n}),
\end{equation}
with 
\begin{equation}
    f(\theta) = \frac{\cos(\theta)}{1 - \kappa \cos(\theta)}.
\end{equation}
Since $f(\theta)$ has a period of $2\pi$, we can find its Fourier modes
\begin{equation}
    f_k = \int_{0}^{2\pi} \frac{d\theta}{2\pi} f(\theta) e^{-i k\theta},~~~\text{for}~k \in \mathbb{Z}.
\end{equation}
The above integral can be done by introducing a complex variable $z \equiv e^{i \theta}$, and we can write the integral for $f_k$ as a contour integral along a unit circle in the complex plane
\begin{equation}
    f_k = \oint \frac{dz}{2\pi i} \frac{z^{-k - 1}(z^2 + 1)}{2z - \kappa z^2 - \kappa}.
\end{equation}
The denominator of the integrand has two poles at
\begin{equation} \label{eq: z_minus}
    z_{\pm} = \frac{1 \pm \sqrt{1 - \kappa^2}}{\kappa},
\end{equation}
where $z_{-}$ lies in the unit circle. For $k \geq 0$, there is also another pole at $z = 0$ coming from the numerator. It is sufficient to calculate $f_k$ for $k \leq 0$, and then the expression for $k \geq 1$ is obtained from $f_k = f_{-k}$. Then, $f_k$ are given by
\begin{equation}
    f_k = -\frac{1}{\kappa} \delta_{k,0} + \frac{z_{-}^{\abs{k}}}{\kappa\sqrt{1 - \kappa^2}}.
\end{equation}
\indent Then
\begin{equation}
\begin{split}
    &\sum_{l=0}^{n-1} f(\frac{2\pi l}{n}) \\
    &= \sum_{k \in \mathbb{Z}} f_k \sum_{l = 0}^{n-1} e^{i k \frac{2\pi l }{n}} \\
    &= n \sum_{k \in \mathbb{Z}} f_{nk} \\
    &= n f_0 + 2n\sum_{k = 1}^{\infty} f_{nk} \\
    &= nf_0 + \frac{2n}{\kappa\sqrt{1 - \kappa^2}} \frac{z_{-}^n}{1 - z_{-}^{n}}
\end{split}
\end{equation}
\begin{equation}
    f(0) = \sum_{k \in \mathbb{Z}} f_k = f_0 + 2\sum_{k = 1}^{\infty} f_k = f_0 + \frac{2}{\kappa\sqrt{1 - \kappa^2}}\frac{z_{-}}{1 - z_{-}}
\end{equation}
$\frac{d\Tilde{\gamma}_n}{d\kappa}$ is then given by
\begin{equation}
    \frac{d\Tilde{\gamma}_n}{d\kappa} =  \frac{2n}{\kappa \sqrt{1 - \kappa^2}} \left( \frac{z_{-}}{1 - z_{-}} - \frac{z_{-}^n}{1 - z_{-}^{n}} \right).
\end{equation}
Noticing that $\frac{dz_{-}}{d\kappa} = \frac{z_{-}}{\kappa \sqrt{1 - \kappa^2}}$ and $\frac{d\ln(1 - z_{-})}{d\kappa} = -\frac{1}{\kappa \sqrt{1 - \kappa^2}} \frac{z_{-}}{1 - z_{-}}$, we can integrate both sides of the equation over $\kappa$ and obtain
\begin{equation}
    \Tilde{\gamma}_n = -2n \ln(1 - z_{-}) + 2 \ln(1 - z_{-}^n).
\end{equation}
\indent The subleading term $\gamma_n$ is finally obtained as
\begin{equation}
    \gamma_n =-\frac{1}{2(n - 1)} \Tilde{\gamma}_n = \frac{n}{n - 1} \ln(1 - z_{-}) - \frac{1}{n - 1} \ln(1 - z_{-}^{n}).
\end{equation}
The replica limit $n \xrightarrow{} 1$ is calculated as
\begin{equation}
    \begin{split}
        \gamma_1 &=  \ln(1 - z_{-}) - \lim_{n \xrightarrow{} 1} \frac{1}{n - 1}\ln(\frac{1 - z_{-}^n}{1 - z_{-}}) \\
        &= \ln(1 - z_{-}) - \frac{z_{-}}{z_{-} - 1} \ln(z_{-}).
    \end{split}
\end{equation}

\section{Details of calculating the correlation matrices} \label{app: calculation of correlation matrices}
To calculate correlation functions of the Hamiltonian in Eq. \eqref{eq: k_space_coupled_HO}, we use the Bogoliubov transformation by introducing bosonic operators $a_k$ and $b_k$. The bosonic operators are related to the harmonic oscillator operators by
\begin{equation}
    \begin{split}
        &a_k \equiv \sqrt{\frac{a^2 \omega_k}{2K_c'}} \left( \widetilde{q}_{c,k} + \frac{iK_c'}{a^2 \omega_k} \widetilde{p}_{c,k} \right),~~~a^{\dagger}_k \equiv \sqrt{\frac{a^2 \omega_k}{2 K_c'}} \left( \widetilde{q}_{c,-k} - \frac{i K_c'}{a^2 \omega_k} \widetilde{p}_{c,-k} \right) \\
        &b_k \equiv \sqrt{\frac{a^2 K_s \omega_k}{2}} \left( \widetilde{q}_{s,k} + \frac{i}{a^2 K_s \omega_k} \widetilde{p}_{s,k} \right),~~~b^{\dagger}_k \equiv \sqrt{\frac{a^2 K_s \omega_k}{2}} \left( \widetilde{q}_{s,-k} - \frac{i}{a^2 K_s \omega_k} \widetilde{p}_{s,-k} \right).
    \end{split}
\end{equation}
The Hamiltonian in second quantized form becomes
\begin{equation}
    \begin{split}
        H = \omega_0 \left( a^{\dagger}_0 a_0 + b_{0}^{\dagger}b_0 \right) + \sum_{k=1}^{N - 1} \left[ \omega_k \left(a^{\dagger}_k a_k + b^{\dagger}_{k} b_k \right) + \Delta \sqrt{\frac{K_c'}{K_s}} \omega_k \left( a_{k}^{\dagger} b_{-k}^{\dagger} + a_{k} b_{-k} \right) \right]
    \end{split}
\end{equation}
Finally we can diagonlize the finite momentum part of the Hamiltonian using a Bogoliubov transformation. The quasiparticle operators are given by
\begin{equation}
\begin{split}
    &a_{k}^{\dagger} =  f_{+} c_{k}^{\dagger} + f_{-} d_{-k},~~b_{-k}^{\dagger} = f_{+}d_{-k}^{\dagger} + f_{-} c_{k},~~ \forall k \neq 0 \\
    &\text{with}~f_{\pm}^2 = \frac{1}{2}\left( (1 -\Delta^2 \frac{K_c'}{K_s})^{-0.5} \pm 1 \right)
\end{split}
\end{equation}
The Hamiltonian in this diagonal basis reads
\begin{equation}
    H = \omega_0 \left( a^{\dagger}_0 a_0 + b_{0}^{\dagger}b_0 \right) +  \sum_{k = 1}^{N - 1} \sqrt{1 -\Delta^2 \frac{K_c'}{K_s}} \omega_k \left(c_k^{\dagger} c_{k} + d_{k}^{\dagger} d_k \right).
\end{equation}
\indent Now we are ready to compute all the correlation functions in this Gaussian theory. Before computing the two correlation matrices $[Q_s]_{rs} \equiv \expval{q_{A,r}q_{A,s}}$ and $[P_s]_{rs} \equiv \expval{p_{A,r}p_{A,s}}$, we first compute the harmonic oscillator correlations in the momentum space. They are given by
\begin{equation}
    \begin{split}
        \expval{\widetilde{q}_{s,0} \widetilde{q}_{s, 0}} &= \frac{1}{2a^2 K_s \omega_0}  \expval{ (b_{0}^{\dagger} + b_0 )(b_{0}^{\dagger} + b_{0})} = \frac{1}{2a^2 K_s \omega_0}  \expval{ b_0 b_0^{\dagger}} = \frac{1}{2a^2 K_s \omega_0}
    \end{split}
\end{equation}

\begin{equation}
    \begin{split}
        \expval{\widetilde{q}_{s,k} \widetilde{q}_{s, k'}} &= \frac{1}{2a^2 K_s} \sqrt{\frac{1}{\omega_k \omega_{k'}}} \expval{ (b_{-k}^{\dagger} + b_k )(b_{-k'}^{\dagger} + b_{k'})} \\
        &= \frac{1}{2a^2 K_s} \sqrt{\frac{1}{\omega_k \omega_{k'}}} \left( \expval{b_{-k}^{\dagger} b_{-k'}^{\dagger}} + \expval{b_{-k}^{\dagger} b_{k'}} + \expval{b_k b_{-k'}^{\dagger}} + \expval{b_k b_{k'}} \right) \\
        &= \frac{1}{2a^2 K_s} \sqrt{\frac{1}{\omega_k \omega_{k'}}}  \left(\expval{b_{-k}^{\dagger} b_{k'}} + \expval{b_k b_{-k'}^{\dagger}} \right) \\
        &= \frac{1}{2a^2 K_s} \sqrt{\frac{1}{\omega_k \omega_{k'}}} \left( \expval{(f_+ d_{-k}^{\dagger} + f_{-} c_k) (f_+ d_{k'} + f_{-} c_{-k'}^{\dagger})} + \expval{(f_+ d_k + f_{-} c_{-k}^{\dagger})(f_+ d_{-k'}^{\dagger} + f_{-} c_{k'})} \right) \\
        &= \frac{1}{2a^2 K_s} \sqrt{\frac{1}{\omega_k \omega_{k'}}} \left( f_{-}^2 \expval{c_k c_{-k'}^{\dagger}} + f_{+}^2 \expval{d_k d_{-k'}^{\dagger}} \right) \\
        &= \frac{1}{2a^2 K_s} \sqrt{\frac{1}{\omega_k \omega_{k'}}} \sqrt{\frac{1}{1 - \Delta^2\frac{K_c'}{K_s}}} \delta_{k, -k'},
    \end{split}
\end{equation}
where to go from the second line to the third line we use $\expval{a_{-k}^{\dagger} a_{-k'}^{\dagger}} = \expval{a_k a_{k'}} = 0$. Similar calculation is done for $\expval{\widetilde{p}_{s,k} \widetilde{p}_{s, k'}}$,
\begin{equation}
    \begin{split}
        \expval{\widetilde{p}_{s,0} \widetilde{p}_{s, 0}} &= -\frac{1}{2}a^2 K_s \omega_0 \expval{(b_{0}^{\dagger} - b_0) (b_{0}^{\dagger} - b_{0})} = \frac{1}{2}a^2 K_s \omega_0 \expval{ b_0 b_0^{\dagger}} = \frac{1}{2}a^2 K_s \omega_0
    \end{split}
\end{equation}

\begin{equation}
    \begin{split}
        \expval{\widetilde{p}_{s,k} \widetilde{p}_{s, k'}} &= -\frac{1}{2}a^2 K_s \sqrt{\omega_k \omega_{k'}} \expval{(b_{-k}^{\dagger} - b_k) (b_{-k'}^{\dagger} - b_{k'})} \\
        &= \frac{1}{2}a^2 K_s \sqrt{\omega_k \omega_{k'}} \left( \expval{b_{-k}^{\dagger}b_{k'}} + \expval{b_k b_{-k'}^{\dagger}} \right) \\
        &= \frac{1}{2}a^2 K_s \sqrt{\omega_k \omega_{k'}} \sqrt{\frac{1}{1 - \Delta^2\frac{K_c'}{K_s}}} \delta_{k, -k'}.
    \end{split}
\end{equation}
Using the above expressions for $\expval{\widetilde{q}_{s,k} \widetilde{q}_{s, k'}}$ and $\expval{\widetilde{p}_{s,k} \widetilde{p}_{s, k'}}$, the matrix elements of $Q_A$ and $P_A$ are thus given by
\begin{equation}
    \begin{split}
        \expval{q_{s,m}q_{s,n}} &= \frac{1}{N} \sum_{k, k'} e^{2\pi i k m / N} e^{2 \pi i k' n / N} \expval{\widetilde{q}_{s, k} \widetilde{q}_{s, k'}} \\
        &= \frac{1}{2a^2 K_s \omega_0 N} + \frac{1}{2a^2 K_s N} \sqrt{\frac{1}{1 - \Delta^2 \frac{K_c'}{K_s}}} \sum_{k, k' \neq 0}\sqrt{\frac{1}{\omega_k \omega_{k'}}}  e^{2\pi i k m / N} e^{2 \pi i k' n / N} \delta_{k, -k'} \\
        &= \frac{1}{2a^2 K_s \omega_0 N} +  \frac{1}{2a^2 K_s N} \sqrt{\frac{1}{1 - \Delta^2 \frac{K_c'}{K_s}}} \sum_{k = 1}^{N - 1} \frac{1}{\omega_k} e^{2\pi i k (m -n ) / N}
    \end{split}
\end{equation}
\begin{equation}
    \begin{split}
        \expval{p_{s,m}p_{s,n}} &= \frac{1}{N} \sum_{k, k'} e^{2\pi i k m / N} e^{2 \pi i k' n / N} \expval{\widetilde{q}_{s, k} \widetilde{q}_{s, k'}} \\
        &= \frac{a^2 K_s \omega_0}{2N} + \frac{a^2 K_s}{2N} \sqrt{\frac{1}{1 - \Delta^2 \frac{K_c'}{K_s}}} \sum_{k,k' \neq 0} \sqrt{\omega_k \omega_k'} e^{2\pi i km / N} e^{2\pi i k'n/N} \delta_{k, -k'} \\
        &= \frac{a^2 K_s \omega_0}{2N} + \frac{a^2 K_s}{2N} \sqrt{\frac{1}{1 - \Delta^2 \frac{K_c'}{K_s}}} \sum_{k=1}^{N - 1} \omega_k e^{2\pi k i (m -n) / N}.
    \end{split}
\end{equation}
\section{Weak charge measurement followed by feedback} \label{app:charge weak measurement}
In a real experimental setup, we would couple ancilla qubits to physical degrees of freedom and then perform measurements to ancilla qubits. Since the coupling with the ancilla qubits is not always perfect, it would be natural to relax the strong projective measurements to weak-measurements\cite{nielsen2002quantum}. For a generic weak-measurement, the wavefunction would not completely collapse toward the measurement outcome. In this case, it is worth studying the stability of the measurement followed by feedback protocol.\newline
\indent We use two approaches to study projective measurement followed by feedback. The first approach is a calculation based on lattice and the second approach is a field-theoretic description of the weak-measurement followed by feedback process. Both approaches show that as long as we slightly tune away from the projective measurement limit, the spin-spin correlation would decay exponentially.
\subsection{Weak-measurement of Fermion Number Parity and Feedback Protocol} \label{sec: weak-measurement lattice protocol}
Consider the input state $\ket{\psi_0}$ is free spinful fermions on a 1d chain. We perform weak-measurements to the fermion number parity $\hat{\Gamma}_i \equiv (-1)^{{\hat{n}_{i}}}$, where $\hat{n}_{i} \equiv \hat{n}_{i,\uparrow} + \hat{n}_{i, \downarrow}$ is the total fermion occupation on each site $i$, and apply single-site Pauli-Z feedback based on the measurement outcomes $m = \{m1, m2,... \}$. Such a protocol can be described by the following weak-measurement operator $ \hat{P}_{\{m\}}(\mu)$ and feedback unitary $\hat{U}_{\{m\}}$
\begin{equation}
    \hat{P}_{\{m\}}(\mu) = \prod_i \frac{e^{-\mu(\frac{1 - m_i  \hat{\Gamma}_{i}}{2})}}{\sqrt{1 + e^{-2\mu}}},~~~~~\hat{U}_{\{m\}} = \prod_{j} (\hat{S}_{j}^{z})^{\sum_{i \leq j} \frac{1 - m_i}{2}}.
\end{equation}Here $\mu$ controls the weak-measurement strength and when $\mu \xrightarrow{} \infty$ we restore the strong projective measurement limit that is discussed in the main text. \newline
\indent Given the initial pure density matrix $\rho_0 \equiv \ket{\psi_0}\bra{\psi_0}$, the entire ensemble obtained from weak-measurements and feedback is given by $\rho = \sum_{\{m\}}\hat{U}_{\{m\}}\hat{P}_{\{m\}}\rho_0 \hat{P}_{\{m\}}\hat{U}_{\{m\}}^{\dagger} $, where we have averaged over all possible measurement outcomes. Then we are ready to compute correlation functions in this mixed state $\rho$. We notice that under the unitary feedback $\hat{S}^{x}$ transforms in the following way: $\hat{U}_{m}^{\dagger}\hat{S}^{x}_j \hat{U}_m = \hat{S}^{x}_j (-1)^{\sum_{i \leq j} \frac{1 - m_i}{2}} =\hat{S}^{x}_j (\prod_{i \leq j}m_i) $. We can then evaluate the $\hat{S}^{x}_i \hat{S}^{x}_j$ correlation with respect to $\rho$ as:
\begin{align}
\begin{split}
    \Tr(\rho \hat{S}^{x}_i \hat{S}^{x}_j) &= \sum_{\{m\}} \Tr(\hat{U}_{\m}^{\dagger}\hat{P}_{\m}(\mu)\rho_0 \hat{P}_{\m}\hat{U}_{\m} \hat{S}^{x}_i \hat{S}^{x}_j) \\
    &= -\sum_{\m} \Tr(\hat{P}_{\m} \rho_0 \hat{P}_{\m} \hat{S}^{x}_i (\prod_{k=i+1}^{j-1} m_k) \hat{S}^{x}_j ) \\
    &= - \Tr( \rho_0  \hat{S}^{x}_{i} \big[\sum_m \hat{P}_{m}(\prod_{k=i+1}^{j-1} m_k) \hat{P}_{m}\big] \hat{S}^{x}_{j} )
    \end{split}
\end{align}
The term in the bracket can be evaluated as
\begin{align}
\begin{split}
    \sum_{\{m\}} \hat{P}_{\{m\}}(\prod_{k=i+1}^{j-1} m_k) \hat{P}_{\m} &=\prod_{k=i+1}^{j-1} \sum_{m_k = \pm 1} m_k \frac{e^{-\mu(1 - m_k  \hat{\Gamma}_{k})}}{1 + e^{-2\mu}} \\
    &= \prod_{k=i+1}^{j-1} \frac{e^{-\mu(1 - \hat{\Gamma}_{k})} - e^{-\mu(1 + \hat{\Gamma}_{k})}}{1 + e^{-2\mu}}\\
    &= \prod_{k=i+1}^{j-1} \frac{1 - e^{-2\mu}}{1 + e^{-2\mu}} \hat{\Gamma}_k \\
    &= (\tanh[{\mu}])^{j - i - 2} \prod_{k=i+1}^{j-1}\hat{\Gamma}_k.
    \end{split}
\end{align}
Therefore the $\hat{S}^{x}_{i} \hat{S}^{x}_{j}$ correlation equals to
\begin{align}
\begin{split}
    &\Tr(\rho \hat{S}^{x}_{i} \hat{S}^{x}_{j})\\
    &= -(\tanh[{\mu}])^{j - i - 2} \Tr(\rho_0 \hat{S}^{x}_{i} (\prod_{k=i+1}^{j-1}{\hat{\Gamma}}_k)  \hat{S}^{x}_{j}) \\
    &\sim \tanh[\mu]^{\abs{j - i}} \frac{1}{\abs{j - i}}
    \end{split}
\end{align}
  where we only keep the leading singularity at finite wavevector. We have used the fact that the string operator $\hat{S}^{x}_{i} (\prod_{k=i+1}^{j-1}{\hat{\Gamma}}_k)  \hat{S}^{x}_{j}$ decays as $\frac{1}{\abs{j - i}}$ in LL\cite{kruis2004geometry}. Therefore, under measurements and feedback, the $\tanh[\mu]^{\abs{j - i}}$ factor signals that the $S^{x}_{i} S^{x}_{j}$ correlation decays exponentially.

\subsection{Continuum field theory description of the mixed state density matrix}
In this section, we study the consequences of weak-measurement and feedback within the continuum field theory of the LL. Notice that instead of weak-measuring the fermion number parity, we consider weak-measuring the fermion number density, which is easier to handle in continuum. We denote such a weak-measurement operator of the fermion number density by $\hat{P}_m$ given the measurement outcome $m(x)$, which has been coarse-grained to a scalar field. The feedback unitary parametrized by $\sin[\pi\delta]$ is given by
\begin{equation}
    \hat{U}_{m}(\delta) \equiv \prod_{j} \exp\left[ i \pi \sin[\pi \delta] \left(\sum_{i<j}m_{i} \right)\hat{S}^{z}_{j}\right].
\end{equation}
\indent In the continuum limit, the continuum action of $\hat{\mathcal{P}}_{m}$ on a basis state of $\pc$ and $\ps$ can be written as
\begin{equation}
    \hat{\mathcal{P}}_{m}(\mu)\ket{\pc, \ps} \propto e^{-\mu \int dy [m(y) - n(y)]^2} \ket{\pc, \ps} \equiv e^{-S_{int}}\ket{\pc, \ps},
\end{equation}
where the scalar field $n(y)$ represents the charge density and is given by replacing the $\hat{\phi}_c$ and $\hat{\phi}_s$ operators in Eq.(\ref{eq:bosonized_density}).\newline

\indent The action of the unitary operator $\hat{\mathcal{U}}_{m}(\delta)$ in the continuum is given by
\begin{align} 
    \hat{\mathcal{U}}_{m}(\delta)\ket{\pc,\vts} = \ket{\pc,\vts + \sqrt{\frac{\pi}{2}}\sin[\delta\pi]\int_{-\infty}^{x}dy~m(y)}.
\end{align}
This is consistent with the following transformation that one would get on the lattice $\hat{U}_{m}^{\dagger} (\delta)\hat{S}^{+}_j \hat{U}_m(\delta) = \hat{S}^{+}_j e^{i \pi \sin[\pi \delta]\sum_{i \leq j} \frac{1 - m_i}{2}} $.\newline
\indent Given $\hat{\mathcal{P}}_{m}(\mu)$ and $\hat{\mathcal{U}}_{m}(\delta)$, the mixed state $\rho$ can be written as
\begin{align}
    \begin{split}
        \rho &= \sum_{\m}\hat{\mathcal{P}}_{\m}\hat{\mathcal{U}}_{\m}\rho_0 \hat{\mathcal{U}}_{\m}^{\dagger} \hat{\mathcal{P}}_{\m} \\
        &\propto \int \mathcal{D}\{...\} e^{-\frac{1}{2}S_{\partial}[\pc', \vts'] - \frac{1}{2}S_{\partial}
        [\pc,\vts]}
        \hat{\mathcal{P}}_{\m} \hat{\mathcal{U}}_{\m}\ket{\pc', \vts'}\bra{\pc, \vts}\hat{\mathcal{U}}_{\m} \hat{\mathcal{P}}_{\m}\\
        &\propto \int \mathcal{D}\{...\} e^{-\frac{1}{2}S_{\partial} - \frac{1}{2}S_{\partial}'} \ket{\pc', \ps'}\bra{\pc', \ps'}\hat{\mathcal{P}}_{\m} \ket{\pc', \vts' + \sqrt{\frac{\pi}{2}} \sin[\delta\pi]\int_{-\infty}^{x}dy~m(y)} \\
        &\bra{\pc, \ts + \sqrt{\frac{\pi}{2}} \sin[\delta\pi]\int_{-\infty}^{x}dy~m(y)} \hat{\mathcal{P}}_{m}\ket{\pc,\ps}\bra{\pc,\ps} \\
        &\propto\int \mathcal{D}\{...\}e^{-\frac{1}{2}S_{\partial} - \frac{1}{2}S_{\partial}' - S_{int} - S_{int}' + i\int dx (\nabla \vts' + \sqrt{\frac{\pi}{{2}}}\sin[\delta\pi]m)\ps' - i\int dx(\nabla \vts + \sqrt{\frac{\pi}{{2}}}\sin[\delta\pi]m)\ps} \ket{\pc', \ps'}\bra{\pc, \ps}
    \end{split}
\end{align}
    where $S_{\partial}$ is the boundary action given in Eq.\eqref{eq:boundary_action}. The symbol $\mathcal{D}\{...\}$ denotes the functional integral over all the fields. In the third line, we act $\hat{\mathcal{U}}_{m}(\delta)$ ($\hat{\mathcal{U}}_{m}(\delta)^{\dagger}$) to the basis state $\ket{\varphi_c', \vartheta_s'}$($\bra{\varphi_c, \vartheta_s}$). In the third line,  we have also inserted two resolutions of identity $\int \mathcal{D}\ps' \ket{\ps'}\bra{\ps'}$ and $\int \mathcal{D}\ps \ket{\ps}\bra{\ps}$. In the fourth line, we have used the fact $\braket{\ps'|\vts' + \sqrt{\frac{\pi}{2}} \sin[\delta\pi]\int_{-\infty}^x dy ~m(y)} = e^{i\int dx (\nabla \vts' + \sqrt{\frac{\pi}{2}} \sin[\delta\pi] m)\ps'}$ and act $\hat{\mathcal{P}}_{m}(\mu)$ on its diagonal basis $\ket{\varphi_c, \vartheta_s}$.
Then we can first integrate out $\vts$, which amounts to write $S_{\partial}$ in terms of $\pc$ and $\ps$. The same also applies to $S_{\partial}'$ when we integrate out $\vts'$. Secondly, we integrate out the measurement outcome field $m(x)$. The part of the action involving $m(x)$ reads
\begin{equation}
    S[m(x)] = \int dx \big[ 2\mu m(x)^2 + m(x) (-2\mu n - 2\mu n' + i\sqrt{\frac{\pi}{2}} \sin[\delta\pi]\ps -i\sqrt{\frac{\pi}{2}} \sin[\delta\pi]\ps' ) + \mu n^2 + \mu n'^2 \big]
\end{equation}
Upon integrating out $m(x)$, the effective action we get is
\begin{equation}
\begin{split}
    S_{\mathrm{eff}} &= -\int dx ~\frac{\big[-2\mu n - 2\mu n' + i\sqrt{\frac{\pi}{2}} \sin[\delta\pi]\ps -i\sqrt{\frac{\pi}{2}} \sin[\delta\pi]\ps'  \big]^2}{8\mu} + \int dx (\mu n^2 + \mu n'^2) \\
    &= \int dx \Big[ \frac{\mu}{\pi}(\nabla \pc)^2 + \frac{\mu}{\pi}(\nabla \pc')^2  -\frac{2\mu}{\pi} \nabla\pc \nabla\pc' + \frac{\mu}{2\pi^2}\cos[\sqrt{2\pi}\ps]^2 + \frac{\mu}{2\pi^2}\cos[\sqrt{2\pi} \ps']^2 \\
    &-\frac{\mu}{\pi^2}\cos[\sqrt{2\pi}(\pc - \pc')]\cos[\sqrt{2\pi}\ps]\cos[\sqrt{2\pi}\ps'] + \frac{i}{2} \sin\left[\delta\pi\right](\nabla \pc + \nabla \pc')(\ps -\ps') + \frac{\pi\sin[\delta\pi]^2}{16\mu} (\ps - \ps')^2 \Big]
\end{split}
\end{equation}
where we neglect all the oscillating terms which would vanish by doing the integral. We see that the charge sector and the spin sector are mixed together. The density matrix can be compactly written as
\begin{equation}\label{eq:weak-measurement_density_matrix}
    \rho \propto \int \mathcal{D}\{\pc,\pc', \ps, \ps'\} e^{-\frac{1}{2}S_{\partial}[\pc,\ps] - \frac{1}{2}S_{\partial}[\pc', \ps'] - S_{\mathrm{eff}}[\pc, \pc', \ps, \ps']} \ket{\pc', \ps'}\bra{\pc, \ps}.
\end{equation}

\subsection{Correlation functions}We first compute the expectation value of the simplest correlation function $e^{i\sqrt{2\pi}(\hat{\theta}_s(x) - \hat{\theta}_s(0))}$ in the spin sector.
\begin{equation}
    \begin{split}
        &\left\langle e^{i\sqrt{2\pi}(\hat{\theta}_s(x) - \hat{\theta}_s(0))}\right\rangle =\int \mathcal{D}\vts\Tr\Big[ \rho \ket{\vts}\bra{\vts} e^{i\sqrt{2\pi}(\vts(x) - \vts(0))} \Big] 
    \end{split}
\end{equation}
where we have inserted a resolution of identity in the $\hts$ eigenbasis, so that we can replace the $\hat{\theta}_s$ operator by its eigenvalue $\vts$. Plugging in the density matrix $\rho$ in Eq.\eqref{eq:weak-measurement_density_matrix}, we get
\begin{equation}\label{eq: F13}
    \begin{split}
        \left\langle e^{i\sqrt{2\pi}(\hat{\theta}_s(x) - \hat{\theta}_s(0))}\right\rangle &\propto \int \mathcal{D}\{...\} e^{-\frac{1}{2}S_{\partial} - \frac{1}{2}S_{\partial}' - S_{\mathrm{eff}} + i\int dx \nabla\vts(\ps - \ps') + i\sqrt{2\pi}\int_{0}^{x} dy \nabla \vts(y)}\braket{\pc|\pc'}
    \end{split}
\end{equation}
where we have used the fact that $\braket{\ps|\vts} = e^{2i\int dx \nabla \vts \ps}$. We then integrate out the $\vts(x)$ field and the effect is to enforce $\ps'(y) = \ps(y) + \sqrt{2\pi} T_{0,x}(y)$, where $T_{0,x}(y) = 1$ for $0 \leq y \leq x $ and $T_{0,x}(y) = 0$ otherwise. We also have the constraint $\pc = \pc'$ from $\braket{\pc|\pc'}$. Then the actions $S_{\partial}$, $S_{\partial}'$, and $S_{\mathrm{eff}}$ appeared on the right-hand side of Eq.\eqref{eq: F13} can be written only in terms of $\pc$ and $\ps$
\begin{equation}
    \begin{split}
        \frac{1}{2}S_{\partial}[\pc, \ps] + \frac{1}{2}S_{\partial}[\pc', \ps'] &= S_{\partial}[\pc, \ps] + \frac{\sqrt{2\pi}}{K_s} \int \frac{dq}{2\pi}\abs{q}\ps \Tilde{T}_{0,x}^{*} +\frac{\pi}{K_s} \int \frac{dq}{2\pi}\abs{q}\abs{\Tilde{T}_{0,x}}^2 \\
        S_{\mathrm{eff}} &=  \int dy \Big[ i\sin[\delta\pi] \sqrt{2\pi}\nabla\pc T_{0,x} + \frac{\sin^2[\delta\pi]}{8\mu}T_{0,x}^2 \Big] \\
        &= -\int \frac{dq}{2\pi}\sin[\delta\pi] \sqrt{2\pi} q \pc \Tilde{T}_{0,x}^{*} + \int dy\frac{ \sin^2[\delta\pi]}{8\mu} \abs{\Tilde{T}_{0,x}}^2 
    \end{split}
\end{equation}
We can then integrating out $\pc$ and $\ps$ in Eq.\eqref{eq: F13}, and the result reads
\begin{equation}
\begin{split}
    \left\langle e^{i\sqrt{2\pi}(\hat{\theta}_s(x) - \hat{\theta}_s(0))}\right\rangle & \propto \exp{\int \frac{dq}{2\pi}\Big[ -\frac{\pi \abs{q}}{2K_s} - \frac{\sin^2[\delta\pi] \pi K_c \abs{q}}{2}\Big]\abs{\Tilde{T}_{0,x}}^2 - \int dy \frac{\sin^2[\delta\pi]}{8\mu}\abs{{T}_{0,x}}^2} \\ 
    &\sim  \exp{-\ln(x)(\frac{1}{K_s} + \sin^2[\delta\pi] K_c - \frac{\sin^2[\delta\pi]}{8\mu}x} \\
    & \sim \frac{e^{-\sin^2[\delta\pi] x / 8\mu}}{x^{1/K_s + \sin^2[\delta\pi]K_c}}
    \end{split}
\end{equation}
where from the first to second line we consider only small $q$ (long wavelength) contribution. Then the integral over $q$ can be computed as 
$\int_{-\Lambda}^{\Lambda} \frac{dq}{2\pi} \abs{q} \abs{\Tilde{T}_{0,x}}^2 = 4\int_{-\Lambda}^{\Lambda} \frac{dq}{2\pi}\frac{\sin^2(qx/2)}{\abs{q}} \simeq \frac{2}{\pi}\ln(x) + \mathrm{const}$. The final result shows that we are getting an exponentially decaying correlation.\newline
\indent The $\hat{O}_{+,2k_F}(x)\hat{O}_{-, 2k_F}(0)$ correlation that represents the $\mathrm{SDW}^{\pm}$ order can be computed in a similar way as $\left\langle e^{i\sqrt{2\pi}(\hat{\theta}_s(x) - \hat{\theta}_s(0))}\right\rangle$ and we will only present the final result here. In the long wavelength limit, it behaves as
\begin{equation}
    \left\langle \hat{O}_{+,2k_F}(x)\hat{O}_{-, 2k_F}(0) + \mathrm{h.c.} \right\rangle \sim \sum_{\sigma =\pm 1} \frac{\cos[2k_F x]e^{-\sin^2[\delta\pi]x / 8\mu}}{x^{(1 +\sigma \sin[\delta\pi])^2 K_c + 1/K_s}},
\end{equation}
which is again an exponentially-decaying correlation that matches the lattice calculation in Appendix. \ref{sec: weak-measurement lattice protocol}. At the strong measurement limit where $\mu \xrightarrow{} \infty$, the exponentially decaying factor becomes $\lim_{\mu \xrightarrow{} \infty} e^{-\sin^2[\delta\pi]x / 8\mu} = 1$. We thus restore the result in Eq .\eqref{eq:spin_up_spin_down_corr}.

\end{document}